\documentclass[apj,twocolumn]{openjournal}
\usepackage{graphicx}
\usepackage{xcolor}
\usepackage{amsmath}
\usepackage{amsfonts}
\usepackage{amssymb}
\usepackage{upgreek}
\usepackage[pdfborder={0 0 0}]{hyperref}
\usepackage{txfonts}
\usepackage{lipsum}
\usepackage{makecell} 
\usepackage{hyperref}
\hypersetup{
     colorlinks   = true,
     citecolor    = blue
}

\newcommand{\orcidauthor}[3]{\author{\href{http://orcid.org/#1}{#2 \openin1 Orcid-ID.png \ifeof1 \else \hskip2pt\includegraphics[width=9pt]{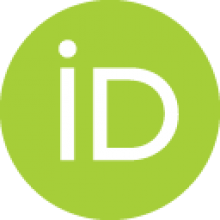}\fi}$^{#3}$}}
\begin{document}
\title{Testing Gravity with Binary Pulsars in the SKA Era}
\orcidauthor{0000-0001-9518-9819}{V.~Venkatraman Krishnan}{1}
\orcidauthor{0000-0002-1334-8853}{L.~Shao}{2,1,3}
\orcidauthor{0000-0003-3244-2711}{V.~Balakrishnan}{1}
\orcidauthor{0000-0002-2965-5911}{M.~Colom~i~Bernadich}{4,1}
\orcidauthor{0000-0000-0000-0000}{A.~Carelo}{4}
\orcidauthor{0000-0000-0000-0000}{A.~Corongiu}{4}
\orcidauthor{0000-0001-9434-3837}{A.~Deller}{5}
\orcidauthor{0000-0003-1307-9435}{P.~C.~C.~Freire}{1}
\orcidauthor{0000-0002-2822-1919}{M.~Geyer}{6}
\orcidauthor{0000-0001-7456-216X}{E.~Hackmann}{7}
\orcidauthor{0000-0002-3407-8071}{H.~Hu}{8,1}
\orcidauthor{0000-0002-3081-0659}{Z.~Hu}{9,2}
\orcidauthor{0000-0001-7990-8713}{J.~Kunz}{10}
\orcidauthor{0000-0002-4175-2271}{M.~Kramer}{1}
\orcidauthor{0000-0002-2953-7376}{K.~Liu}{11,12,1}
\orcidauthor{0000-0001-9208-0009}{M.~E.~Lower}{5}
\orcidauthor{0000-0000-0000-0000}{X.~Miao}{3,1}
\orcidauthor{0000-0001-5902-3731}{A.~Possenti}{4}
\orcidauthor{0000-0000-0000-0000}{D.~Perrodin}{4}
\orcidauthor{0000-0002-1602-4168}{D.~S.~Pillay}{1}
\orcidauthor{0000-0001-5799-9714}{S.~Ransom}{13}
\orcidauthor{0000-0002-9142-9835}{I.~Stairs}{14}
\orcidauthor{0000-0001-9242-7041}{B.~Stappers}{15}
\author{The SKA pulsar science working group}

\affiliation{$^{1}$Max-Planck-Institut f{\"u}r Radioastronomie, auf dem H{\"u}gel 69, 53121, Bonn, Germany }
\affiliation{$^{2}$Kavli Institute for Astronomy and Astrophysics, Peking University, Beijing
  100871, China}
\affiliation{$^{3}$National Astronomical Observatories, Chinese Academy of Sciences, Beijing 100012, China}
\affiliation{$^{4}$INAF-Osservatorio Astronomico di Cagliari, Via della Scienza 5, I-09047 Selargius, Italy}
\affiliation{$^{5}$Center for Astrophysics and Supercomputing, Swinburne University of Technology, Post Office Box 218, Hawthorn, VIC 3122, Australia}
\affiliation{$^{6}$High Energy Physics, Cosmology \& Astrophysics Theory (HEPCAT) Group, Department of Mathematics and Applied Mathematics, University of Cape Town,  Rondebosch 7701, South Africa}
\affiliation{$^{7}$Center of Applied Space Technology and Microgravity (ZARM), University of Bremen, Am Fallturm 2, 28359 Bremen, Germany}
\affiliation{$^{8}$Lohrmann Observatory, Technische Universit\"at Dresden, Mommsenstraße 13, 01062 Dresden, Germany}
\affiliation{$^{9}$Department of Astronomy, School of Physics, Peking University, Beijing 100871, China}
\affiliation{$^{10}$Institute of  Physics, University of Oldenburg,
D-26111 Oldenburg, Germany}
\affiliation{$^{11}$Shanghai Astronomical Observatory, Chinese Academy of Sciences, 80 Nandan Road, Shanghai 200030, China}
\affiliation{$^{12}$State Key Laboratory of Radio Astronomy and Technology, A20 Datun Road, Chaoyang District, Beijing, 100101, P. R. China}
\affiliation{$^{13}$National Radio Astronomy Observatory, 520 Edgemont Road, Charlottesville, VA 22901, USA}
\affiliation{$^{14}$Department of Physics and Astronomy, University of British Columbia, 6224 Agricultural Road, Vancouver, BC V6T 1Z1 Canada}
\affiliation{$^{15}$Jodrell Bank Centre for Astrophysics, Department of Physics and Astronomy, The University of Manchester, Manchester M13 9PL, United Kingdom}

\begin{abstract}
Binary (and trinary) radio pulsars are natural laboratories in space for understanding gravity in the strong field regime, with many unique and precise tests carried out so far, including the most precise tests of the strong equivalence principle and of the radiative properties of gravity. The Square Kilometre Array (SKA) telescope, with its high sensitivity in the Southern Hemisphere, will vastly improve the timing  precision of recycled pulsars, allowing for a deeper search of potential deviations from general relativity (GR) in currently known systems. A Galactic census of pulsars will, in addition, will yield the discovery of dozens of relativistic pulsar systems, including potentially pulsar -- black hole binaries, which can be used to test the cosmic censorship hypothesis and the ``no-hair'' theorem. Aspects of gravitation to be explored include tests of strong equivalence principles, gravitational dipole radiation, extra field components of gravitation, gravitomagnetism, and spacetime symmetries. In this chapter, we describe the kinds of gravity tests possible with binary pulsar and outline the features and abilities that SKA must possess to best contribute to this science.
\end{abstract}


\maketitle

\section{Introduction}

Even in the this era of routine gravitational wave (GW) detections from merging neutron stars (NSs) and black holes (BHs), pulsars have a unique and complementary role to play in the understanding of neutron star interiors and gravity. Pulsars stand a class apart in being the only strong-field gravity experiment that provides tests of gravity theories with precisions that are orders of magnitude better than their counterparts in many observables. The Square Kilometer Array (SKA) telescope will open up the next frontier for these tests by providing improved timing of known pulsars, and more excitingly, by finding more relativistic binaries that will improve the precision, nature and extent of these tests. 

In November 1915, Albert Einstein published the final paper that completed his theory of gravity and spacetime \citep{Einstein1915}, now known as general relativity (GR), fundamentally changing our view of gravity and spacetime. During its first fifty years, GR was considered a theoretical tour-de-force but one with little observational evidence. Only in the 1960s did technology usher in the remarkable field of experimental gravity (cf. reviews in \cite{mtw73,Will2018}). Over the subsequent sixty years, GR has passed all experimental tests in laboratories, in the Solar System and in various stellar systems \citep{Berti2015,Will2018}, particularly with state-of-the-art radio observations near black holes \citep{EHT2019_M87,EHT2022_SgrA} and high precision observations of radio pulsars  \citep{LorimerStairsLRR,FreireWex2024} -- the topic of this work.

The surprising 1967 discovery of radio pulsars, and the subsequent demonstration that they were neutron stars, was an extraordinary breakthrough, for several reasons. First, it demonstrated the existence of neutron stars. Although their structure had been calculated in detail in the late 1930s \citep{tol_1939,ov_1939}, it had not been fully established until then. Second, the association of fast-spinning pulsars with the Crab and Vela supernova remnants demonstrated that neutron stars (and pulsars) are the end stages of the lives of at least some very massive stars; indeed suggesting that the instantaneous loss of the enormous binding energy of the newly formed neutron star is the power source of supernovae, as had been hypothesized by \cite{bz_1934}. Third, their radio beams, which, akin to the sweeping signals of a rotating lighthouse creates apparent pulses, allow for the extremely precise monitoring of the spin of the star, and for the measurement of its precise position and proper motion. Critically for tests of gravity, the observed radio pulsations from NS in binaries also lead to precise measurements of the orbital motion.

The years 2024 and 2025 mark the 50th anniversary of the discovery of the first binary pulsar (a double neutron star system known as PSR B1913+16) and its publication \citep{ht75}. 
In the following years, precise radio timing of this system provided the first tests of GR for strongly self-gravitating objects, in this case neutron stars (NSs), and provided experimental confirmation of the existence of gravitational waves \citep{tfm79,TaylorWeisberg1982}. By the end of the 1980s, this experiment had confirmed that the system was losing orbital energy due to gravitational wave emission within 1\% of the rate predicted by GR \citep{TaylorWeisberg1989}.

The Hulse-Taylor system opened a new window on the Universe: not only did it confirm the existence of GWs and verify that they conform (within measurement accuracy) to the predictions of GR, but also the inevitable mergers among systems like PSR~B1913+16\footnote{Because of the energy loss from GW emission, the two neutron stars are now 180 m closer to each other than at the time of discovery in 1974; in 300 million years they will inevitably collide.} in the Universe provided guaranteed a source population for ground-based GW detectors, facilitating the construction of the Laser Interferometer Gravitational Wave Observatory (LIGO), Virgo, and KAGRA. This year, 2025, marks the 10th anniversary of the first direct GW detection, GW150914 \citep{GW150914} from two merging BHs. In the mean time, there is also a confirmed detection of GWs from two NS-NS mergers \citep{GW170817,GW170817_Integral,GW190425}.

Following the discovery of the first pulsar binary, discovery and tests of gravity theories from such systems have also advanced significantly. In particular, the timing of the double pulsar \cite{Burgay2003,Lyne2004} has resulted in a leap in precision of the measured orbital decay of a binary pulsar \cite{Kramer2021}. The discovery of a millisecond pulsar in a triple star system has provided an improvement of the test of the universality of free fall for strongly self-gravitating objects \citep{Archibald2018,Voisin2020}, which is three orders of magnitude more precise than any such previous tests. A set of  pulsar - white dwarf systems has significantly constrained the possibility of dipolar gravitational wave emission; these fundamental tests of the nature of gravitational waves have introduced tight constraints on alternative theories of gravity and have largely ruled out the phenomenon of  matter induced spontaneous scalarisation \citep{Zhao2022}. Further details on this can be found in section~\ref{sec:state_of_the_art}. 
The high sensitivity of SKA will allow the more precise  timing of known pulsars and will discover dozens of rare relativistic binary systems that will facilitate new and novel tests of gravity. The science case presented here builds upon the previous descriptions of SKA Key Science presented by
\citet{Cordes:2004,Kramer:2004} and \cite{Shao2015}.

While we limit ourselves to Galactic field stellar mass binaries in this chapter, complementary tests of gravity can be performed with a pulsar orbiting the supermassive black hole at the centre of our Galaxy  (See \citealt{Abbate2025_SKA_GalCen} ) and exotic gravitational laboratories can be unearthed through targeted searches of globular clusters  (See \citealt{Bagchi2025_SKA_GlobClust} ). Many techniques used here can also be used to obtain a measurement of the Moment of Inertia of neutron stars (See \citealt{Basu2025_SKA_EOS}).

\section{General Relativity and its extensions}
\label{sec:GR}

\subsection{Effects predicted by general relativity}

Although General Relativity (GR) is in a sense the most simple and consistent theory of gravity, it predicts a wealth of effects that can be confronted with experiments, and in particular with pulsar observations \citep{Will:2018bme}. As binary pulsars generate strong gravitational fields, such tests are complementary to weak field tests conducted in say, the Solar System. This circumstance is true for both the common (neutron star–white dwarf or double neutron star) systems and, even more so, for pulsar–black hole (PSRBH) binaries. In the latter case, the SKA will provide constraints on GR that are better than such as gravitational wave observations, Solar System experiments, stellar orbits around Sgr A*, or imaging with the Event Horizon Telescope (see also Figure \ref{fig:curvature}).

GR effects in binary systems can be generally divided into the following categories: effects on the orbits of the two masses, gravitational radiation, effects due to the non-negligible spin of the component stars, and effects on the signal propagation. For the particular case of PSRBH binaries, predictions on the nature of black holes, such as the cosmic censorship conjecture can also be tested\citep{Liu:2014uka}. We next explain these effects briefly.

\subsubsection{Effects on the orbits of the two component stars}
While in Newtonian gravity, binaries move on ellipses in a single orbital plane, this is no longer true in GR. The most prominent effect is the advance of the argument of periastron $\omega$, causing a relativistic precession of the ellipse along its orbital plane. For example, in the double neutron star system PSR J1946+2052, $\dot{\omega}$ amounts to about 25.6 degree per year \citep{Stovall:2018ApJ, Meng:2024yyh}, illustrating the enormous impact of relativistic effects in compact, eccentric binaries.\footnote{For comparison, Mercury's relativistic precession is about 43 arcseconds per century, at least four orders of magnitude lower than that has been observed with DNS systems!}

\subsubsection{Effects due to non-negligible spin of the component stars}

At higher orders, the spin of the component stars also has a gravitational influence on the orbit. First derived by \citet{Lense1918}, it causes an additional periastron advance as well as a precession of the orbital plane around the total angular momentum of the system; this results in a change of the orbital inclination $i$ and on the longitude of the ascending node $\Omega$. In the double pulsar system, the Lense-Thirring contribution to the advance of periastron is of the same order of magnitude as the second order post-Newtonian correction to $\dot{\omega}$, but they have not yet been measured independently \citep{Kramer2021}. Presently, the only binary pulsar system where evidence for the Lense-Thirring frame-dragging has been found is the system containing PSR J1141$-$6545 and a massive WD \citep{VenkatramanKrishnan2020}. Due to its unusual binary evolution, the WD was spun up to spin periods of only a few hundred seconds, which contributed to a precession of the orbit that is detectable via the variation of $i$ and its effect on the projected semi-major axis of the pulsar orbit, $x \equiv a_{\rm p} \sin i / c$ where $a_{\rm p}$ is the semi-major axis of the pulsar orbit \citep{VenkatramanKrishnan2020}.  The Lense-Thirring precession of a binary system consisting of a pulsar and a stellar-mass black hole was investigated in \citet{wk99}. Given some favourable configurations, the precession can be measured and used to derive the spin and even the quadrupole moment of the black hole \citep{lewk14}. The spin might be measurable in the near future if the high-mass companion to PSR~J0514$-$4002E \citep{Barr2024} turns out to be a fast-spinning black hole, {\em but only with the sensitivity of the SKA}.

Another relativistic effect due to the spin of the pulsar is the de-Sitter effect, also called geodetic precession. It originates from parallel transport of the spin vector of the pulsar along its orbit, which causes a precession if the spin is misaligned from the orbital angular momentum. In the double pulsar system this effect was responsible for pulsar B's radio emission beam to precess out of our line of sight in 2008. Although tests of geodetic precession are currently not competitive with the Gravity Probe B result of $0.28\%$ \citep{Everitt2011}, it provides a complementary test involving strongly self-gravitating objects 
\citep{Breton2008, Fonseca2014,Desvignes2019,Lower2024}. In addition to the geodetic precession, the spin-spin coupling, also called the Schiff effect \citep{Schiff1960}, will introduce another precession in a generally different direction; this effect, which is about two orders of magnitude smaller than the de-Sitter effect, has so far not been measured with pulsars. Geodetic precession also provides the possibility of providing a 2D view of the pulsar's emission geometry which is helpful in understanding pulsar emission and the magnetosphere. This prospect is discussed in \citep{Oswald2025_MAG}.

\subsubsection{Gravitational radiation}

The first ever (indirect) detection of GW emission in the Hulse-Taylor system was a major breakthrough in astrophysics and a triumph for GR \citep{tfm79,TaylorWeisberg1982,TaylorWeisberg1989}. After decades of observations, the orbital decay due to GW emission in this system has validated GR at the $0.16\%$ level \citep{WeisbergHuang2016}. Currently the best results are obtained from the timing of the double pulsar (see Section~\ref{sec:state_of_the_art}).

\subsubsection{Effects on signal propagation}

The gravitational field of the two-body system also affects the signal propagation. Special relativistic effects include the Doppler shift and aberration \citep{Stairs2003}. In GR, in addition, the gravitational redshift as seen by a distant observer has to be considered, which is (together with the second-order Doppler effect) encoded in the Einstein delay $\gamma$ \citep{DD86,DamourTaylor1992}. A prominent GR effect on the signal is the Shapiro delay, which accounts for 

the extra path length of the trajectory near a massive object.
In the pulsar timing model, the Shapiro delay is usually split into a range and a shape parameter, which are measured separately, if the timing precision, companion mass and inclination angle are favourable. The Shapiro delay is largest during superior conjunction, when both masses are approximately on our line of sight. For example, in the double pulsar system, which is seen almost edge-on from the Earth, the Shapiro delay amounts to $130\, \rm \mu s$ \citep{Kramer2021}. For such edge-on systems, the deflection of light due to the companion's gravitational field can also become relevant \citep{Schneider1990,Rafikov2005}, as well as the retardation effect due to the motion of the companion during the propagation of the signal \citep{Kopeikin1999,Rafikov2006}. The deflection of light also adds corrections to the above mentioned aberration \citep{Doroshenko1995,Rafikov2006b,Hu2022}. Delays on the signal propagation in PSRBH systems, due to the spin of the black hole, have been investigated for stellar masses \citep{wk99} and in extreme mass ratio systems \citep{Ben-Salem2022}, but are challenging to disentangle from the bending delay. 

\subsection{Testing GR and alternative gravity theories using pulsar timing}

In order to test for effects beyond GR, a set of
theory-independent ``post-Keplerian'' (PK) parameters are introduced
\citep{DD86,DamourTaylor1992} (see Table~\ref{tab:PK} for a collection
of the most important PK parameters relevant for this work).  These are phenomenologically added to the pulsar timing model, and hence have the advantage of being theory agnostic. Hence, PK parameters that are used to specify relativistic effects are at times also contaminated by other astrophysical and astrometric effects. For example, the observed orbital decay
parameter, $\dot P_\mathrm{b}$, has ``kinematic'' contributions, which result
from the proper motion of the system and the difference of Galactic
accelerations between that system and the Solar System Barycentre.
Here, unless otherwise stated, we assume that non-gravitational
contributions to the PK parameters are adequately corrected for or demonstrably negligible in
effect. If so, in the case of two well-separated masses with
negligible spin contributions, the PK parameters are functions of the
well-measured Keplerian parameters, the component masses, the equation
of state (EOS) of stellar matter, and the parameters describing the
gravitational theory \citep{DamourTaylor1992}.

\begin{figure}[t]
  \centering
  \includegraphics[width=0.5\textwidth]{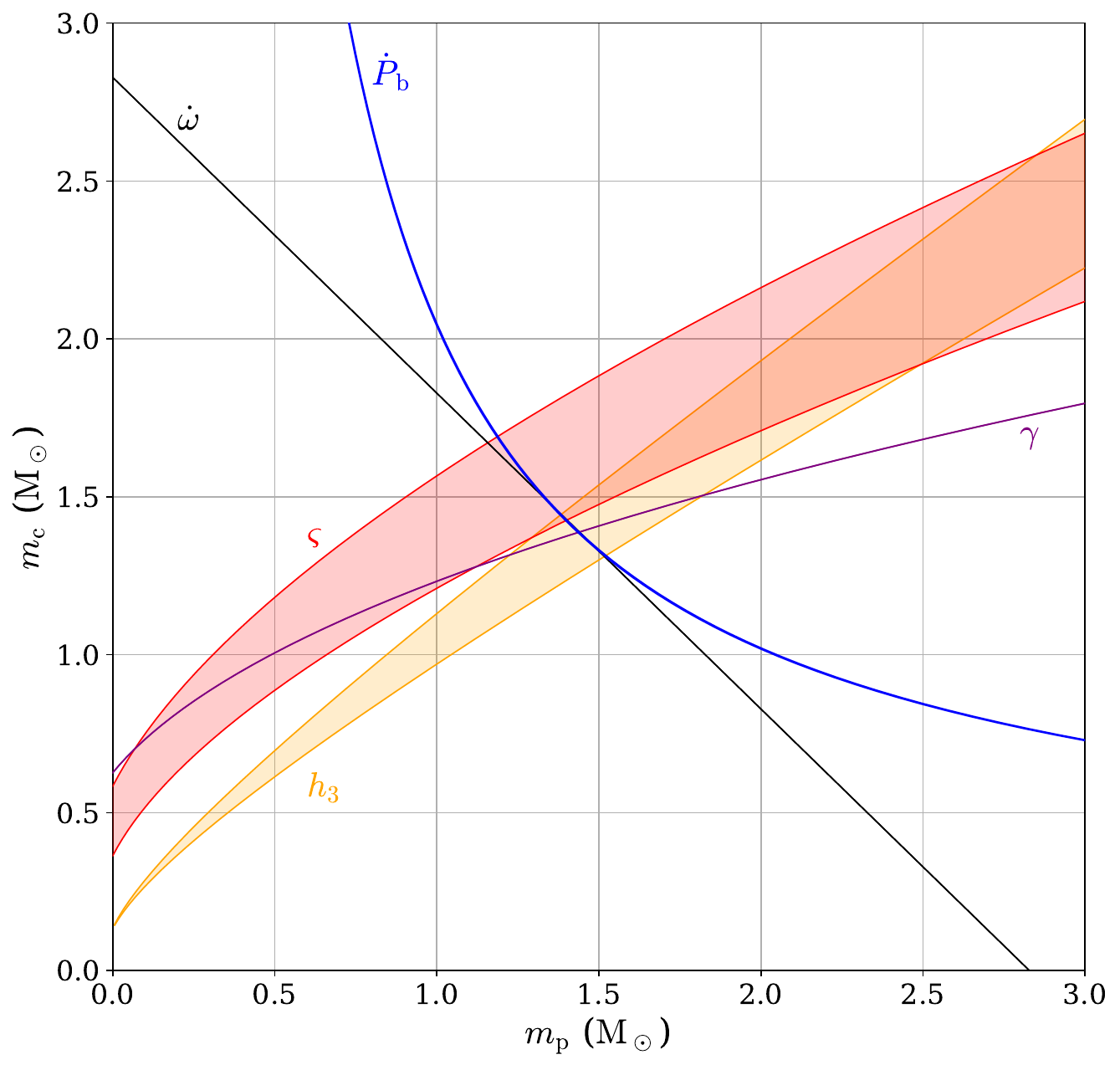}
  \caption{The mass-mass diagram of the Hulse-Taylor pulsar, PSR~B1913+16
    based on the PK parameters measured by \cite{WeisbergHuang2016}. In the figure the
    underlying gravitational theory is assumed to be GR. Under this theory,
    we can calculate bands denoting the $1-\sigma$ uncertainties in the component masses inferred from various relativistic effects. The fact that all bands meet at the same region in the 
    diagram implies that GR passes these tests and provides a self-consistent description of the masses. Figure from \cite{FreireWex2024}, reused without changes under a \href{https://creativecommons.org/licenses/by/4.0/}{Creative Commons Attribution 4.0 International License}\label{fig:mass_mass}.}
\end{figure}

Measuring two PK parameters, we can use a specific theory of gravity
to determine the component masses of the system, which are important
to study stellar evolution theories and, in some cases, to constrain the NS
EOS \citep{wxe+14}.  If more PK parameters are measured, then the
theory can be tested by a self-consistency argument --- using the
masses derived in the first stage one should be able to predict the
subsequent PK measurements.  Another way of viewing this is the
so-called ``mass-mass diagram'' (see Figure~\ref{fig:mass_mass} for a
the first example of such a test with the Hulse-Taylor pulsar, \citealt{WeisbergHuang2016}). For a
gravitational theory to pass the test(s), all curves in the diagram
should intersect in some region, {\it i.e.}, the theory must be able
to describe the component masses in a self-consistent way
\citep{TaylorWeisberg1982}.

\begin{table*}[t]
  \caption{The most important PK parameters that could be used in
    pulsar timing of binaries \citep{BT1976,DD86,DamourTaylor1992,wk99,lk04,ehm06}.  In
    practice, for a specific binary pulsar, only some PK parameters
    are measured, depending on the characteristics of the pulsar
    timing experiment.\label{tab:PK}}
    \centering
  \begin{tabular}{ll}
    \hline\hline
    Parameter &  \\
    \hline
    $\dot\omega$ & time derivative of the longitude of periastron
    $\omega$ \\
    $\gamma$ & amplitude of the Einstein delay\\
    $\dot{P}_\mathrm{b}$ & time derivative of the orbital period $P_\mathrm{b}$ \\
    $r$ & range of the Shapiro delay\\
    $s$ & shape of the Shapiro delay \\
    $\Omega_{\rm SO}$ & precession rate of the pulsar spin \\
    $\delta_\theta$ & mismatch in eccentricities (see text) \\ 
    $\dot{e}$ & time derivative of the orbital eccentricity $e$ \\
    $\dot{x}$ & time derivative of the projected semimajor axis of the
    pulsar orbit $x$\\
    $\ddot{\omega}$ & second time derivative of the longitude of
    periastron  \\
    $\ddot{x}$ & second time derivative of the projected semimajor axis of the
    pulsar orbit \\
    \hline
  \end{tabular}
\end{table*}

Of all double neutron star (DNS) systems, the most relevant for tests of gravity theories is the double pulsar. This system has a unique combination of characteristics that make it, in many respects, the best pulsar laboratory for gravity. 
At the time of discovery, it was by far the most compact double neutron star (DNS) system known in our Galaxy \citep{Burgay2003}\footnote{This is no longer true since the discovery of PSR~J1946+2052 \citep{Stovall:2018ApJ}.}.
Additionally, the detection of radio pulsations from its NS companion \citep{Lyne2004} made it the first (and thus far the only) double pulsar known.
Furthermore, it is also the most edge-on system known \cite{Kramer2006}, providing the most precise measurement of Shapiro delay \citep{Shapiro1964} as predicted by GR. 
While the current results from its timing analysis are discussed in detail in the next section, we also refer to \cite{Kramer:2009} who used the double pulsar to demonstrate the use of a Lagrangian that generalizes the Lagrangian of the post-Newtonian orbital dynamics to the strong-field regime to test fully conservative theories of gravity with a modified Einstein–Infeld–Hoffmann formalism.

\subsection{Scalar-Tensor theories of gravity}

Scalar-tensor theories of gravity can be considered as the simplest extensions to GR. Many of them, including the Jordan–Fierz–Brans–Dicke (JFDB) theory and Damour-Esposito-Far\`{e}se (DEF) theory, are metric theories that satisfy the 
Einstein Equivalance Principle (EEP). However, the additional scalar fields break the 
Strong Equivalence Principle (SEP), which is a general feature of such alternative gravity theories; GR is perhaps the only theory that fully embodies the SEP.

Breaking the SEP leads to some observable effects. Two discussed in detail here are the emission of dipolar radiation and the violation of the universality of free fall (UFF) for self-gravitating bodies. Asymmetric systems like pulsar-WD (PSRWD) binaries are particularly suitable for constraining these theories
as these effects generally depend on the difference of the sensitivity of each body, which is a quantity that is related to the self-gravity or compactness of the object, and hence can be very different for NSs and WDs~\citep{Will2018}. 

The emission of dipolar GWs was first predicted to occur for the JFBD theory by \cite{Eardley1975}; but it is a general feature of alternative theories of gravity. Limits on their detection, especially by pulsar - white dwarf binaries, are especially powerful probes of DEF gravity. As discussed more in the next section, dipolar GWs have never been detected, and this has introduced very stringent constraints on SEP violation, and on many alternative theories of gravity.

Regarding tests of the UFF, a way of detecting it is via the Nordtvedt effect \citep{Nordtvedt1968}. This is similar to the Stark effect, where a neutral atom is polarised by a strong external electrical field. In its gravitational equivalent, a binary system consisting of two objects that fall in an external gravitational field with slightly differing accelerations will see their orbit "polarised", i.e., acquire a secularly evolving eccentricity. UFF test was one of the motivations for the lunar Laser Ranging (LLR) experiments \citep{LLR}.

In the experiments discussed here, instead of the Earth and Moon falling in the field of the Sun, we have a pulsar and its binary companion (preferably a WD) falling in the field of the Milky Way Galaxy \citep{ds:1991}, the latter's gravitational acceleration provides the external ``polarising" force.
The best way of detecting this effect  currently is via a detection of the variation of the orbital eccentricity \citep{fkw:2012}. This method resulted in the limit from PSR~J1713+0747 \citep{Zhu:2018etc}, which has been used to constrain a possible long-range fifth force (additional force to the 4 fundamental forces of nature) induced by dark matter \citep{Shao:2018klg}. Compared to the lunar laser ranging test, pulsar experiments have the advantage of the much stronger binding energy of NSs but also the disadvantage of the much weaker gravitational acceleration caused by the Galaxy. As discussed by \cite{fkw:2012}, the ideal system would be a millisecond pulsar (MSP) in a hierarchical triple star system, where the distant member of the system would cause a much stronger gravitational acceleration of an ``inner" binary containing the MSP. Such a system containing the pulsar J0337+1715 and two white dwarfs was soon discovered soon after \citep{Ransom2014}, with the results confirming the expectation of \cite{fkw:2012}. As we will see in the next section, no violation of the UFF has been detected, providing especially stringent limits on many alternative theories of gravity.

\subsection{Quadratic gravity theories}

Theories like the DEF theory predict an interesting phenomenon called {\it  matter induced spontaneous scalarization}~\citep{DEF93}. Such theories can show large deviations from GR in the non-perturbative regime of a pulsar binary while still satisfying all constraints from weak-field experiments. Hence this effect cannot be tested in the Solar System. Currently, based on the constraints of the dipolar GW emission of 7 binary pulsar systems, the neutron star's effective scalar coupling $\alpha_A$ has been limited to a level of $|\alpha_A|\lesssim 6\times 10^{-3}$, which excludes the possibility of matter induced spontaneous scalarization of NSs in the original DEF theory~\citep{Shao:2017gwu, Zhao2022}. Nevertheless, other varieties of DEF-like theories still remain unconstrained if the scalar field has a mass which suppresses the dipolar radiation at infinity. In this case, depending on the wavelength of the scalar field, the theory may leave detectable imprints in the pulsar orbital dynamics or the properties of the pulsar such as its mass-radius relationship and moment of inertia \citep{Doneva:2016xmf,AltahaMotahar:2017ijw,Xu:2020vbs}. These effects can be important targets for future timing observations given that the current timing of the double pulsar system already shows the potential of measuring some of them \citep{Hu:2020ubl}.

Quadratic gravity theories are well-motivated from an effective field theory point of view with Einstein-scalar-Gauss-Bonnet (EsGB) theories and Chern-Simons (CS) gravity theories representing well studied classes, that involve scalar fields that are coupled to an invariant.

In EsGB theories the choice of the coupling function is crucial for the resulting pulsar properties.
For linear or dilatonic coupling functions the pulsars are always scalarized.
However, in the linear (shift-symmetric) case, the pulsars do not carry scalar charge and therefore are not constrained by orbital decay due to dipole radiation \citep{Yagi:2015oca}.
Constraints arise due the observed maximum neutron star mass \citep{Pani:2011xm,Yordanov:2024lfk}.
In contrast, a dilatonic coupling function allows for a small scalar charge \citep{Kleihaus:2016dui}.
Here the maximum mass and the binary pulsar orbital decay due to dipole radiation both significantly constrain the coupling parameters in different regimes \citep{Yordanov:2024lfk}.
The present strongest bounds are obtained from the gravitational wave signals during BH-NS and BH-BH inspirals \citep{Corman:2024vlk,Sanger:2024axs,Julie:2024fwy}.
On the other hand, when the coupling function allows for curvature induced spontaneous scalarization of neutron stars (analogous to matter induced spontaneous scalarization in DEF), the orbital decay of WD-NS binaries provide the strongest constraints (Especially PSRs J0348+0432, J1012+5307, J2222$-$0137, J1141$-$6545 and J1738+0333) \citep{Yordanov:2024lfk}.

In CS gravity theories a non-dynamical or dynamical pseudoscalar field is coupled to the parity violating CS term. For a non-dynamical field, the periastron precession of the double pulsar PSR~J0737$-$3039A yields a very strong constraint on the scalar field that is $10^{11}$ times stronger than solar system constraints \citep{Yunes:2008ua}. Binary pulsar observations in widely separated binaries do not place stringent constraints on dynamical CS gravity, since the scalar field source is too small \citep{Yagi:2013mbt}. Here the most stringent current constraint is from NS-NS mergers \citep{Silva:2020acr}.

\subsection{Massive gravity}
In GR, gravity is mediated by a massless spin-2 graviton. In extensions of GR, some alternative gravity theories propose that the graviton has a nonzero mass, and these theories remain viable as they have not been entirely ruled out. Consequently, various approaches exist to study the mass of the graviton. Binary pulsar systems also play an important role in tests of massive gravity.

\citet{Finn:2002} introduced a phenomenological term in the Lagrangian for linearized gravity with a mass term of the form $\sim m_{g}^{2}(h_{\mu\nu}^{2}-\frac{1}{2}h^{2})$. This choice ensures that the theory reduces to GR when $m_{g}\rightarrow 0$ and yields the wave equation in its standard form.
In this formulation, the massive graviton induces additional gravitational wave radiation.

The observed $\dot{P}_\mathrm{b}$ from binary pulsars can be employed to limit the graviton mass. \citet{Finn:2002} used PSRs B1913+16 and B1534+12 to first constrain $m_{g}$ using pulsars. Subsequently, \citet{Miao:2019}, using 9 well-timed binary pulsars and Bayes' theorem, provided an improved limit of $m_{g} < 5.2\times 10^{-21}\,{\rm eV/c^{2}}$ at the $90\%$ confidence level.

\cite{Finn:2002} focuses solely on the two tensor modes. However, for the spin-2 massive graviton, there could be five degrees of freedom in the propagation, including two additional vector modes and one scalar mode. 
The vector modes can be ignored, because they usually do not couple to matter in the decoupling limit \citep{deRham:2017}. 
The extra scalar mode can couple to matter and induce a fifth force.
By introducing the 
Vainshtein mechanism, the scalar mode can be decoupled, suppressing the fifth force, allowing massive gravity theories to recover GR when $m_{g}\rightarrow0$.
The simplest model describing this is called the cubic Galileon, a Lorentz-invariant massive gravity model.
The extra scalar mode can lead to extra GW radiation, which includes not only quadrupole radiation but also monopole and dipole radiation.
However, in the vast majority of cases, the quadrupole radiation dominates \citep{de:2013, Shao:2020PhRvD}, and only when $e\geq0.9$ does monopole radiation become dominant and dipole radiation becomes visible. Thus, in general, we only need to consider quadrupole radiation. Following this,

\cite{Shao:2020PhRvD} chose 14 well-timed binary pulsars and placed a bound on the graviton mass, $m_{g}\leq2\times10^{-28}\,{\rm eV/c^{2}}$ at the $95\%$ confidence level.


\subsection{Preferred-frame effects (PFEs)}

Apart from these theory-specific discussions, we can probe the symmetries of the gravitational interaction in a more theory-independent way.

\begin{table*}
\centering
\caption{List of generic parameters subject to gravity tests (including the ten PPN parameters; first group); their physical meaning and predictions for them in GR and fully conservative and semi-conservative gravity theories (under ``c/sc''; the parameters $\alpha_1$ and $\alpha_2$ are 0 in fully conservative theories). Reused without changes under a \href{https://creativecommons.org/licenses/by/4.0/}{Creative Commons Attribution 4.0 International License} from the review by \cite{FreireWex2024}.
\label{tab:PPN}}
\begin{scriptsize}
\begin{tabular} { l l l l }
\hline
Parameter & GR & c/sc & Physical meaning \\
 & & & \\
\hline\hline
$\gamma_\mathrm{PPN}$ & 1 & $\gamma$ & Space curvature produced by unit rest mass \\
$\beta_\mathrm{PPN}$ & 1 & $\beta$ & Non-linearity in superposition law for gravity \\ 
$\xi$ & 0 & $\xi$ & Preferred-location effects \\
$\alpha_1$ & 0 & $\alpha_1$ & Preferred-frame effects \\
$\alpha_2$ & 0 & $\alpha_2$ & Preferred-frame effects \\
$\alpha_3$ & 0 & 0 & Preferred-frame effects and non-conservation of momentum \\
$\zeta_1$ & 0 & 0 & Non-conservation of momentum \\
$\zeta_2$ & 0 & 0 & Non-conservation of momentum \\
$\zeta_3$ & 0 & 0 & Non-conservation of momentum \\
$\zeta_4$ & 0 & 0 & Non-conservation of momentum \\ \hline
$\eta$    & 0 & $\eta$ & Nordtvedt effect; combination of seven PPN parameters \citep{will93} \\
$\kappa_\mathrm{D}$ & 0 & $\kappa_\mathrm{D}$ & Dipolar radiation coupling \\
$\dot G/G$ [${\rm yr}^{-1}$] & 0 & $\dot G/G$ & Variation of Newton's gravitational constant \\
\hline
\end{tabular}
\end{scriptsize}
\end{table*}

In certain alternative gravity theories, the gravitational interaction may violate the local Lorentz invariance (LLI) in the gravitational interaction, implying the existence of a preferred reference frame. 
Apart from the PK parameters,\citep{will1972} proposed the parametrized post-Newtonian (PPN) framework through a post-Newtonian expansion, assuming weak-field and slow-motion conditions.
Within the PPN parametrization, LLI violation is characterized by the parameters $\hat{\alpha}_{1}$, $\hat{\alpha}_{2}$ and $\hat{\alpha}_{3}$, which are listed in Table \ref{tab:PPN}. The parameter
$\hat{\alpha}_{3}$ has been tightly constrained to very high precision
and is associated with the energy-momentum conservation violation, see Table \ref{tab:PPN}. Consequently, studies of LLI typically focus on $\hat{\alpha}_{1}$ and $\hat{\alpha}_{2}$,
as their values might be non-zero in conservative or semi-conservative gravity theories \citep{Will1993}.
A nonzero $\hat{\alpha}_{1}$ can cause the additional motion of a binary pulsar system relative to a preferred reference frame, typically chosen as the Cosmic Microwave Background \citep{Damour:1992ah}. 
A nonzero $\hat{\alpha}_{2}$ induces spin precession in a pulsar around its direction of motion relative to the preferred frame \citep{Shao:2013}. In a binary pulsar system, a nonzero $\hat{\alpha}_{2}$ causes the orbital angular momentum of a binary pulsar to precess around the direction of the velocity of the binary system with respect to the preferred frame \citep{Shao:2012xb}.
For low eccentricity binary systems, the effects induced by $\hat{\alpha}_{1}$ and $\hat{\alpha}_{2}$ decouple such that they can be tested independently \citep{Shao:2012eg}.

\citet{Damour:1992PhRvD} provided a geometrical way to constrain $\hat{\alpha}_{1}$ with eccentricity binary pulsars.
The eccentricity vector ${\bf e}(t)$ can be expressed as a vectorial superposition of a `rotating eccentricity' ${\bf e}_{R}(t)$ and a fixed `forced eccentricity' ${\bf e}_{F}$. The component
${\bf e}_{R}(t)$ has a constant magnitude and rotates in the orbital plane with the relativistic periastron advance.
The component ${\bf e}_{F}$ is parallel to the binary orbital plane and perpendicular to the velocity of the binary system with respect to the preferred frame, $\bf{w}$. 

Because ${\bf e}(t)$ is a sum of ${\bf e}_{R}(t)$ and ${\bf e}_{F}$, and the orientation of ${\bf e}_{R}(t)$ is unknown, we cannot directly constrain $\hat{\alpha}_{1}$ using the observed value of $e$.
However, as ${\bf e}_{R}(t)$ rotates with a rate of the relativistic precession of periastron, the orientation of ${\bf e}_{R}(t)$ changes with time. 
When the timing observation span is long enough to accumulate a significant angle $\Delta{\theta}$ and break the possible cancellation of ${\bf e}_{R}(t)$ and ${\bf e}_{F}$, we can then constrain $\hat{\alpha}_{1}$ with binary pulsars, as discussed in Table \ref{tab:PPN_params_experimental}. 

If $\hat{\alpha}_{2}$ is non-zero, a solitary pulsar undergoes precession around its velocity $\bf{w}$ relative to the preferred frame, and the precession rate is given by \cite{Nordtvedt:1987},
\begin{equation}\label{eq:alpha2}
    \Omega_{\hat{\alpha}_2}^{\text {prec }}=-\hat{\alpha}_2 \frac{\pi}{P}\left(\frac{w}{c}\right)^2 \cos \psi\,, 
\end{equation}
where $\psi$ is the angle between the spin direction and $\bf{w}$.
In binary pulsar systems, a nonzero $\hat{\alpha}_{2}$ induces the orbital angular momentum to precess around the direction of $\bf{w}$. In Eq.~(\ref{eq:alpha2}), the spin period $P$ is changed to the orbital period $P_{\rm b}$. \citet{Shao:2012} have used binary pulsar systems J1012+5307 and J1738+0333 to derive the limit, $\left|\hat{\alpha}_2\right| < 1.8 \times 10^{-4}$ at $95\%$ confidence.
However, this constraint from binary pulsar systems is five orders of magnitude weaker than the current limit from solitary pulsars discussed in the next section.

As an example of Lorentz-violating gravity theories let us now briefly address 
Einstein-{\AE}ther 
gravity, where the preferred frame is modelled through a timelike unit vector, the {\ae}ther field.
Here the strong constraints on $\hat\alpha_1$ and $\hat\alpha_2$ can be exploited to reduce the four coupling constants of the theory to just two independent coupling constants \citep{Yagi:2013qpa,Yagi:2013ava}.
Since Einstein-{\AE}ther gravity also allows for dipole radiation, these two coupling constants have been strongly constrained by the observed orbital decay of binary pulsars (PSR J1141$-$6545, PSR J0348+0432, and PSR J0737$-$3039) \citep{Yagi:2013qpa,Yagi:2013ava},
as well as pulsars (PSR J1738+0333, PSR J0348+0432, PSR J1012+5307, PSR J0737$-$3039) and the triple system PSR J0337+1715 \citep{Gupta:2021vdj}.

For binary pulsars that exhibit a high rate of periastron advance, \cite{Wex:2007} presented an alternative phenomenological methodology for measuring PFEs. They showed that, in such systems, the existence of a preferred frame for gravity leads to an observable signature in the timing data. In the presence of PFEs, one can expect a set of new timing parameters describing these signatures to exhibit a unique relationship that can be incontrovertibly measured and tested. While the Double Pulsar was the only system available for such an experiment at the time, it was predicted that a combination of several suitable systems in a ``PFE antenna array'', along with new sources to be discovered by the SKA, would provide full sensitivity to possible violations of local Lorentz invariance in strong gravitational fields in all directions of the sky.

\section{Current observational constraints}
\label{sec:state_of_the_art}

\begin{figure}[t]
    \vspace{-2pt}
    \centering    \includegraphics[width=0.45\textwidth]{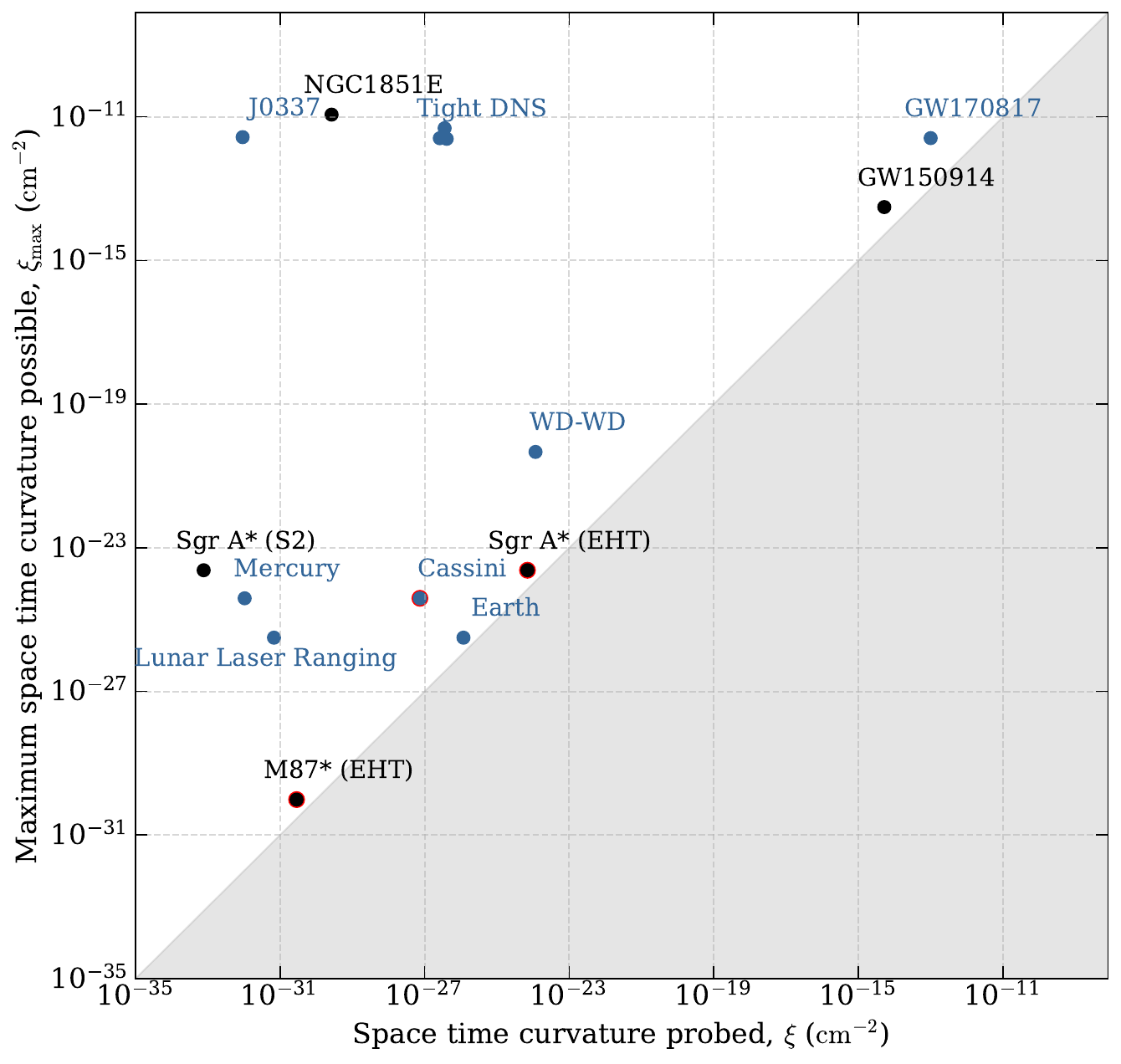}
    \caption{Comparison of different gravity experiments in terms of spacetime curvature probed and maximum spacetime curvature possible. The y-axis gives the maximum spacetime curvature in the system. Since the Y axis is the maximum value of the X- axis, the lower diagonal is greyed out as impossible. The curvature is calculated as the square-root of the Kretschmann scalar $R_{\alpha\beta\gamma\delta} R^{\alpha\beta\gamma\delta}$ (full contraction of the Riemann tensor). `Earth' stands for near-Earth orbit experiments, like Gravity Probe B. Figure adapted from \cite{Wex2020Univ}.}
    \label{fig:curvature}
\end{figure}

In the 50 years since the discovery of the Hulse-Taylor pulsar (PSR B1913+16), hundreds of new binary pulsars have been discovered. The last decade - corresponding to the time since the previous version of this chapter  \citep{Shao2015} - has seen major progress in tests of gravity theories with these systems \citep[for a detailed review, see][]{FreireWex2024}.
Some highlights include:

First, the timing update on PSR~B1913+16 has confirmed that the orbital decay agrees well with the GR prediction with a relative precision of 0.16\% \citep{WeisbergHuang2016}. 

\begin{figure}[t]
  \centering
  \includegraphics[width=0.45\textwidth]{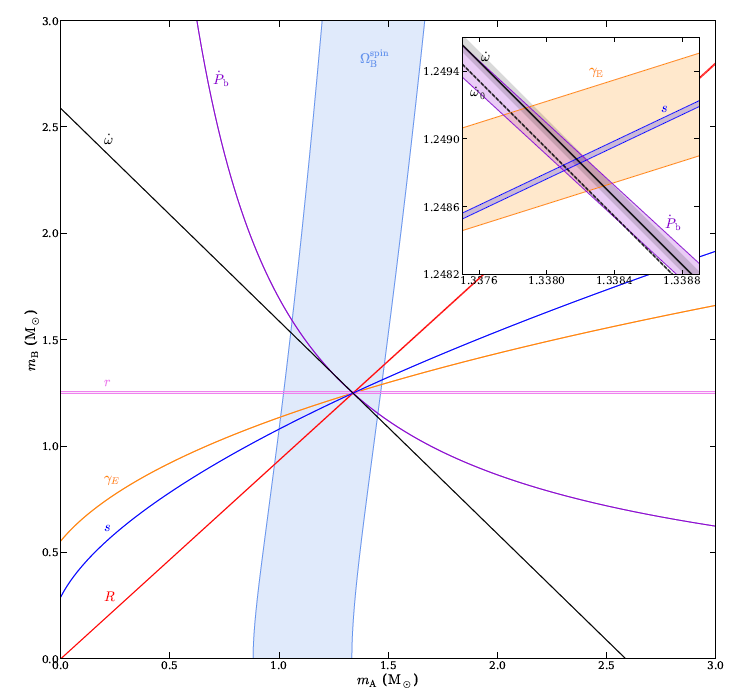}
  \caption{The mass-mass diagram of PSR~J0737$-$3039A/B, also known as
    the double pulsar \citep{Kramer2021}. In the figure the
    underlying gravitational theory is assumed to be GR.  The inset is an
    expanded view of the region of principal interest, where the
    intersection of all curves within a small region within measurement
    uncertainties means that GR has passed all these tests. For more details, see \cite{Kramer2021}, figure from \cite{FreireWex2024}, reused without changes under a \href{https://creativecommons.org/licenses/by/4.0/}{Creative Commons Attribution 4.0 International License}.\label{fig:mass}}
\end{figure}

 The 2021 update on the timing of the double pulsar system by \cite{Kramer2021}  and modelling of its unique eclipse phenomenon by \cite{Lower2024} has provided a record of 
 11 independent tests of GR in this system. Some of these can be visualised in its mass-mass diagram (Figure~\ref{fig:mass}). These tests include:
\begin{itemize}
\item The most precise test ever of the radiative properties of gravity: the orbital decay matches the GR prediction within a 1-$\sigma$ uncertainty of 0.0063\%, which is 25 times more precise than the test by \cite{WeisbergHuang2016}.
\item Four of the tests are of next-to-leading order relativistic effects detected in this system, representing a first in pulsar timing. 
One example of this is the Shapiro delay in this system, which probes spacetimes with curvatures that are
$10^9$ and $10^3$ times larger than those probed by black hole imaging around M87 \citep{EHT2019_M87} and Saggitarius A* \citep{EHT2022_SgrA}, and about $10^6$ times larger than in similar experiments in the Solar System \citep{Bertotti2003}. 
A comparison with various pulsar systems and other gravity experiments is shown in Figure~\ref{fig:curvature}. While Shapiro delay has in general been seen in other binaries before, this system required the accounting of post-Newtonian corrections at the 1.5PN level caused by the moving companion and aberration effects due to light bending had to be taken into account for the first time. 
These next-to-leading-order (NLO) signal propagation effects have been independently tested and further improved with MeerKAT observations \citep{Hu2022}, which {\it clearly show the potential for improvements provided by the SKA}.
\item Another example is the advance of periastron, where second-order effects including the Lense-Thirring effect, had to be taken into account to calculate the masses correctly. With continued timing provided by MeerKAT and the SKA, these observations have a clear potential for measuring the moment of inertia of PSR~J0737$-$3039A \citep{Hu:2020ubl}.
\item Finally, independent timing of both pulsars combined with tracking of secular changes in the radio eclipses of PSR~J0737$-$3039A by the plasma-filled magnetosphere of PSR~J0737$-$3039B offers a novel means of testing spin-orbit coupling in strong-field gravity \citep{Breton2008}. Current constraints from modelling the eclipses detected over a 3\,yr period by MeerKAT match the prediction from GR at the 6.1\% level \citep{Lower2024}.
\end{itemize}

These impressive tests have been complemented with a host of experiments on
other systems that have probed --- and in one case will probe --- additional aspects of gravity theories, in particular the aforementioned effects from the violation of the SEP, dipolar GW emission in the orbital decay of binary pulsars or UFF violation via the detection of Nordtvedt effect.
As discussed above, these effects are expected to be especially strong in asymmetric systems, like PSRWD systems. Searching for them is especially important for two reasons: 1) They are generally predicted by alternative theories of gravity, and
2) Detecting them would falsify GR. The first reason implies that a non-detection constrains alternative theories of gravity that predict that effect.
\begin{itemize}
\item A set of radiative tests with pulsar - WD systems, like
PSR~J1738+0333 and PSR~J2222$-$0137 \citep{Freire2012,Guo2021} and the asymmetric DNS J1913+1102 \citep{Ferdman2020}
have introduced stringent limits on DGW emission.
This represents one of the most fundamental tests of the nature of gravitational waves; the results show that they are, within the measurement precision, purely quadrupolar as predicted by GR.
\item Because of the high masses of PSR~J2222$-$0137 and J1913+1102, these tests have also largely ruled out the phenomenon of matter induced
spontaneous scalarization \citep{Zhao2022}, a highly non-linear, non-perturbative enhancement of the scalar
charge of NS that is generally predicted for some NS masses by scalar-tensor theories of gravity \citep{DEF93}. 
\item The discovery of PSR~J0337+1715, the first millisecond pulsar in a triple stellar system, where the other two components are WDs \citep{Ransom2014}
has allowed a limit on the UFF violation parameter for neutron stars from the lack of detection of the Nordtvedt effect ($\Delta$) of $\Delta < 2.0 \times 10^{-6}$ \citep{Archibald2018,Voisin2020,Voisin2024}, which is 3 orders of magnitude better than any previous limits on the UFF of neutron stars. In particular, this results in the most restrictive limit on the $\omega$ parameter of the JFBD theory of gravity,
$\omega_{BD} > 150,000$ \citep{Voisin2020,FreireWex2024}, which is stronger than the Solar System limit from Cassini~\citep{Voisin2020,Bertotti2003}. 
\item In 2022, \cite{Ridolfi2022} used MeerKAT to discover 13 new pulsars in the globular cluster NGC~1851. One of these, PSR~J0514$-$4002E,
is in a binary system with a total mass of $3.887 \pm 0.004 \, \rm M_{\odot}$ \citep{Barr2024}, which is significantly larger than for any other DNS systems known in the Galaxy. The companion mass is somewhere between 2.1 and $2.7 \, \rm M_{\odot}$; this means it is in the mass gap between the most high-mass NSs and least massive BHs and it could be either. If the companion is a BH, it will likely be spinning fast \citep{Barr2024}, opening the prospect for the detection of the Lense-Thirring effect in this system and a test of the cosmic censorship hypothesis. {\it The latter might only be achievable with the precise timing to be provided by the SKA.}
\end{itemize}

The experimental limits on the PPN parameters listed in Table~\ref{tab:PPN} (and their effective strong-field counterparts) are summarised in Table~\ref{tab:PPN_params_experimental}. As we can see, there is a strong complementarity between pulsar and Solar System tests, but in several cases the best (and in two cases, the only) constraints come from pulsar experiments.

\begin{table*}
\centering
\begin{tabular} {| l | l l | l l |}
\hline
Parameter & Solar-system test & Limit & Pulsar test (eff.~PPN) & Limit \\
 & & & & \\
\hline\hline
$\gamma_\mathrm{PPN}$ & Cassini, Shapiro delay & $2\times 10^{-5}$ $^\dagger$ & double pulsar, Shapiro delay \citep{Hu2022} & $7 \times 10^{-3}$ $^\dagger$ \\
$\beta_\mathrm{PPN}$ & Perih. shift of Mercury; MESSENGER & $2\times 10^{-5}$  $^\dagger$ &---&---\\ 
$\xi$ &  Solar alignment with ecliptic  & $4\times10^{-7}$ & Two solitary pulsars \citep{Shao_Wex:2013} & $4 \times 10^{-9}$ \\
$\alpha_1$ & LLR & $2 \times 10^{-4}$ & J1909$-$3744 \citep{Liu:2020} & $2 \times 10^{-5}$ \\
$\alpha_2$ & Solar alignment with ecliptic  & $4\times10^{-7}$ & Two solitary pulsars \citep{Shao:2013} & $2 \times 10^{-9}$ \\
$\alpha_3$ & Perihelion shift of Earth and Mercury & $2\times10^{-7}$ & J1713$+$0747 \citep{Zhu:2018etc} & $4\times 10^{-20}$ \\
$\zeta_1$ & Combined PPN limits & $2\times10^{-2}$ &---&---\\
$\zeta_2$ &---&---& Multiple binary pulsars \citep{Miao:2020} & $10^{-5}$ \\
$\zeta_3$ & Lunar acceleration & $10^{-8}$ &---&---\\
$\zeta_4$ & Not independent &---&---&---\\
$\eta$    & LLR  & $7 \times 10^{-5}$ & J0337+1715 \citep{Voisin2020,Voisin2024} & $2 \times 10^{-5}$ \\
$\kappa_\mathrm{D}$ &---&---& J1738+0333 \citep{Freire2012} & $2 \times 10^{-4}$ \\
$\dot G/G$ [${\rm yr}^{-1}$] & LLR  & $ 10^{-14}$ & J1713+0747 \citep{Zhu:2018etc} & $5 \times 10^{-13}$ \\
\hline
\end{tabular}
\caption{Comparison of Solar System and binary pulsar tests for the parameters listed in Table~\ref{tab:PPN}. Binary pulsars test strong-field ``effective'' PPN parameters.
For the sake of simplicity, we only give leading order values for all limits in this table. Table adapted from review by \cite{FreireWex2024}. \protect\\
 $^\dagger$ Limit on deviation from 1.
\label{tab:PPN_params_experimental}}
\end{table*}

As discussed in Section \ref{sec:GR}, some of these parameters deserve a more detailed mention:

\begin{itemize}
\item \citet{Shao:2012} have used 10 years of timing results of PSR J1738+0333 and derived the constraint, $\hat{\alpha}_1=-0.4_{-3.1}^{+3.7} \times 10^{-5}$ at $95\%$ confidence.
In 2020, \citet{Liu:2020} have used 15-yr timing of PSR J1909$-$3744 and improve the limit to, $\left|\hat{\alpha}_1\right| < 2.0 \times 10^{-5}$ at $95\%$ confidence.

\item \citet{Shao:2013} have used millisecond pulsars B1937+21 and J1744$-$1134 with 15 years of timing data and placed a constraint, $\left|\hat{\alpha}_2\right| < 1.6 \times 10^{-9}$ at $95\%$ confidence.

\item The limits on $k_{\rm D}$ and the Nordtvedt parameter $\eta$ are derived from the aforementioned limits on dipolar GW emission and the violation of the UFF.

\end{itemize}

\section{The Role of the SKA}
\label{sec:SKA}

The SKA Phase 1 efforts at low and mid frequencies are of extreme importance to testing gravity, not only in improving tests of gravity with presently timed systems, but also in discovering new binaries that are in tighter, more relativistic orbits. In this section, we outline the most important features and characteristics needed for the SKA that will be crucial for this science.

\begin{table}
    \caption{Summary of observing cadence $\rm{T_{cad}}$, length of each observation $\rm{T_{obs}}$, and sub-integration time $\rm{T_{sub}}$ assumed for the simulation. For PSR~J0514$-$4002E, an additional 20-hr campaign every year is assumed.} \label{tab:obs}
    \vspace{5pt}
    \centering
    {\small
    \begin{tabular}{l c c c}
    \hline\hline
    Pulsar name &$\rm{T_{cad}}$[days] & $\rm{T_{obs}}$[min]  & $\rm{T_{sub}}$[s] \\ \hline
    PSR~J0737$-$3039A & 30 & 180 & 30 \\ 
    PSR~J1946+2052 & 60 & 120 & 30 \\ 
    PSR~J1913+1102 & 30 & 60 & 240 \\ 
    PSR~J1036$-$8317 & 90 & 480 & 300 \\ 
    PSR~J0514$-$4002E & 30 & 120 & 1800 \\ \hline
    \end{tabular} 
    \vspace{-5pt}
    }
\end{table}

\subsection{Instantaneous sensitivity}

\label{sec:inst_sensitivity}

Contrary to often popular consensus about astronomical observations, one cannot just ``integrate longer" to achieve the desired sensitivity in a binary pulsar experiment. This is because each radio frequency resolved time of arrival estimate (ToA) requires enough frequency and orbital resolution, while also having enough timing precision to unambiguously detect relativistic effects and to disentangle interstellar medium effects with the pulsar's orbital dynamics. A rule of thumb is to obtain at least an orbital resolution of $< 10^{-2}$ in orbital phase for a clear, unambiguous measurement of orbital/relativistic parameters. Even for a pulsar with a relatively long orbit of about 5 days, the ideal maximum integration time per ToA is only about 7 minutes. For a pulsar like the double pulsar, this is roughly 30 seconds. These small integration times place enormous importance on the instantaneous sensitivity of the telescope. The difference  between the sensitivities of AA* and AA4 for example, could make or break the detection of orbital effects such as Shapiro delay for some systems. This is perhaps best seen in the double pulsar, where adding just 5-years of MeerKAT data to a 16-year historic dataset, we can already measure the Shapiro delay with $3\times$ the significance, including higher order contributions to the effect such as gravitational retardation and light bending, as described in the previous section \citep{Hu2022}. 

In order to further emphasize the improvements in the double pulsar as our current flagship pulsar for this science, we perform simulations of ToAs observed with the SKA for a 10-year experiment from 2028 with both AA* and AA4 configurations. Here, and in all other simulations presented in this chapter, unless explicitly stated otherwise, we use historic observations of the source to determine the desired cadence/orbital campaign and full-orbit observations and we determine the expected ToA precision from observations using MeerKAT and/or FAST telescopes. The details of telescope parameters and expected timing precision are summarised in Table \ref{tab:sim}. Table~\ref{tab:obs} lists the observation details considered for these pulsars for the simulations. 

\begin{figure}
    \centering
    \includegraphics[width=0.5\textwidth]{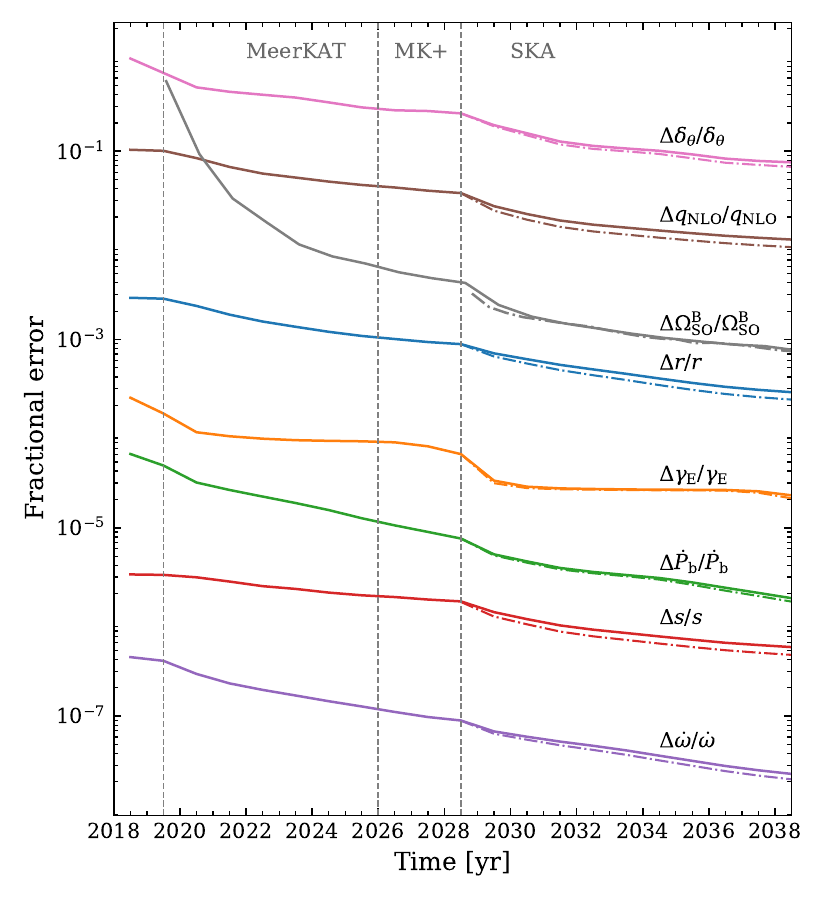}
    \caption{Fractional error of PK parameters for J0737$-$3039A with simulated future data. The solid lines represent results using AA* and the dash-dotted lines represent results using AA4.}
    \label{fig:0737frac}
\end{figure}

The expected improvements in the precision of the PK parameters of the double pulsar, based on nominal simulations that span up to 2038 in the future, and combines the 16-year \citep{Kramer2021} and 21-year dataset (including MeerKAT; Hu et al., in prep.) for the Double Pulsar, is reported in Table~\ref{tab:0737}. The fractional improvements are also shown in Figure \ref{fig:0737frac}: We obtain improvements of a factor of 12--38 and 5--13 for the various PK parameters with respect to these two datasets. 

\begin{table*}
\centering
{\small
\caption{Precision of PK parameters for PSR~J0737$-$3039A with simulated data from MeerKAT+SKA AA4 combined with previous data (``Comb AA4'', 2003-2038) compared to results using 16yr data \citep{Kramer2021} and 21yr data (16yr + MeerKAT 5yr; Hu et al. in prep.). The parameter $q_\mathrm{NLO}$ is the scaling factor of NLO signal propagation effects.
}
\vspace{5pt}
\begin{tabular}{c | c  c c }
    \hline\hline
        & Comb AA4 / 16yr &  Comb AA4 / 21yr & Fractional error \\ \hline
        $\dot{\omega}$ & 36.2 & 12.6 & $2.1\times10^{-8}$ \\ 
        $\gamma_\mathrm{E}$ & 11.8 & 4.5 & $2.1\times10^{-5}$ \\ 
        $\dot{P}_\mathrm{b}$ & 38.0 & 10.9 & $1.6\times10^{-6}$\\ 
        $s$ & 21.5 & 6.9 & $7.2\times10^{-4}$ \\ 
        $q_\mathrm{NLO}$ & 13.5 & 4.8 & $1.0\times10^{-2}$ \\ 
        $m_\mathrm{c}$ & 14.8 &  5.9 & $2.3\times10^{-4}$ \\ 
        $\delta_\theta$ & 15.1 & 5.5 & $6.9\times10^{-2}$ \\ \hline
\end{tabular}
\label{tab:0737}
}
\end{table*}

Additionally, there will also be improvements to the timing parallax and proper motion of 9--100 times the present measurements from a combination of Very Long Baseline Interferometry (VLBI) and timing \citep{Kramer2021}, assuming that DM variations can be adequately modelled. A complementary result that can be obtained from the same observations is the measurement of the moment of inertia of pulsar A which provides constraints on the neutron star EOS (\citealt{Hu:2020ubl,Kramer2021}, Hu et al. in prep.). 

We also simulated the anticipated improvement in geodetic precession rate of pulsar B in the double pulsar through joint MeerKAT+SKA observations. Off-eclipse `noise' from intrinsic flux density fluctuations in pulsar A were sampled from existing MeerKAT-UHF light curves, while the radiometer noise in the eclipse troughs were scaled by a factor of 1/3 for SKA-Mid AA* and 1/4 for AA4. Using an assumed monthly observing cadence, we then fit the resulting eclipse light curves using the hierarchical inference approach in \cite{Lower2024}. The resulting evolution in $\Omega_{\rm SO}^{\rm B}$ is shown by the grey trace in Figure~\ref{fig:0737frac}. Note, the simulation assumes that systematic differences between observed and simulated eclipse light curves will have been overcome. This is currently the dominant source of systematic uncertainty in the eclipse-derived precession rate \citep{Breton2008, Lower2024}.

As a final point, we note that it is important to obtain a good, localised sampling of the orbit at every observing session of a binary pulsar -- hence the observing time per session is not solely a function of flux density, and even bright pulsars need to be observed for a sufficiently long time per session.

\begin{table*}[t]
\caption{Summary of simulation setup for different observing phases and ToA uncertainty $\sigma_{\rm{ToA}}$ assumed in the simulation of five pulsars (for the sub-integration time given in Table ~\ref{tab:obs} over the full bandwidth). For PSR~J0737$-$3039A, the ToA uncertainties for both UHF and L band are listed, whereas for other pulsars only L band is shown. For MeerKAT+, 20 SKA antennas are added to 64 MeerKAT antennas, however a maximum of 64 antennas can be used for observations.
}\label{tab:sim}
\vspace{5pt}
\centering
{\small
\begin{tabular}{l|ccccc} 
\hline \hline
Telescope & MeerKAT & FAST & MeerKAT+ & SKA AA* & SKA AA4 \\ \hline 
Number of antennas & 64 & 1 & 64 of 84 & 144 & 197 \\ 
Effective diameter [m] & 108 & 300 & 111 & 172 & 204 \\ 
Observing period [yr] & \makecell{2025/Q1- \\2025/Q4} & \makecell{2025/Q1- \\2028/Q2} & \makecell{2026/Q1-\\2028/Q2} & \makecell{2028/Q3-\\2038/Q2} & \makecell{2028/Q3-\\2038/Q2} \\ 
Observing band & UHF/L & L & UHF/L & 1/2 & 1/2 \\ \hline
\hline 
Pulsar name & \multicolumn{1}{c}{}& \multicolumn{1}{c}{}& \multicolumn{1}{c}{$\sigma_{\rm{ToA}}[\mu\rm{s}$]} \\ \hline
PSR~J0737$-$3039A & 1.6/2.8 & - & 1.5/2.7 & 0.6/1.1 & 0.5/0.8 \\ 
PSR~J1946+2052 &- &8.0&-& 19.1& 13.6 \\ 
PSR~J1913+1102 &-& 26.0 & - & 62.1& 44.2 \\ 
PSR~J1036$-$8317 & 0.60 &- & 0.57 & 0.24 & 0.17 \\ 
PSR~J0514$-$4002E & 16.0 & - &15.1 & 6.3 & 4.8 \\ \hline
\end{tabular}
}
\end{table*}

\subsection{Observing strategy and duration}

While all pulsar experiments need regular observations (henceforth ``cadence observations") to maintain their timing baseline, it is vital for binary pulsar timing observations to also regularly sample the entire orbit. To simplify scheduling constraints, this is periodically done (every few months) either via ``full-orbit observations" if the orbital period is sufficiently short, or via so-called orbital campaigns wherein we perform long observations covering specific orbital phases such as superior conjunction and periastron passage along with daily observations covering the rest of the orbit. Incomplete sampling of the orbit (especially for highly eccentric systems where random observing slots are more likely to end up on just one side of the orbit) has been seen to significantly bias the estimates of some relativistic parameters \citep{Kramer2021}. Hence, the SKA needs to have a flexible scheduler that is capable of conducting these observations at the required times.

\subsection{Targeted pulsar searches for short period binaries with compact companions}

Although the SKA will deliver prompt refinements to the timing of known relativistic binaries, transformative advances will surely come from new discoveries either through ongoing surveys with existing facilities or through the SKA itself. To understand this better, let's revisit Figure \ref{fig:curvature} that contrasts the space-time curvature currently sampled by each experiment with the theoretical maximum reachable for the same class of sources. From this, we can infer (i) binary-pulsar timing and gravitational-wave detections of compact mergers have the best potential to probe the deepest space-time curvatures, and (ii) binary pulsars retain by far the greatest headroom, with existing systems still $\sim15$ orders of magnitude shy of their ultimate potential. Because pulsars permit tracking of relativistic effects at precisions unmatched by any other strong-field probe, pushing them toward higher curvature remains essential—even if ground-based interferometers have otherwise already sampled more extreme gravitational potentials. Moreover, the statistical power of gravity tests in pulsar binaries scales steeply with orbital period, viz $P_{\rm b}^{-\frac{8}{3}}$ \citep{Batrakov2024}; thus, every newly discovered short-period system not only probes deeper curvature but also will overtake the constraints set by all currently timed binaries at an accelerated pace. Although the presently known binaries have wildly different orbital parameters, this general trend can be observed: PSR J0737$-$3039A with $P_{\rm b} \simeq 2.5$~h and a 16-yr timing baseline, now can test GR $25\times$ better than the original binary binary PSR B1913+16 ($P_{\rm b} \simeq 8$ h), while the even more compact PSR J1946+2052 ($P_{\rm b} \simeq 1.8$ h) achieves a factor of $2\times$ improvement with only 7 years of observations. Based on the these discussions, it is evident that the next frontier of binary pulsar tests must be to find binaries with shorter orbital periods and/or with heavier compact companions such as high mass NSs or black holes. 

The presently envisioned pulsar search pipeline for the SKA's Galactic plane surveys (assuming the envisioned 10-minute observing times) is unfortunately largely insensitive to binaries in this regime. The main reason for this is that the orbital corrections for SKA pulsar searches assume that within an observation the change in the apparent pulse period can be adequately modelled by a constant ``acceleration". While this assumption has been used in most of the historic surveys so far and have obtained a plethora of binaries including the double pulsar, short orbit binaries found via this method can be attributed to either the pulsar being bright enough to be detected even with residual orbital smear or be in highly eccentric orbits that facilitated catching them at an orbital phase when the acceleration approximation is still valid. With 10 minute observing times, acceleration searches are nominally sensitive to $> 100$ minute orbits. In the mean time, PSR J1946+2052 which has an orbit of ca. 100 minutes will have accumulated $\sim15$ yr of timing baseline and hence would have already provided competitive strong-field tests at the very curvature scale that the standard SKA acceleration search is poised to be sensitive to. The best way for SKA to remain competitive in this space is to change the search methodology to be sensitive to orbits shorter than 100 minutes, where discoveries can test gravity at steeper curvatures and reach better precisions than PSR J1946+2052 at an accelerated space.

Searches with SKA-low benefit from the fact that the large field of view facilitates multiple surveys of the sky that somewhat mitigates this issue. However, searches at these frequencies are will be limited by interstellar dispersion and scattering for mildly recycled pulsars except for the top part of the band, which is unfortunately contaminated with higher radio frequency interference. SKA-mid in this case offers the best chance at detection. However, it is reasonable to expect that a full survey of the Galactic plane will only happen once. To maximise our sensitivity to short orbit binaries, it is imperative to perform additional targeted searches that retain sensitivity to short orbits. This can be achieved via template bank searches, where we model the orbital motion with a full Keplerian circular or elliptical orbit (c.f. \cite{Balakrishnan:2022}). 

\begin{figure}
    \centering
    \includegraphics[width=0.5\textwidth]{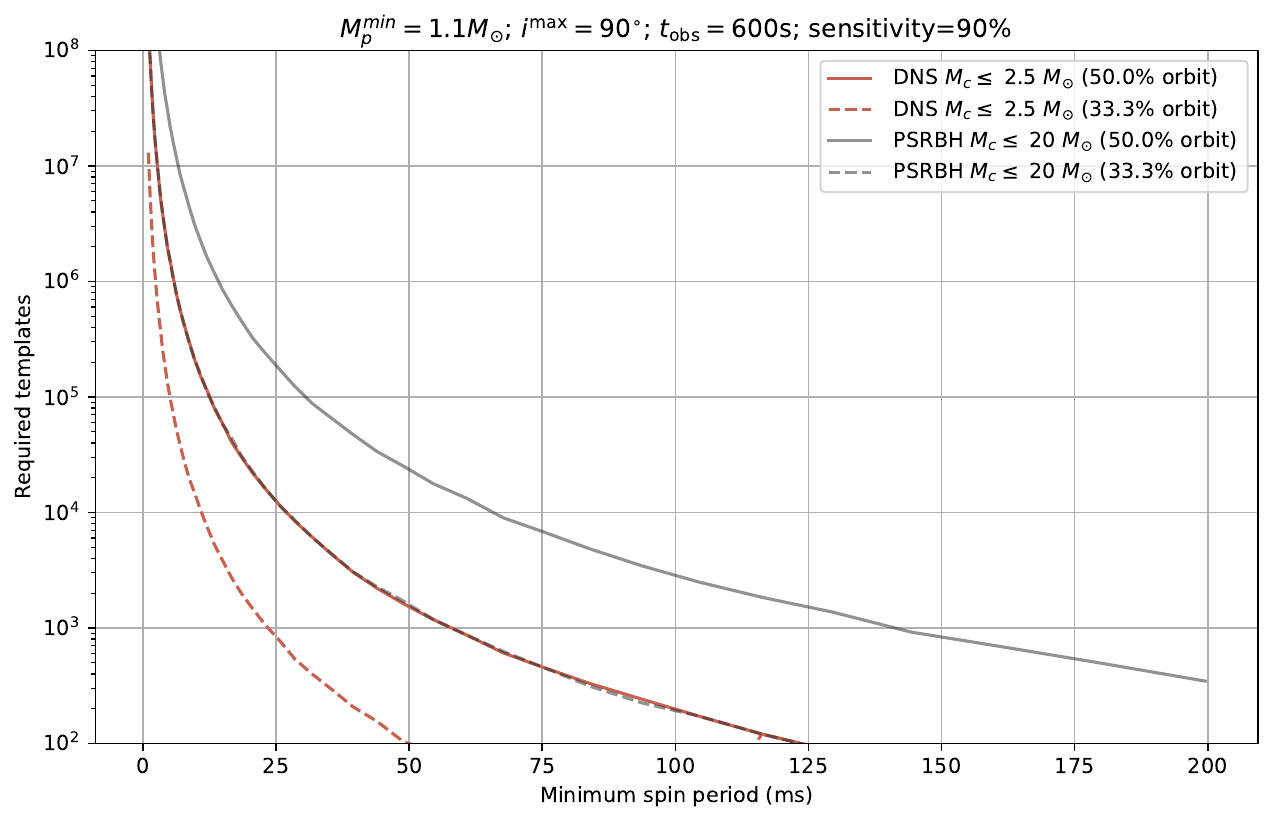}
    \caption{Required number of circular orbit templates as a function of minimum spin period of the pulsar for a SKA-mid survey for DNS and PSRBH systems denoted as red and gray lines respectively. The dotted and solid lines assume that there is 33.3\% and 50\% of the orbit within the observation respectively.}
    \label{fig:templates}
\end{figure}

\begin{figure}
    \centering
    \includegraphics[width=0.5\textwidth]{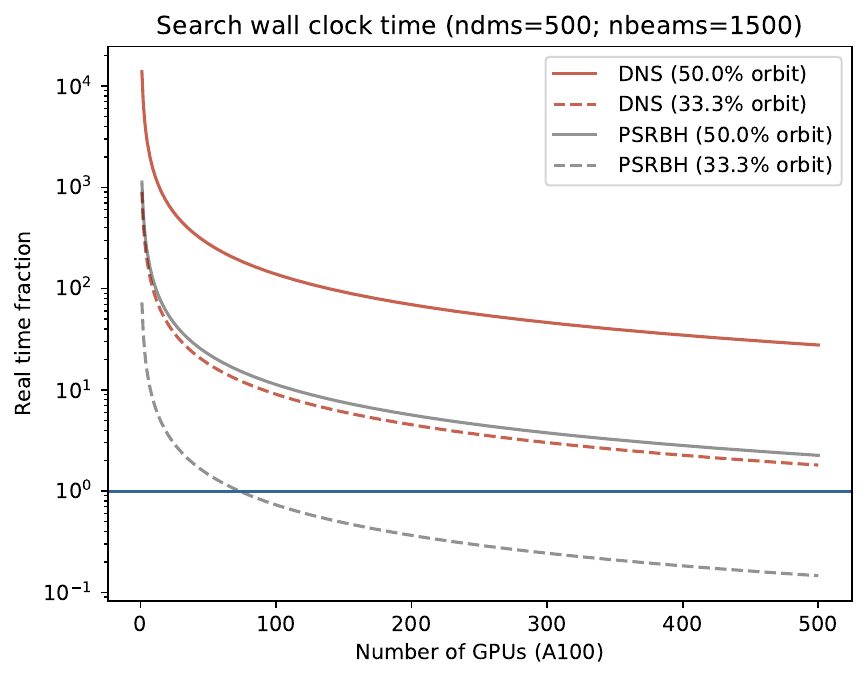}
    \caption{Wall clock pulsar search run time as a function of number of A100 GPUs for circular orbit searches for DNS and PSRBH binaries. The colors and line styles are same as \ref{fig:templates}.}
    \label{fig:search_time}
\end{figure}

Keplerian bank searches generally take orders of magnitude longer compared to traditional searches, which is partly why they were never widely adopted in historic surveys. However, the rapid growth in computing, especially using graphics processing units, have already made this a possibility for targeted searches of locations such as Globular clusters and the Galactic Centre. In order to understand the computation need required to carry out a circular orbit search with SKA-mid, we consider a scenario where we have 10-minute observing times, and we limit ourselves to finding 0.5-2 hour orbits. We restrict the minimum pulsar mass to be $1.17  M_{\odot}$, a maximum inclination of 90 degrees and a search sensitivity (equivalent to the ``tolerance" in acceleration searches) of 90\%. Figure \ref{fig:templates} shows the total number of templates needed for this search as a function of the fastest spin period $P_{\rm spin}^{\rm min}$ to be sensitive to, assuming the maximum companion mass to be $2.5 M_{\odot}$ and $20 M_{\odot}$ for DNS and PSRBH searches respectively. As it can be seen from the figure, the number of templates are a steep function of the minimum spin period that we choose to be sensitive to. In case of dynamic locations such as globular Clusters, we do not possess any prior information of the distribution of pulsar spin periods in DNS and PSRBH binaries. However, for the Galactic plane, one can use the wisdom from binary evolution: We expect DNS systems to have mildly recycled pulsars of periods $P_{\rm spin}^{\rm min} \simeq 20$~ms. We also expect pulsars in Galactic PSRBH systems are likely not recycled at all; their spin periods, at our target orbital periods are expected to be well beyond $P_{\rm spin}^{\rm min} > 100$~ms. Assuming a conservative $P_{\rm spin}^{\rm min} = 15$ ms for DNS and $P_{\rm spin}^{\rm min} = 85$ ms for DNS and PSRBH searches respectively, we benchmark how long would take in a computing cluster with Nvidia A100 GPUs. This is represented in Figure \ref{fig:search_time}. This search is already feasible with a relatively miniscule additional investment to the pulsar search subsystem. We note that the caveat to the non-recycled PSRBH searches that a significant fraction of those pulsars might have their radio emission switched off at this point of their isolated binary evolution. However, as it can be seen Figure \ref{fig:search_time}, the PSRBH searches are cheaper than the DNS searches.

We also note that targeted searches of such binaries in high probability regions for finding such binaries, viz globular clusters and the Galactic centre need to be highly prioritised. These locations are small enough to use the 16 available pulsar timing beams to record coherently dedispersed filterbank data, that can be saved to disk and sent to regional centres for post processing.

\subsubsection{Potential impact of short period binary discoveries}
\label{sec:timing}

In this section, we explore the potential benefits of finding binaries with short orbital periods than those presently known, whose discovery will accelerate our ability to test gravity with better precision. To this effect, we compare the expected performance of the SKA AA$^\ast$ and the SKA AA4 configurations by simulating ToAs from potential new discoveries and evaluating the evolution of timing measurements in both scenarios. As specific examples, we choose to simulate time-evolved versions of PSR J0737$-$3039A and PSR J0514$-$4002E, which are likely to be part of the population of the Galactic field and globular clusters respectively. The simulated orbital parameters were derived by propagating from the current-day orbital parameters and masses presented in \cite{Hu2022} and \cite{Barr2024} via the orbital decay from gravitational-wave emission \citep{Pet64}. For PSR J0737$-$3039A, we simulate its expected orbital configuration in 72 Myr from now, while for PSR J0514$-$4002E, we simulate its expected orbital configuration in  380 Gyr from now. While the latter is unrealistic, this exercise purely serves to provide a starting point for our simulations. After obtaining the updated orbital parameters, we also update the newly expected PK parameters. The resulting Keplerian and PK parameters (Orbital period $P_\mathrm{b}$, eccentricity $e$, projected semi-major axis $x$, range of Shapiro delay $r$, shape of Shapiro delay $s$, rate of periastron advance $\dot\omega$, amplitude of the Einstein delay $\gamma_\textrm{E}$, and the orbital period derivative $\dot P_\textrm{b}$) are presented in Table~\ref{tab:J07J05_orb}. Given the strong  influence of the globular cluster on PSR J0514$-$4002E,  we assumed for this system  $\dot{P}_\mathrm{b} = 1.79\times10^{-11}$ s s$^{-1}$, i.e. the current  observed value \citep{Barr2024},  and not the predicted general relativity value, as kinematics effects are dominant.

\begin{table*}
\centering
{\small
\caption{Propagated Keplerian and PK parameters of PSR J0737$-$3039A and PSR J0514$-$4002E after their respective time of evolution. The Shapiro delay parameters of PSR J0737$-$3039A remain unchanged, while inclination angle of J0514$-$4002E is too low to detect any Shapiro delay. The quoted values of $\dot P_\textrm{b}$  are the theoretical predictions according to general relativity theory.}
\vspace{5pt}
\begin{tabular}{c  c  c  c  c  c  c   c  c}
    \hline\hline
    Evolved form          & Time  & $P_\textrm{b}$ & $e$   & $x$   &  $\dot\omega$ & $\gamma_\textrm{E}$ & $\dot P_\textrm{b}$ \\ 
          of PSR          & Gyr   & hours          &       & ls     & deg\,yr$^{-1}$ & ms & s\,s$^{-1}$\\ \hline
   J0737$-$3039A & 0.072 & 1.08           & 0.017 & 0.600  & 65.950 & 0.124 & $-1.908\times10^{-12}$ \\ 
  J0514$-$4002E & 380   & 7.29           & 0.08 & 3.2955    & 3.603 & 0.0009 & $-3.7782\times10^{-13}$ \\ \hline
\end{tabular} 
\label{tab:J07J05_orb}
}
\end{table*}

\begin{table*}
\centering
{\footnotesize
\caption{ToA measurement scheme and ToA uncertainness used in the simulation of the evolved versions of J0737$-$3039A and J0514$-$4002E shown in Table~\ref{tab:J07J05_orb}. All observations are assumed to be at L-band, with a central frequency of 1.3GHz and a bandwidth of 0.8 GHz. The ToA integration times, sub-bands, and observing strategy are kept as close as possible to the most recent MeerKAT studies of J0737$-$3039A and J0514$-$4002E \citep{Hu2022,Barr2024}.}
\vspace{5pt}
\begin{tabular}{c  c  c  c  c  c  c  c}
    \hline\hline
      Evolved form            & Integration  &  & MeerKAT & SKA AA$^\ast$ & SKA AA4 & Observation & Observing \\
     of PSR           & time  & Sub-bands & uncertainty  & uncertainty & uncertainty & length & cadence\\
    & (s) &  & (\textmu s) & (\textmu s) & (\textmu s) & (hours) & (days) \\ \hline
    J0737$-$3039A & 32    & 16 & 10 & 4 & 3 & 3 & 30 \\ 
    J0514$-$4002E & 1,200 & 4  & 10 & 4 & 3 & 3 & 30 \\ \hline
\end{tabular} 
\label{tab:J07J05_obs}
}
\end{table*}

\begin{figure*}[t]
  \centering
  \includegraphics[width=7.1cm,height=5.4cm]{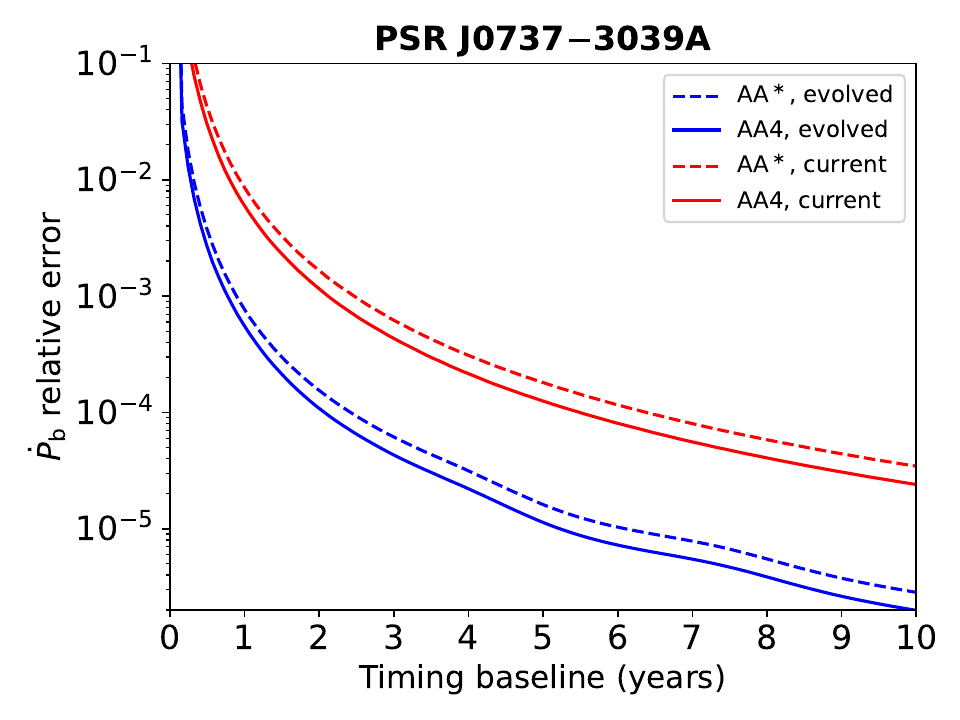}
  \includegraphics[width=7.1cm, height=5.4cm]{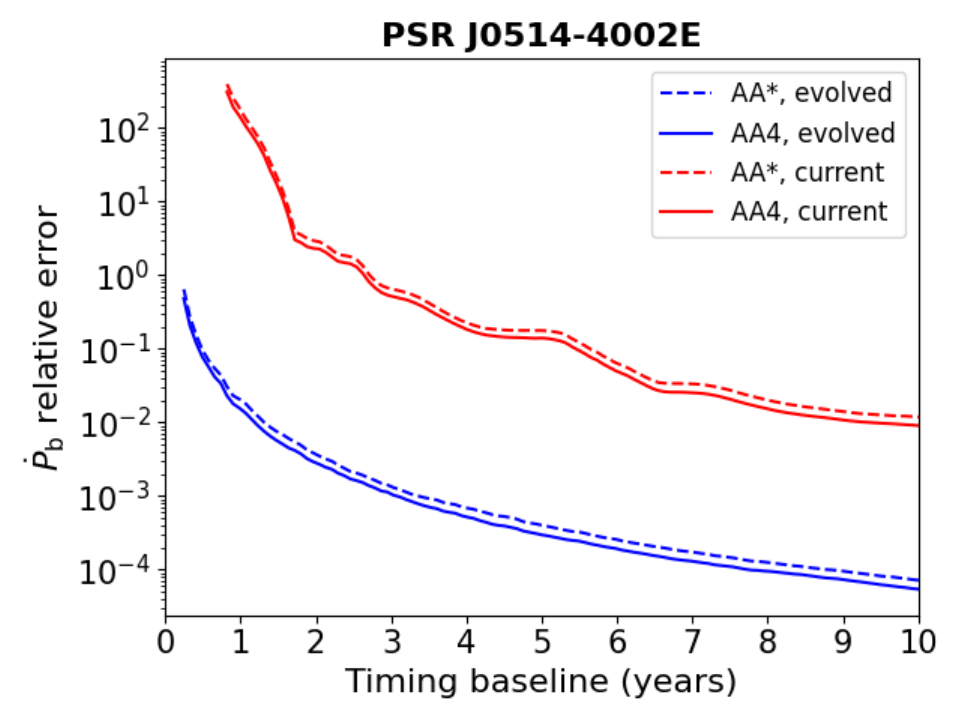}
  \caption{Evolution of the relative error for the derivative of the orbital period   as obtained by simulations over ten years of evolved versions of PSR J0737$-$3039A (with $P_\textrm{b}\simeq1$~hour) and PSR J0514$-$4002E (with $P_\textrm{b}\simeq7$~hours), assuming the SKA AA$^\ast$ and SKA AA4 configurations. We also plot the evolution of the $\dot P_\textrm{b}$ measurement assuming the current orbital configurations for comparison. The observing cadence emulates that of the latest published timing experiments on those systems \citep{Hu2022,Barr2024}. The expected sensitivity of the SKA AA4 consistently reduces the PK measurement uncertainty to 60\% of the one derived with the SKA AA$^\ast$ sensitivity.
  \label{fig:sims}}
\end{figure*}

\begin{figure*}[th]
    \centering
    \includegraphics[width=\linewidth]{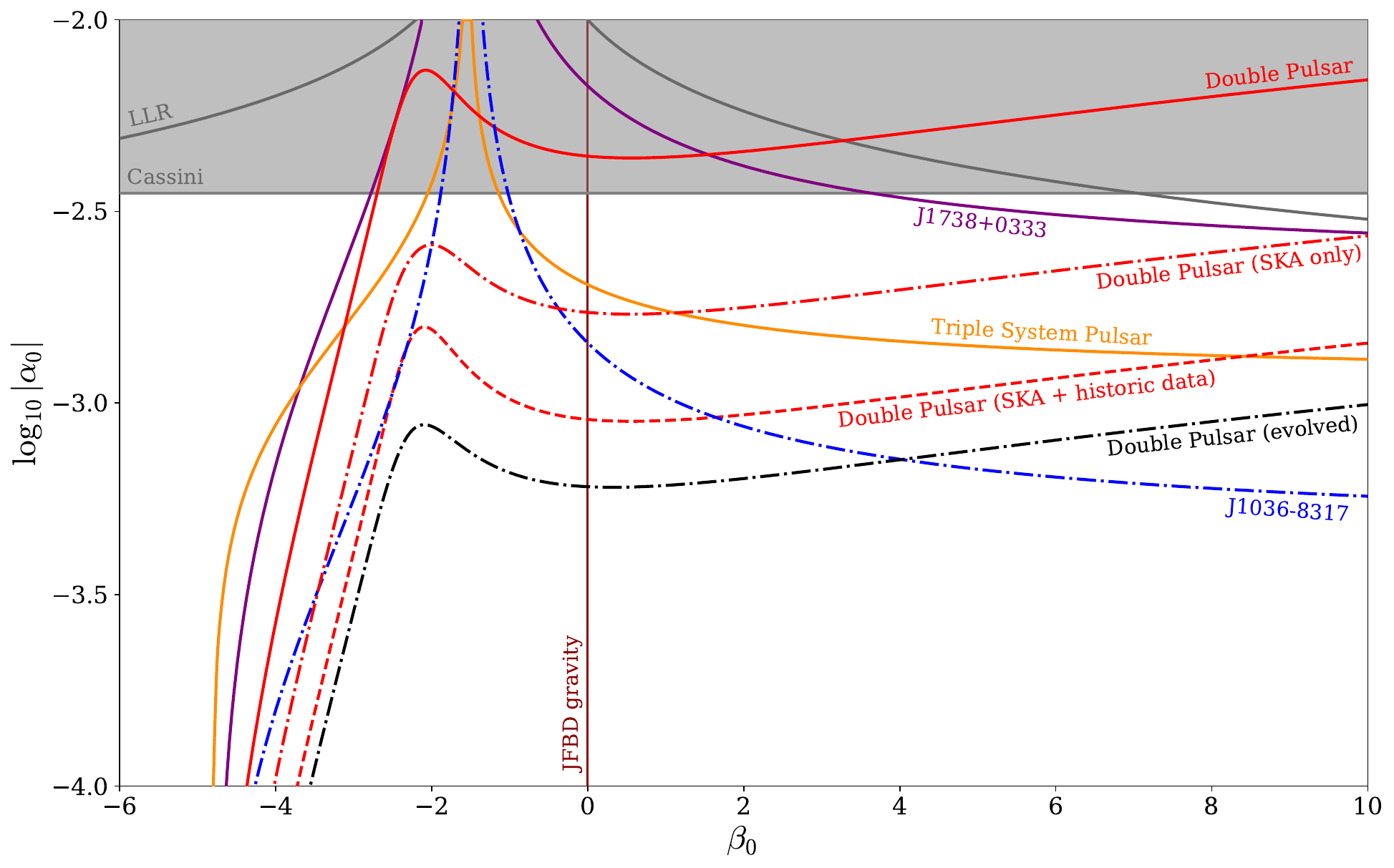}
    \caption{Present and potential future constraints on Damour-Esposito Farésé (DEF) gravity from binary pulsars for the theory's linear ($\alpha_0$) and quadratic ($\beta_0$) coupling coefficients of the scalar field. When $\alpha_0 = \beta_0 = 0$, then the theory reduces to General Relativity. The solid red line is the constraints from the 16-year dataset of \cite{Kramer2021}. The dash-dotted red line is the double pulsar, if it were discovered and timed solely by the SKA-mid AA4. The dashed red line is the improvement when adding 10-years of SKA-mid AA4 data to the 16-year historic dataset. The black dash-dotted line is a putative evolved form of the double pulsar of orbital period $\simeq 1$ hour, that is discovered and and timed by SKA-mid AA4. Potential constriants from J1036$-$8317 is given in blue, assuming 10-years of timing with SKA-mid AA4 and constraints on the mass ratio from optical observations of its companion. The present best constraints from J1738+0333 and the triple system are given in purple and orange for reference. Figure produced by Norbert Wex. }
    \label{fig:a0b0}
\end{figure*}

We simulate ToAs based on previous MeerKAT campaigns and estimating the SKA timing precision. We estimate the sensitivity of the SKA configurations for each of these two systems assuming the same received flux as in the real counterparts of PSR J0737$-$3039A and PSR J0514$-$4002E. Both of these systems have had timing experiments with MeerKAT \citep{Hu2022,Barr2024}, the SKA precursor, constituting a very good basis for scaling the ToA precision from one facility to the other. Based on the increase of surface area from MeerKAT to SKA AA$^\ast$ and SKA AA4, and assuming the same integration time and bandwidth as in the MeerKAT timing experiments, the ToA uncertainty is reduced to 39\% in the SKA AA$^\ast$ and to 28\% in the SKA AA4. In Table~\ref{tab:J07J05_obs}, the used ToA uncertainties and observing strategies used for the SKA AA$^\ast$ and SKA AA4 configurations are shown.

We use the configurations stated above to compare the precision of PK measurements in the SKA AA$^\ast$ and the SKA AA4. Assuming discovery in any of these stages of the SKA, the observing schemes are simulated for 10 years for SKA AA$^\ast$ and SKA AA4, and the ToAs are fit with baselines increasing from 0 to 10 years. The results are shown in Figure~\ref{fig:sims}, where the relative uncertainty of the $\dot P_\textrm{b}$ measurements are drawn over the increasing timing baseline, for both the evolved versions of PSR J0737$-$3039A and PSR J0514$-$4002E in the SKA AA$^\ast$ and SKA AA4 configurations, and with their current orbital configurations as well.

The SKA AA4 configuration has significant advantage over the SKA AA$^\ast$ configuration. Both the the SKA AA$^\ast$ and SKA AA4 configurations achieve very significant results early on, with the SKA AA4 configuration halving uncertainties with respect to the SKA AA$^\ast$ configuration. In long-term timing, when the reduction of $\dot P_\textrm{b}$ measurement uncertainty slows down over increasing timing baseline, but it still takes 1-2 years longer for SKA AA$^\ast$ to achieve the same sensitivity as SKA AA4. This demonstrates that the need for SKA AA4 especially for binary pulsar experiments. It is also worth noting that the evolved versions of these two systems yield $\dot P_\textrm{b}$ measurements of the same significance as their current versions approximately five years prior. Given the capacity of the SKA to discover more compact systems akin to the evolved version of PSR J0737$-$3039A and PSR J0514$-$4002E, this implies an increased capacity for faster tests of gravity. As an example, we show how the present and evolved form of PSR J0737$-$3039A improve constraints on scalar tensor gravity in Figure \ref{fig:a0b0}.

\subsubsection{BH spin measurement from a pulsar---stellar-mass black hole system} \label{sssec:psr-bh}

\begin{figure}[ht]
    \centering
    \includegraphics[width=0.45\textwidth]{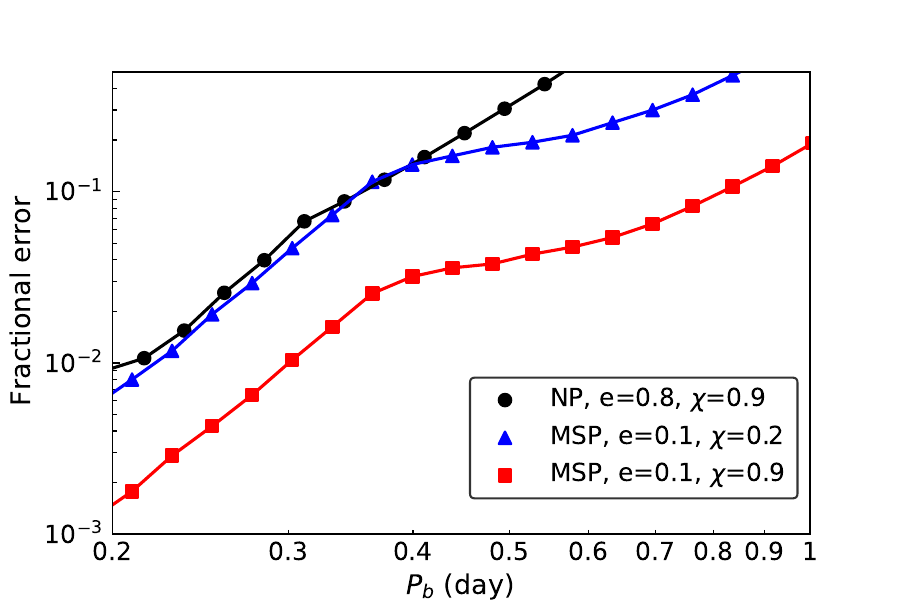}
    \caption{Fractional error of the BH spin measurement as a function of orbital period of a PSR-SBH system. Here we assumed timing observations with weekly cadence and 10-yr time span. For NP, we used 100\,$\mu$s timing precision from each observation. For MSP, we assumed AA4 sensitivity, and 1-hr observations per epoch. }
    \label{fig:pbhb-spin}
\end{figure}

A PSRBH system is considered as the holy grail of gravity tests, providing unprecedented opportunity to allow for precision tests of BH physics. If the black hole rotates significantly, by modelling the Lense-Thirring precession of the orbit, both the amplitude and the orientation of the BH spin can be determined (e.g. \citealt{wk99,lewk14}). This, in combination with the BH mass obtained from other PK parameters, will deliver a direct examination of the Cosmic Censorship Conjecture. In GR, this hypothesis requires the spacetime singularity to be hidden within the event horizon \citep{pen79}. For a Kerr BH, it means
\begin{equation}
  \chi \; \equiv \; \frac{c}{G}\,\frac{S}{m^2_{\rm BH}} \; \le 1 \;,
\end{equation}
where $c$ is the speed of light, $G$ the gravitational constant, $S$
the angular momentum and $M_{\rm BH}$ the BH mass. Therefore, a measurement of the mass and the spin of a black hole can  be used to test this inequality.

From pulsar surveys with the SKA, a PSRBH system is expected to be discovered. In the Galactic plane, a PSRBH system can form by following the standard binary evolution procedures. Here, the pulsar is formed from the supernova explosion of the secondary star, and would be a normal pulsar (NP) without experiencing any recycling. The orbit is expected to be eccentric caused by the kick of the supernova explosion \citep{yp98,vt03}. The canonical binary evolution does not produce a PSRBH system where the pulsar is recycled, though exotic mechanisms exist \citep[e.g.,][]{spn04,ppr05}. In contrast, in regions of high stellar density, such as globular clusters and the Galactic Centre, a PSRBH system can be formed through stellar capture during a multiple body encounter \citep{khm93,fl10}. Figure~\ref{fig:pbhb-spin} shows the anticipated measurement precision of BH spin in both types of systems. For the case of a NP, the BH spin can be measured for compact orbits and fast spinning black holes. In contrast, if the pulsar is an MSP, the anticipated higher timing precision would allow BH spin to be measured for wider orbits or black holes with a low spin.

\subsubsection{Detectability of new binary pulsars with the SKA}
\label{sec:searches}

For completeness, we evaluate the current plans of pulsar searches with SKA for its detectability of binaries in general.  We simulate an all sky pulsar survey as envisioned to be done with the SKA-Mid (with the standard telescope configuration) and the estimated pulsar search procedure using the envisioned pulsar search subsystem (PSS). Table\,\ref{tab:binpars} presents the range and distribution of pulsar and binary parameters from which we randomly select pulsars. 

\begin{table*}
\centering
\begin{tabular}{|l|c|c|c|}
\hline
Parameter & Symbol & Range/Value & Distribution \\
\hline
Profile shape & & Gaussian & \\
Pulse intrinsic equivalent width (pase) & $W_0$ & 0.1 & Fixed \\ 
Minimum source elevation (degrees) & $h_{\rm min}$ & $15.0$ & \\
Dispersion measure (pc/cm$^3$) & $DM$ &$\leq3000.0$ & \\
Spin period (s) & $P_0$ & $0.001\,-\,1.0$ & Uniform in ${\rm log}_{10}$ \\
Flux (mJy) & $f$ & $0.01\,-\,1.0$ & Uniform in ${\rm log}_{10}$ \\
Pulsar mass ($M_\odot$) & $M_{\rm P}$ & $1.17\,-\,2.50$ & \\
Companion mass ($M_\odot$) & $M_{\rm C}$ & $0.01\,-\,10.0$ & Uniform in ${\rm log}_{10}$ \\
Orbital period (h) & $P_{\rm B}$ &$1.0\,-\,10.0$ & Uniform in ${\rm log}_{10}$ \\
Orbital inclination (degrees) & $i$ & $0.0\,-\,90.0$ & Uniform in ${\rm cos} i$\\
Orbital eccentricity & $e$ & $0.0\,-\,1.0$ & \\
Mean anomaly (phase) & $\mu$ & $0.0\,-\,1.0$ & \\
Longitude of periastron (degrees) & $\omega$ & $0.0\,-\,360.0$ &\\
\hline
\end{tabular}
\caption{Pulsar parameters of our population synthesis. Their values have been randomly generated from a linearly uniform distribution, unless explicitly specified.}
\label{tab:binpars}
\end{table*}

For each simulated line of sight we drew a DM uniformly between 0\,pc/cm$^3$ and $2\times$ the maximum value predicted by the YMW16 model for that direction, capping the upper limit at  3000\,pc/cm$^3$ as dictated by the capability of the PSS. Scattering in the ionised interstellar medium was included by computing the pulse-broadening time, $\tau_{\rm scatt}$ at 1.0 GHz from YMW16 and scaling it to the centre of the observing band with a scattering spectral index $\alpha_{\rm scatt}=-3.5$. To mimic a typical observation of an eccentric binary sampled at an arbitrary epoch, we drew a random mean anomaly—rather than the true anomaly—so that systems are statistically more likely to be simulated away from periastron. Each observation is assumed to consist of a 10-min L-band dwell time taken with an array of 100 antennas, a compromise that balances raw sensitivity against survey speed: adding antennas increases gain but narrows the primary beam, and the resulting smaller field of view would otherwise inflate the number of pointings needed to tile the visible sky. We did not apply any additional sensitivity correction for a source’s off-axis position within a beam. All telescope-specific parameters adopted in the simulations are summarised in Table \ref{tab:ac-telescope-pars}.

\begin{table}[ht]
\centering
\begin{tabular}{|l|c|}
\hline
Parameter & Value \\
\hline
Number of antennas     & 100 \\
$T_{\rm sys}$ (K)    & 17.0 \\
Gain (K/Jy) & 5.68 \\
Central frequency (MHz)    & 1284.0 \\
Bandwidth  (MHz) & 300.0 \\
\hline
\end{tabular}
\caption{Telescope configuration parameters}
\label{tab:ac-telescope-pars}
\end{table}

The PSS limits the corrections for accelerations to a maximum of 85 acceleration trials for a maximum of 500 dispersion trials. As a trade-of between reducing radio frequency interference, and increasing the pulsar flux, we decided on an acceleration search scheme for $50\leq \rm  DM\leq 250$\,pc/cm$^3$ for the directions in the sky for which $\rm DM_{\rm max}\geq250$\,pc/cm$^3$. For the others we used $\rm DM_{\rm max}-200$\,pc/cm$^3\leq  \rm DM \leq \rm DM_{\rm max}$. A uniform spacing of  0.4\,pc/cm$^3$ keeps the total number of DM trials within the 500-trial budget. 

For a pulsar of spin period $P_0$ we determined the acceleration step $\delta a_l = c P_0 /T^2$, where $c$ is the speed of light and $T$ is the observing time. This relation comes from the requirement that the pulsar spin frequency in the Fourier spectrum should not drift more than one Fourier bin because of the pulsar acceleration. In our simulations we assumed a maximum acceleration along the line of sight $a_l$ of $\pm500$\,m\,s$^{-2}$. Given the limit on the number of acceleration trials, namely 85, we set $\delta a_l = 1000$\,m\,s$^{-2}/84=11.9$\,m\,s$^{-2}$ for those spin periods that require a minimum acceleration step larger than this value.
Table\,\ref{tab:search-strategy} summarizes all relevant parameters in the adopted search strategy.

\begin{table*}[ht]
\centering
\begin{tabular}{|l|c|}
\hline
Parameter & Value \\
\hline
Minimum dispersion measure for dedispersion (pc/cm$^3$) & 50.0 \\
Maximum dispersion measure for dedispersion (pc/cm$^3$) & 250.0 \\
Number of dispersion measure steps & 500 \\
Step in dispersion measure (pc/cm$^3$) & 0.4 \\
Minimum orbital acceleration (m/s$^2$) & -500.0 \\
Maximum orbital acceleration (m/s$^2$) & +500.0 \\
Number of acceleration steps & 85 \\
\hline
\end{tabular}
\caption{Current proposed search parameters, partly derived from the requirements of the pulsar search sub-system for the SKA. Given the hard limit of only 500 dispersion trials that can be searched with acceleration correction, we choose a nominal DM range that can be covered with these trials.}
\label{tab:search-strategy}
\end{table*}

A simulated pulsar was deemed detectable only if two conditions were met simultaneously: (i) its observed equivalent pulse width, $W$, did not exceed one-half of the spin period, $P$, and (ii) ts signal-to-noise ratio, computed from the radiometer equation, surpassed the adopted detection threshold. Objects failing criterion (i) were labelled as ``smeared out", whereas those satisfying criterion (i) but not criterion (ii) were tagged as ``low S/N". The width $W$ was taken to be the quadrature sum of the intrinsic width, $W_0$, and three broadening terms: (a) interstellar scattering, $\Delta W_{\rm scatt}$, assumed to be $1.5 \tau_{\rm scatt}$ (b) residual dispersion smearing $\Delta W_{\rm DM}$,arising from the difference between the true dispersion measure and the nearest trial DM; and (c) acceleration smearing, $\Delta W_{\rm ACC}$, set to one-half of the broadening produced by the mismatch between the true and trial line-of-sight accelerations. To quantify the impact of each effect, we applied the detection test sequentially—first using only $W_0$ and  $\Delta W_{\rm scatt}$ to estimate losses due to scattering which is completely not accounted for, then adding $\Delta W_{\rm DM}$ and $\Delta W_{\rm ACC}$, in that order. A pulsar that survived all three stages was classified as detected.

The pulsar distribution in the Galaxy, which we assumed uniform in both longitude and latitude, is a critical assumption, since the free electrons distribution, responsible for both the Ionized ISM frequency dependent signal dispersion and scattering, is not. For this reason we performed three independent simulations, for the low- ($|b_{\rm Gal}|\leq5^\circ$), then mid- ($5^\circ<|b_{\rm Gal}|\leq50^\circ$), and high-latitude ($|b_{\rm Gal}|>50^\circ$) objects, respectively. We generated $10^5$ realizations of 1000 random objects for each segment, determined the fraction of objects in our simulations that fall in the aforementioned classes, and then averaged these values across all realizations into these categories. Our results are summarized in Table\,\ref{tab:simpop1-lats}.

\begin{table*}[ht]
\centering
\begin{tabular}{|l|r|r|r|}
\hline
Class &  Low-lat \% & Mid-lat \% & High-lat \% \\
\hline
IISM scattering smeared  &  37.94$\pm$1.53  &  2.33$\pm$0.48  &  0.00$\pm$0.00 \\
IISM scattering low SNR  &  1.70$\pm$0.41  &  0.42$\pm$0.20  &  0.00$\pm$0.01 \\
Dedispersion smeared  &  22.77$\pm$1.33  &  4.83$\pm$0.68  &  0.00$\pm$0.00 \\
Dedispersion low SNR  &  0.98$\pm$0.31  &  0.71$\pm$0.27  &  0.56$\pm$0.24 \\
Acceleration smeared  &  1.01$\pm$0.32  &  2.00$\pm$0.44  &  1.96$\pm$0.44 \\
Acceleration low SNR  &  0.58$\pm$0.24  &  2.14$\pm$0.46  &  2.46$\pm$0.49 \\
Detected  &  35.02$\pm$1.51  &  87.58$\pm$1.04  &  95.02$\pm$0.69 \\
\hline
\end{tabular}
\caption{Simulation results: low- mid- and high- Galactic latitudes.}
\label{tab:simpop1-lats}
\end{table*}

The differences in the fraction of detected objects, namely $35.0\%$, $87.6\%$ and $95.0\%$ for the low-, mid- and high- latitude objects respectively, are related to the loss of objects due to the unavoidable ISM scattering, and to the residual pulse smearing due to the difference between the true pulsar DM and the trial DM. We verified this by recomputing the fraction of detections  obtained if only the objects that passed the first two criteria are considered. In doing so, obtained  95.6\%,  95.4\%, and 95.5\% for the low-, mid- and high- latitude objects respectively. The resulting consistency of these values, which we can interpret as the efficiency of the acceleration search strategy according to the assumed parameters' space and to our assumptions, clearly support our interpretation. Nevertheless these numbers are still too optimistic and need further investigation. 

\begin{figure*}[t]
  \centering
  \includegraphics[width=0.5\textwidth,angle=270.0]{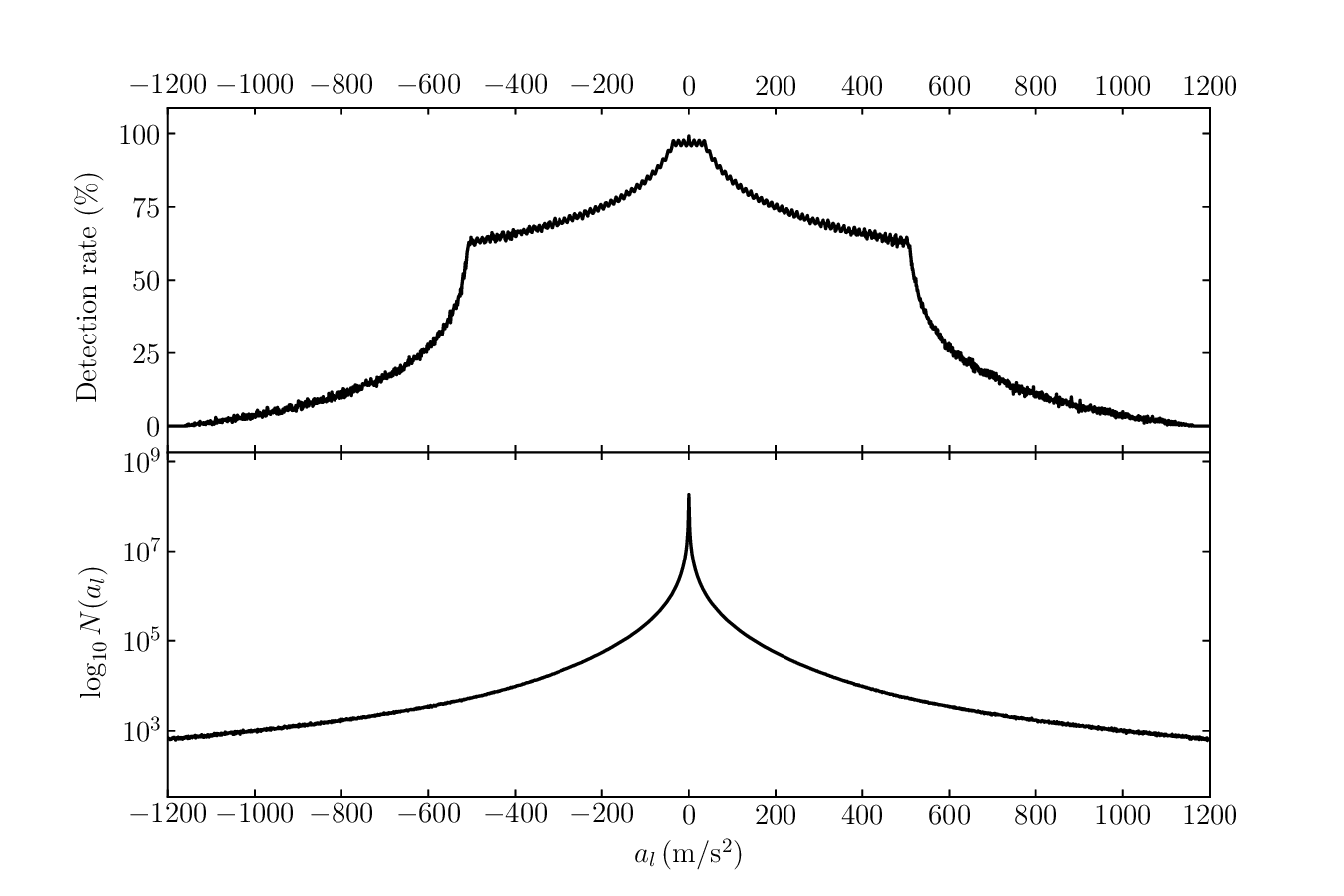}
  \caption{Upper panel: binary pulsar detection rate against the observed acceleration. Lower panel: number of generated pulsar against the observed acceleration. Both panels: the x-axis range is limited to the values for $a_l$ that resulted in at least one detection.}
  \label{fig:accelplot}
\end{figure*}

\begin{figure*}[t]
  \centering
  \includegraphics[width=7.4cm,angle=270.0]{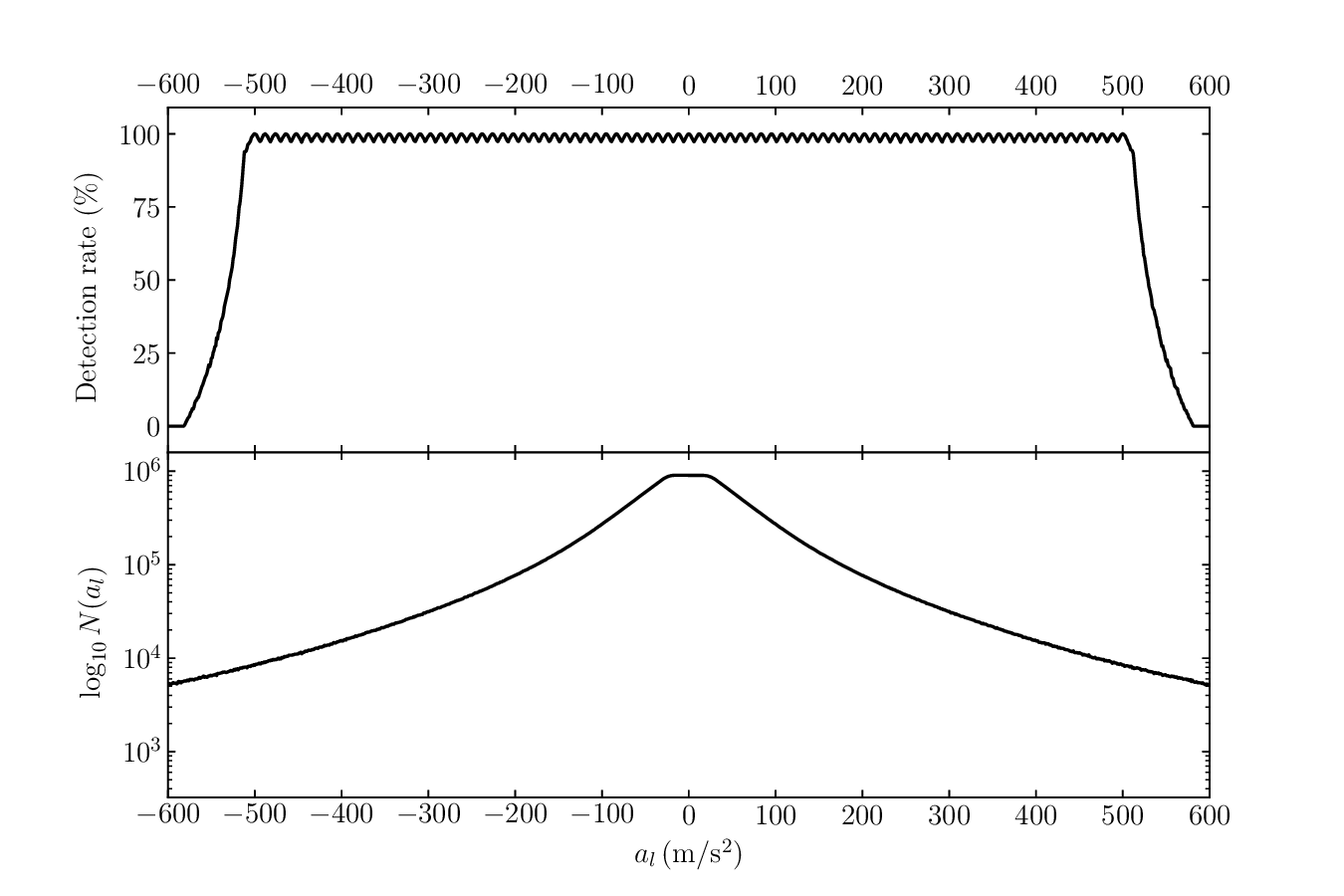}
  \caption{Same as figure\,\ref{fig:accelplot},  but for the DNS systems simulations.}
  \label{fig:accelplot-dns}
\end{figure*}

In order to understand how many binaries are missed purely due to mis-specified binary corrections, we derived the detection rate as a function of the observed acceleration. We extrapolated all simulated objects that passed the ISM and DM scrutinies, regardless of their Galactic latitude, we binned them with respect to the observed acceleration using a step of 1\,m,s$^{-2}$, and we calculated the detection rate for each bin. The upper panel of Figure\,\ref{fig:accelplot} shows that the detection rate is between $\sim62\%$ and $\sim98\%$ for $|a_l|\leq 500$\,m\,s$^{-2}$, and monotonically decreases to negligible percentages, as expected, for $|a_l|> 500$\,m,s$^{-2}$. Moreover, the detection rate is above the $95\%$ only in the range $|a_l|\lesssim 40$\,m\,s$^{-2}$, so the overall efficiency of the acceleration search mentioned above could be considered as too optimistic. The lower panel of Figure\,\ref{fig:accelplot} plots the number of generated objects against their observed acceleration. The distribution is hugely peaked at $a_l=0$\,m\,s$^{-2}$ (the vertical axis scale is logarithmic), therefore these objects dominate this statistics. For this reason we performed a second set of simulations focused on short orbital period DNS systems. We constrained the pulsar spin period to be between 15 and 100\,ms (uniformly in ${\rm log}_{10}P_0$), the orbital period between 30\,minutes and 2\,hours, and the companion mass in the same range of the pulsar mass. The statistics of this second simulation, for which we generated 50,000 realizations, are presented in table\,\ref{tab:simpop1-msps}. The detection rates are $\sim32\%$, $\sim89\%$, and $\sim96\%$ for the low-, mid- and high-latitude surveys, respectively. While in the mid- and high-latitude simulations the object loss due to the dedispersion plan can be considered acceptable, in the low-latitude survey such loss is quite important. Infact, if one recomputes the percentage of objects classified as ``dispersion smeared", among the ones that survive the IISM scattering screening, one obtains that the hypothesized dedispersion plan has an {\it inefficiency} factor of the $\sim50\%$ for the search of DNS systems as described above. The average efficiency of the assumed acceleration search strategy is now $\sim96\%$ ($95.9\pm1.1$\% for the low-, $95.9\pm0.6$\% for the mid- and high-latitude surveys respectively). The plot of the detection rate against the observed acceleration for DNS systems helps in investigating this optimistic result. The upper panel of figure\,\ref{fig:accelplot-dns} shows that the detection rate is constantly around the $100\%$ for $|a_l|\lesssim 500$\,m,s$^{-2}$. This is due to the fact that we are considering spin periods of at least 15\,ms. In fact, spin periods in the considered range require maximum acceleration steps of at least 12.5\,m\,s$^2$, i.e. always larger than the 11.9\,m\,s$^2$ value that results by dividing the considered accelerations range ($-500$\,m\,s$^2\leq a_l \leq +500$\,m\,s$^2$) into 85 equal steps. Also in this case, the statistics are dominated by the objects whose observed acceleration is around 0\,m\,s$^{-2}$, as indicated in the lower panel of figure\,\ref{fig:accelplot-dns}.

We also explored the potential of searching PSRBH systems. This time we constrained the pulsar spin period to be between 100\,ms and 10\,s, assuming that in such systems the pulsar is a young one, the orbital period between 30\,minutes and 20\,hours, as a result of the peculiar evolutionary path of their progenitors (both uniformly in ${\rm log}_{10}P_0$), and the companion mass between 3 and 20\,$M_\odot$, i.e. the companion is a stellar mass black hole. Also in this case we generated 50,000 realizations. We obtained (table\,\ref{tab:simpop1-nsbh}) detection rates of the $\sim76\%$ for the low-latitude objects, and of the $\sim97\%$ mid- and high- latitude ones, respectively. In this case the loss of objects because of the assumed dedispersion plan, the $\sim11\%$, is less severe for the low-latitude objects. This is not unexpected because now we are considering longer spin periods. The acceleration plot for these class of systems is very similar to the one for the DNS systems, as expected because of the larger spin periods now considered, namely the detection rate is of the 100\% for $|a_l|\leq500$\,m\,s$^{-2}$, and it decreases at higher values. The main difference between the DNS systems is that we now obtain detections also for $|a_l|$ values of the order of a few 1000\,m\,s$^{-2}$, and this result is given by both the larger spin and orbital periods now considered.

In summary, we have explored the potentialities of a search of pulsars in binary systems. In fact, on one side we assumed a very efficient strategy for the acceleration search. On the other side, the considered dedispersion plan resulted quite inefficient for the low-latitude objects, i.e. the ones placed at positions that result in DMs of the order of serveral hundreds of pc\,cm$^{-3}$. On top of that we cannot neglect that the aforementioned results clearly depend on the assumed parameters' space, despite the fact that we tried to choose reasonable ranges for each considered parameter. The most significative case, in this respect, is given by the adopted interval for the pulsars' fluxes: we considered only those values that ensure the detectability of the pulsar with a SNR of 10 or higher, thus neglecting those objects that wouldn't meet this requirement, given the assumed telescope setup. A similar argument can be clearly applied to all other physical parameters, but a discussion on the impact of the adopted ranges is beyond the scope of this section.

\begin{table*}[ht]
\centering
\begin{tabular}{|l|r|r|r|}
\hline
Class &  Low-lat \% & Mid-lat \% & High-lat \% \\
\hline
IISM Scattering smeared  &  31.56$\pm$1.47  &  0.76$\pm$0.27  &  0.00$\pm$0.00 \\
IISM Scattering low SNR  &  1.61$\pm$0.40  &  0.18$\pm$0.13  &  0.00$\pm$0.00 \\
Dedispersion smeared  &  33.38$\pm$1.49  &  6.41$\pm$0.78  &  0.00$\pm$0.00 \\
Dedispersion low SNR  &  0.61$\pm$0.25  &  0.29$\pm$0.17  &  0.00$\pm$0.01 \\
Acceleration smeared  &  1.01$\pm$0.32  &  2.74$\pm$0.51  &  2.93$\pm$0.53 \\
Acceleration low SNR  &  0.31$\pm$0.18  &  1.03$\pm$0.32  &  1.15$\pm$0.34 \\
Detected  &  31.52$\pm$1.47  &  88.60$\pm$1.01  &  95.91$\pm$0.63 \\
\hline
\end{tabular}
\caption{Same as table\,\ref{tab:simpop1-lats}, but for the DNS systems simulations.}
\label{tab:simpop1-msps}
\end{table*}

\begin{table*}[ht]
\centering
\begin{tabular}{|l|r|r|r|}
\hline
Class &  Low-lat \% & Mid-lat \% & High-lat \% \\
\hline
IISM Scattering smeared  &  9.09$\pm$0.91  &  0.05$\pm$0.07  &  0.00$\pm$0.00 \\
IISM Scattering low SNR  &  0.76$\pm$0.27  &  0.02$\pm$0.04  &  0.00$\pm$0.00 \\
Dedispersion smeared  &  11.32$\pm$1.00  &  0.71$\pm$0.27  &  0.00$\pm$0.00 \\
Dedispersion low SNR  &  1.55$\pm$0.39  &  0.19$\pm$0.14  &  0.00$\pm$0.00 \\
Acceleration smeared  &  1.49$\pm$0.38  &  2.26$\pm$0.47  &  2.30$\pm$0.47 \\
Acceleration low SNR  &  0.11$\pm$0.11  &  0.18$\pm$0.14  &  0.19$\pm$0.14 \\
Detected  &  75.68$\pm$1.36  &  96.58$\pm$0.58  &  97.50$\pm$0.49 \\
\hline
\end{tabular}
\caption{Same as table\,\ref{tab:simpop1-lats}, but for the pulsar$-$black hole systems simulations.}
\label{tab:simpop1-nsbh}
\end{table*}

\begin{figure*}[t]
  \centering
  \includegraphics[width=7.4cm,angle=270.0]{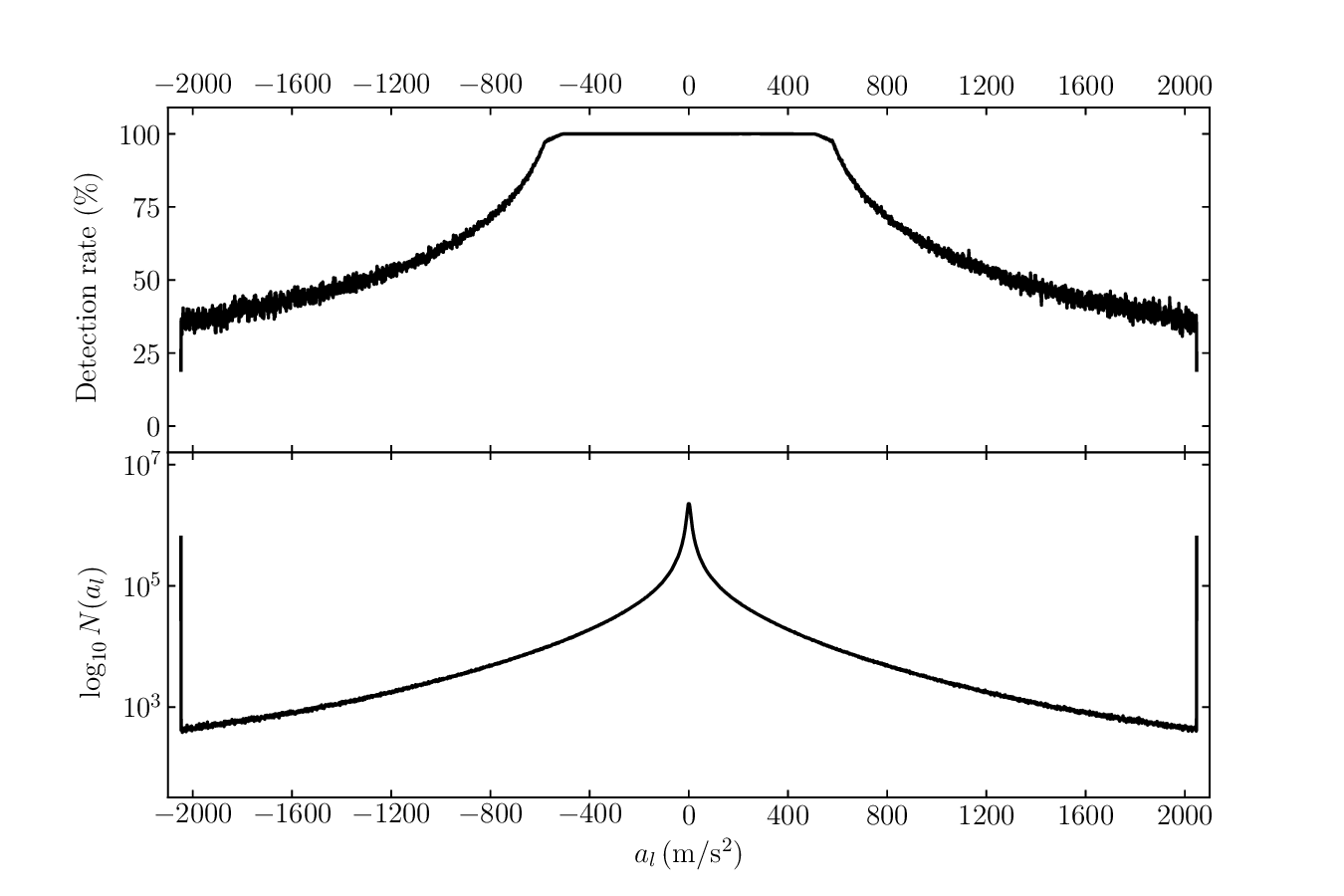}
  \caption{Same as figure\,\ref{fig:accelplot},  but for the pulsar$-$black hole systems simulations.}
  \label{fig:accelplot-nsbh}
\end{figure*}

\subsection{Importance of observing bands}

\begin{table*}[t]
\centering
{\small
\caption{Precision of PK parameters for PSR J1913+1102 with simulated data from SKA AA* and SKA AA4 (2028-2038) combined with previous data (using simulated data) compared to results using 12 yr data (Miao et al. in prep.). The last column of the table represents the parameter precision obtained from the simulation, assuming that the SKA AA4 can provide observations at frequencies ranging from 1650 to 3050 MHz.}
\vspace{5pt}
\begin{tabular}{c|cccc}
    \hline\hline
        & Comb AA* / 12yr & Comb AA4 / 12yr & Comb AA4 (S Band)\,/12yr \\ \hline
        $\dot{\omega}$ & 3.8 & 4.5 & 9.1\\ 
        $\gamma_\mathrm{E}$ & 1.8 & 1.9 & 2.9\\ 
        $\dot{P}_\mathrm{b}$ & 5.9 & 6.7 & 16.2\\ 
        $h_{3}$ & 2.0 & 2.3 & 5.4 \\ \hline
\end{tabular}
\label{tab:J1913}
}
\end{table*}

Apart from instantaneous sensitivity, it is also imperative to choose the right observing band to maximise the scientific outcome from SKA observations. Although it is generally true that pulsars have higher flux at lower frequencies covered by the UHF and L-band receivers of SKA, the flux of a pulsar is not the only determinant of timing precision. The width of the pulse and presence of high frequency structures in the pulse profile also play an important role, as they directly affect the precision of the template matching technique used to extract ToAs. Pulsars generally tend to narrower pulse profiles and multiple components at higher frequencies partly owing to the radius-to-frequency mapping of pulsar emission, but also to the much mitigated effects of interstellar medium such as pulse scattering. For many relativistic binaries, S-band is more suited as a trade-off between spectral index and profile shape to obtain the best possible ToA precision. A classic example is the pulsar PSR J1913+1102, a $\sim 5$ hour double neutron star system \citep{Lazarus2016ApJ,Ferdman2020}. Its pronounced mass asymmetry makes it a prime target for testing dipolar gravitational-wave emission in scalar–tensor gravity. Following Arecibo’s closure, FAST has provided continued timing of this source since November 2021. We simulated 10 yr of SKA AA*/AA4 ToAs (2028–2038) taken at L-band and combined them with Arecibo + FAST data; The resultant parameters (Table~\ref{tab:J1913}) show no significant improvement, reflecting FAST’s $\sim5\times$ higher antenna gain relative to SKA-mid even at its AA4 configuration. However, keeping everything else fixed and just changing the observations to be at S-band instead, we obtain an improvement of $>2\times$ for most of the parameters, as seen in  Table~\ref{tab:J1913}. 
The above serves as a poster child for why the inclusion of an S-band receiver to SKA Phase 1 will provide the best returns for the time invested on some observed pulsars. S-band also serves as a good trade off for pulsar flux and interferometric precision for VLBI experiments as described in detail later.

\subsection{Multiwavelength contributions to gravity tests}

While for most pulsars, measuring relativistic effects and testing gravity involve decades of radio observations, the presence of optically bright WD companions has allowed the determination of additional mass constraints for these rare systems via the combination of radio and optical data. The optical spectroscopy adds an independent WD mass estimate from measurements of the width of the Balmer lines \cite{Antoniadis2013}, although there can be systematic uncertainties that need to be carefully accounted for(c.f. \cite{Saffer2025}). By combining the radial velocity measurements of the WD and pulsar, a precise mass ratio can also be obtained; combining all of these, the component masses of the system can be precisely estimated. If the system has a short enough orbital period, it exhibits relativistic effects that can be measured in a theory-independent way from pulsar timing. These effects are then compared with predictions of
theories of gravity, computed using the optical mass measurements, thereby performing a test of those theories.

A prime example of this technique is PSR J1738+0333 which has a $V = 21.7$ WD companion \citep{Freire2012}. Optical spectroscopy of this companion allowed precise mass measurements: the companion mass is $0.181^{+0.007}_{-0.005} \, \rm M_{\odot}$ and the pulsar mass is $1.47^{+0.07}_{-0.06} \, \rm M_{\odot}$. For these masses, GR predicts a decay of the orbital period ($\dot{P}_{\rm b}$) due to gravitational wave (GW) emission of $-27.7^{1.5}_{-1.9}\times10^{-15}ss^{-1}$. The actual measurements from pulsar timing of $-25.9\pm 3.2\times10^{-15}ss^{-1}$ are consistent with GR's prediction, and hence GR passes this test. However, this is not the case for many alternative gravity theories. For instance, a significant part of the parameter space of Damour-Esposito-Far\`ese (DEF) gravity \citep{Damour:1992PhRvD,Freire2012} was ruled out by this experiment. This system continues to be the best test for some parts of the parameter space of DEF gravity to date. This is possible because the PSR J1738+0333 system is gravitationally asymmetric: it has a strongly gravitating NS with a weakly gravitating WD companion. Such an arrangement enhances the amplitude of DGW emission predicted by a number of alternative theories of gravity (such as DEF gravity). This additional DGW emission was not observed in the orbital decay of PSR J1738+0333, thus ruling out such theories.

A similar pulsar where SKA could potential play a significant role is PSR J1036$-$8317, a 3.4~ms pulsar in an 8 hr circular orbit around an optically bright HeWD. Observations with the MeerKAT telescope showed a $10 \times$ increase in precision compared to earlier Parkes observations, and have already hinted at a 2-3$\sigma$ measurement of the orbital period decay. The position of this pulsar is coincident with a \textit{Gaia} source, inferred to be a WD based on its magnitude and color. Optical observations of this with the NTT telescope have already been done; we expect a measurement of its radial velocity and surface gravity, from which we can infer the binary mass ratio and the mass of the WD to a precision that is twice that of PSR J1738+0333. SKA observations of this system would soon solidify the measurement of $\dot P_\mathrm{b}$ and also produce independent mass measurement via the Shapiro delay. Our nominal simulations show a $100\times$ improvement in $\dot P_\mathrm{b}$ and an independent 14$\sigma$
measurement of the masses. All these measurements together have the potential to surpass PSR J1738+0333's records and provide the most stringent, radiative test of scalar tensor gravity. Assuming nominal values for the pulsar and companion masses of the PSR J1036$-$8317 system, potential constraints from SKA timing are plotted in Fig \ref{fig:a0b0}. It can be seen that the pulsar has the potential to become the most constraining system for positive $\beta_0$ values. 

\subsection{The importance of SKA VLBI}
\label{sec:VLBI}
VLBI observations of pulsars contribute to gravitational tests primarily by measuring the parallax and/or proper motion more precisely than can be achieved using pulsar timing, reducing the uncertainty in timing parameters that depend on these quantities. High precision VLBI astrometry to facilitate these goals requires a combination of high sensitivity (to minimise statistical uncertainty on the position measurement, especially for faint pulsars) and excellent calibration (to minimise systematic position shifts caused by unmodelled propagation delays). The latter requirement is best fulfilled by observing calibrator sources that are closer to the target pulsar (to minimise spatial extrapolation errors) and/or to observe multiple calibrator sources, enabling calibration solutions to be interpolated to the target pulsar position (c.f. \citealt{Chatterjee2009}). 

To date, the second requirement (nearby calibrator sources) has meant that the largest and most precise VLBI astrometry campaigns on pulsars have been performed using the Very Long Baseline Array, as its relatively large field of view means that nearby calibration sources can generally be observed contemporaneously at L-band frequencies \citep[e.g.][]{Deller19,Ding23}. Pulsars that are not visible from the northern hemisphere have been much more challenging to enact VLBI, as the large dishes and phased arrays used in the Southern Long Baseline Array have precluded the use of in-beam calibration and hence highly precise astrometry for pulsars further South.

The inclusion of SKA-VLBI as an element of southern VLBI arrays will facilitate a substantial improvement for these southern pulsars, since the SKA will provide multiple tied beams enabling contemporaneous calibration to be derived for multiple sources within $\sim$30 arcminutes of the target pulsar. For pulsars south of a declination of $\sim$ $-$25 degrees, the theoretical astrometric precision achievable with VLBI will jump from well below that achievable in the North currently, to better than the current state of the art. The point source sensitivity of a VLBI array containing SKA1-mid (AA4) plus other southern antennas providing at least a 30 arcminute field of view at L band is a factor of four better than the current VLBA, driving an astrometric improvement of the same order and making it possible to make precise parallax measurements for sources up to distances of 10 kpc and even beyond (as current-generation surveys can already provide 10\% level parallax accuracy out to 2.5 kpc routinely; \citealp{Deller19}) . This is a result of the improved sensitivity both directly impacting the target pulsar signal--to--noise (and hence nominal astrometric precision) and the calibration fidelity (enabled by the transition from calibration extrapolation from a single nearby in-beam calibator, to interpolation between solutions provided by multiple in-beam calibrators, reducing systematic sources of error).

The most significant impact of this capability will be for pulsars in southerly regions inaccessible to northern hemisphere facilities, and the faintest pulsars where point source sensitivity is the limiting factor to astrometric precision such as PSR J1141$-$6545. Astrometric precision for SKA-VLBI can also be improved by observing at S band rather than L band, as the impact of ionospheric modelling errors decreases. This is yet another reason to consider the inclusion of S band receivers in SKA-mid Phase 1.

\subsection{Impact of multimessenger observations}
\label{sec:multimessenger}
In the future, with the continued construction of more telescopes and gravitational wave detectors, multimessenger detection will undoubtedly become a cornerstone of astrophysics research. The SKA will play an important role in multimessenger detection of observed relativistic binaries. To date, more than 20 DNS systems have been discovered by radio telescopes. Among these DNS systems, PSR J1946+2052 has the shortest orbital period, namely $P_\mathrm{b}=1.88\,{\rm hr}$. GW observations from LIGO/Virgo have detected two DNS systems that are in the merger phase. Therefore, the orbital period of the discovered DNSs has a gap between $\sim2\,$hr and $\sim1\,{\rm ms}$ at the phase of merger. 
Detecting DNS systems with orbital periods on the order of minutes would provide an even more relativistic laboratory for testing theories of gravity. But we have not yet discovered DNS systems with $P_{b}\sim{\rm min}$ by radio observations yet, because the fast-changing Doppler shift caused by the orbital acceleration of the pulsar smears the pulsar signal in the Fourier domain, and correcting the smearing will be computationally prohibitive. Some techniques such as the sideband search method of \textsc{PRESTO} \citep{Ransom2003} partly overcomes this problem. 

Laser Interferometer Space Antenna (LISA), a future space-based GW detector, which is expected to be operational in the 2030s, is sensitive to mHz bands, making DNS systems with orbital periods of approximately minutes prime candidates for detection. \citet{Kyutoku:2019} proposed a multimessenger strategy combining SKA and LISA to search for radio pulsars in orbits with periods shorter than 10 minutes. LISA achieves precise sky localization, reducing the search area for the SKA to a very small region of a few degrees on the sky. This significantly decreases the number of required pointings and improves the efficiency of SKA observations. 
Additionally, LISA provides high-precision measurements of orbital frequency and binary parameters, such as chirp mass and inclination angle. These parameters are critical for correcting Doppler smearing in the SKA’s radio observations of pulsars in tight binary systems. 
By combining these capabilities, LISA dramatically enhances the SKA's ability to detect faint radio pulsars in short-period binary systems, especially those with orbital periods shorter than 10 minutes.

\section{Summary}

As discussed in section~\ref{sec:GR}, GR and its extensions predict a rich array of phenomenology that is observable in the timing of radio pulsars. As discussed in section~\ref{sec:state_of_the_art}, and more extensively by \cite{FreireWex2024}, the timing of radio pulsars has provided some of the best tests of gravity theories.  These are strong-field effects in the sense that the pulsars themselves have substantial gravitational binding energies. This implies that radio timing makes binary pulsars extremely sensitive to violations of symmetries of the gravitational interaction, like the strong equivalence principle, even if the orbits themselves are not very compact, as in the case of the pulsar in a triple star system. Compact binary pulsars provide the most precise tests known of the quadrupole formula for the orbital decay induced by gravitational wave emission, which are fundamental tests of the radiative properties of gravity.

As discussed in section~\ref{sec:SKA}, the prospects for improvements in these precision for the near future are excellent, especially with the sensitivity of the SKA:

\begin{itemize}

\item The continuation of some of the timing experiments of known systems with the higher sensitivity of the SKA will greatly improve many of the tests done with these systems. For instance, continued timing of the double pulsar, which assumes the use of the SKA, might constrain the MoI of PSR J0737−3037A to within 10\% by 2030 \citep{Hu:2020ubl}, apart from significantly improving the precision in the measurement of the orbital decay and other gravity tests (section~\ref{sec:inst_sensitivity}). Although such a determination of the MoI assumes GR to provide the correct description of the needed PK parameters (and will eventually help to constrain the EoS), interpreted as a LT test it still allows to probe for significant short-range deviations from GR that only affect pulsar A locally (see discussion in \citealt{Hu:2020ubl}). As an example of the impact of sensitivity, just 3 years of MeerKAT data yielded a photon propagation test in the double pulsar \citep{Hu2022} that is a factor of two better than the previous one based on 16 years of data from 6 different telescopes \citep{Kramer2021}.

\item The pulsar field in general has been driven, from
the start, by the discovery of more relativistic binary systems. The rate of pulsar discoveries has recently increased significantly, with 1000 new pulsars having been found by FAST and MeerKAT already. As described in section~\ref{sec:searches}, the rate will increase further with the sensitivity of the SKA. Furthermore, the rate of discovery of recycled pulsars - and especially recycled pulsars in very compact orbits - is increasing even faster, because of the much improved time and spectral resolution of the search data, the much improved computing capabilities and search algorithms.

All this will very likely lead to the discovery not only of more extreme versions of the currently known systems, which will allow new leaps in the precision for the types of tests described above, but also of completely new types of systems, such as pulsar-BH binaries, perhaps more massive and compact than NGC 1851E \citep{Barr2024}. Such systems will allow gravity tests that were until now beyond the testing power of pulsar timing \citep{WexKopeikin1999,Liu:2014uka,SeymourYagi2018}. The sensitivity of the SKA will be very important for timing these new discoveries, which are likely to be weak (see Sect.~\ref{sec:timing}).

The prospect of detecting very compact binary pulsars, especially DNSs (or pulsar-BH systems), is of paramount importance for tests of gravity theories. A general reason is the attainable significance of the radiative test in the presence of contaminants, which improves as $P_{\rm b}^{-8/3}$. Such systems would also allow the measurement of the full precession cycle of relativistic spin-orbit coupling on reasonable timescales: for instance, a DNS with an orbital period of 30 min would have a geodetic precession period of about 5 years, which would then be measured precisely from the repeating changes in the pulse profile of the system.  Hence all-sky or Galactic plane searches for these systems must be the sensitive to these orbital periods. 

\item As discussed in section~\ref{sec:VLBI} tests will be further improved if the SKA has a VLBI capability. For many pulsars - including PSR~J0737$-$3039A - the limiting factor in the precision of gravity tests is the lack of knowledge of the distance to the pulsar \citep{Kramer2021}. The VLBI capability should greatly improve the measurement of pulsar distance

\item As discussed in section~\ref{sec:multimessenger}, the SKA will have several synergies with other wavelengths and with GW observatories. In particular, {\em it os extremely important that it operatee at the same time as the Laser Interferometry Space Antenna (LISA)}. This opens up a great synergy: Very compact binary pulsars to be found by the SKA will be detectable at good S/N by LISA mission if they are not too distant from Earth \citep{Thrane2020}. This mission will also find, independently, the most compact NS–NS, NS–WD or NS–BH systems of our Galaxy \citep{Lau2020}. Perhaps some of these NSs will be detectable as pulsars in targeted radio surveys. In either case, binary pulsar experiments would become“multi-messenger” experiments, allowing entirely new tests of gravity theories \citep{Thrane2020,Miao2021}.

\end{itemize}

Finally, the discovery of a pulsar in a relativistic orbit around the supermassive BH at the centre of our Galaxy would allow unprecedented tests of BH physics, in particular in combination with tests from other observations in this extreme gravity environment (see e.g., \citealt{Psaltis2016} and references therein).

\section*{Acknowledgements}
We thank Norbert Wex for comments on the manuscript and for producing the figure on scalar tensor gravity constraints. We thank the anonymous referee for a thorough reading of the manuscript and providing comments to improve it. V.~V.~K. and V.~B. acknowledges financial support from the European Research Council (ERC) starting grant ``COMPACT" (Grant agreement number 101078094). L.S.\ and Z.H.\ were supported by the National SKA Program of China (2020SKA0120300), the Beijing Natural Science Foundation (1242018), and the Max Planck Partner Group Program funded by the Max Planck Society. E.H. is grateful for support from the Deutsche Forschungsgemeinschaft (DFG, German Research Foundation) under Germany’s Excellence Strategy – EXC-2123 QuantumFrontiers – 390837967.
H.~H. acknowledges support from the PRIME programme of the German Academic Exchange Service (DAAD) with funds from the German Federal Ministry of Education and Research (BMBF).
M.E.L. is supported by an Australian Research Council (ARC) Discovery Early Career Research Award DE250100508. X.M. is supported by the National Natural Science Foundation of China (12203072). A.C., A.C. and D.P. acknowledge financial support under the National Recovery and Resilience Plan (NRRP), Mission 4, Component 2, Investment 1.1, Call for tender No. 104 published on 2.2.2022 by the Italian Ministry of University and Research (MUR), funded by the European Union - NextGenerationEU - Project Title GUVIRP-Gravity tests in the UltraViolet and InfraRed with Pulsar timing – CUP C53D23000910006 - Grant Assignment Decree No. 962 adopted on 30/06/2023 by the Italian Ministry of University and Research (MUR).  Pulsar Research at UBC is supported by an NSERC Discovery Grant and by the Canadian Institute for Advanced Research.

\bibliographystyle{aasjournal}

\bibliography{chapter} 

\begin{thebibliography}{}
\expandafter\ifx\csname natexlab\endcsname\relax\def\natexlab#1{#1}\fi
\providecommand{\url}[1]{\href{#1}{#1}}
\providecommand{\dodoi}[1]{doi:~\href{http://doi.org/#1}{\nolinkurl{#1}}}
\providecommand{\doeprint}[1]{\href{http://ascl.net/#1}{\nolinkurl{http://ascl.net/#1}}}
\providecommand{\doarXiv}[1]{\href{https://arxiv.org/abs/#1}{\nolinkurl{https://arxiv.org/abs/#1}}}

\bibitem[{{Abbate} {et~al.}(2025){Abbate}, {Chatterjee}, {Cordes}, {Demorest},
  {Desvignes}, {Eatough}, {Hackmann}, {Hu}, {Kramer}, {Lazio}, {Lee}, {Liu},
  {Rammala}, {Ransom}, {Saowanit}, {Shao}, {Torne}, {Wharton},
  {Wongphechauxsorn}, {Zhu}, \& {{The SKA Pulsar Science Working
  Group}}}]{Abbate2025_SKA_GalCen}
{Abbate}, F., {Chatterjee}, S., {Cordes}, J., {et~al.} 2025

\bibitem[{{Abbott} {et~al.}(2016){Abbott}, {Abbott}, {Abbott}, {Abernathy},
  {Acernese}, {Ackley}, {Adams}, {Adams}, {Addesso}, {Adhikari}, {Adya},
  {Affeldt}, {Agathos}, {Agatsuma}, {Aggarwal}, {Aguiar}, {Aiello}, {Ain},
  {Ajith}, {Allen}, {Allocca}, {Altin}, {Anderson}, {Anderson}, {Arai},
  {Arain}, {Araya}, {Arceneaux}, {Areeda}, {Arnaud}, {Arun}, {Ascenzi},
  {Ashton}, {Ast}, {Aston}, {Astone}, {Aufmuth}, {Aulbert}, {Babak}, {Bacon},
  {Bader}, {Baker}, {Baldaccini}, {Ballardin}, {Ballmer}, {Barayoga},
  {Barclay}, {Barish}, {Barker}, {Barone}, {Barr}, {Barsotti}, {Barsuglia},
  {Barta}, {Bartlett}, {Barton}, {Bartos}, {Bassiri}, {Basti}, {Batch},
  {Baune}, {Bavigadda}, {Bazzan}, {Behnke}, {Bejger}, {Belczynski}, {Bell},
  {Bell}, {Berger}, {Bergman}, {Bergmann}, {Berry}, {Bersanetti}, {Bertolini},
  {Betzwieser}, {Bhagwat}, {Bhandare}, {Bilenko}, {Billingsley}, {Birch},
  {Birney}, {Birnholtz}, {Biscans}, {Bisht}, {Bitossi}, {Biwer}, {Bizouard},
  {Blackburn}, {Blair}, {Blair}, {Blair}, {Bloemen}, {Bock}, {Bodiya}, {Boer},
  {Bogaert}, {Bogan}, {Bohe}, {Bojtos}, {Bond}, {Bondu}, {Bonnand}, {Boom},
  {Bork}, {Boschi}, {Bose}, {Bouffanais}, {Bozzi}, {Bradaschia}, {Brady},
  {Braginsky}, {Branchesi}, {Brau}, {Briant}, {Brillet}, {Brinkmann},
  {Brisson}, {Brockill}, {Brooks}, {Brown}, {Brown}, {Brown}, {Buchanan},
  {Buikema}, {Bulik}, {Bulten}, {Buonanno}, {Buskulic}, {Buy}, {Byer},
  {Cabero}, {Cadonati}, {Cagnoli}, {Cahillane}, {Bustillo}, {Callister},
  {Calloni}, {Camp}, {Cannon}, {Cao}, {Capano}, {Capocasa}, {Carbognani},
  {Caride}, {Diaz}, {Casentini}, {Caudill}, {Cavagli{\`a}}, {Cavalier},
  {Cavalieri}, {Cella}, {Cepeda}, {Baiardi}, {Cerretani}, {Cesarini},
  {Chakraborty}, {Chalermsongsak}, {Chamberlin}, {Chan}, {Chao}, {Charlton},
  {Chassande-Mottin}, {Chen}, {Chen}, {Cheng}, {Chincarini}, {Chiummo}, {Cho},
  {Cho}, {Chow}, {Christensen}, {Chu}, {Chua}, {Chung}, {Ciani}, {Clara},
  {Clark}, {Cleva}, {Coccia}, {Cohadon}, {Colla}, {Collette}, {Cominsky},
  {Constancio}, {Conte}, {Conti}, {Cook}, {Corbitt}, {Cornish}, {Corsi},
  {Cortese}, {Costa}, {Coughlin}, {Coughlin}, {Coulon}, {Countryman},
  {Couvares}, {Cowan}, {Coward}, \& {Cowart}}]{GW150914}
{Abbott}, B.~P., {Abbott}, R., {Abbott}, T.~D., {et~al.} 2016, \prl, 116,
  061102, \dodoi{10.1103/PhysRevLett.116.061102}

\bibitem[{{Abbott} {et~al.}(2017){Abbott}, {Abbott}, {Abbott}, {Acernese},
  {Ackley}, {Adams}, {Adams}, {Addesso}, {Adhikari}, {Adya}, {Affeldt},
  {Afrough}, {Agarwal}, {Agathos}, {Agatsuma}, {Aggarwal}, {Aguiar}, {Aiello},
  {Ain}, {Ajith}, {Allen}, {Allen}, {Allocca}, {Altin}, {Amato}, {Ananyeva},
  {Anderson}, {Anderson}, {Angelova}, {Antier}, {Appert}, {Arai}, {Araya},
  {Areeda}, {Arnaud}, {Arun}, {Ascenzi}, {Ashton}, {Ast}, {Aston}, {Astone},
  {Atallah}, {Aufmuth}, {Aulbert}, {AultONeal}, {Austin}, {Avila-Alvarez},
  {Babak}, {Bacon}, {Bader}, {Bae}, {Bailes}, {Baker}, {Baldaccini},
  {Ballardin}, {Ballmer}, {Banagiri}, {Barayoga}, {Barclay}, {Barish},
  {Barker}, {Barkett}, {Barone}, {Barr}, {Barsotti}, {Barsuglia}, {Barta},
  {Barthelmy}, {Bartlett}, {Bartos}, {Bassiri}, {Basti}, {Batch}, {Bawaj},
  {Bayley}, {Bazzan}, {B{\'e}csy}, {Beer}, {Bejger}, {Belahcene}, {Bell},
  {Berger}, {Bergmann}, {Bernuzzi}, {Bero}, {Berry}, {Bersanetti}, {Bertolini},
  {Betzwieser}, {Bhagwat}, {Bhandare}, {Bilenko}, {Billingsley}, {Billman},
  {Birch}, {Birney}, {Birnholtz}, {Biscans}, {Biscoveanu}, {Bisht}, {Bitossi},
  {Biwer}, {Bizouard}, {Blackburn}, {Blackman}, {Blair}, {Blair}, {Blair},
  {Bloemen}, {Bock}, {Bode}, {Boer}, {Bogaert}, {Bohe}, {Bondu}, {Bonilla},
  {Bonnand}, {Boom}, {Bork}, {Boschi}, {Bose}, {Bossie}, {Bouffanais}, {Bozzi},
  {Bradaschia}, {Brady}, {Branchesi}, {Brau}, {Briant}, {Brillet}, {Brinkmann},
  {Brisson}, {Brockill}, {Broida}, {Brooks}, {Brown}, {Brown}, {Brunett},
  {Buchanan}, {Buikema}, {Bulik}, {Bulten}, {Buonanno}, {Buskulic}, {Buy},
  {Byer}, {Cabero}, {Cadonati}, {Cagnoli}, {Cahillane}, {Calder{\'o}n
  Bustillo}, {Callister}, {Calloni}, {Camp}, {Canepa}, {Canizares}, {Cannon},
  {Cao}, {Cao}, {Capano}, {Capocasa}, {Carbognani}, {Caride}, {Carney},
  {Carullo}, {Casanueva Diaz}, {Casentini}, {Caudill}, {Cavagli{\`a}},
  {Cavalier}, {Cavalieri}, {Cella}, {Cepeda}, {Cerd{\'a}-Dur{\'a}n},
  {Cerretani}, {Cesarini}, {Chamberlin}, {Chan}, {Chao}, {Charlton}, {Chase},
  {Chassande-Mottin}, {Chatterjee}, {Chatziioannou}, {Cheeseboro}, {Chen},
  {Chen}, {Chen}, {Cheng}, {Chia}, {Chincarini}, {Chiummo}, {Chmiel}, {Cho},
  {Cho}, {Chow}, {Christensen}, {Chu}, {Chua}, \& {Chua}}]{GW170817}
---. 2017, \prl, 119, 161101, \dodoi{10.1103/PhysRevLett.119.161101}

\bibitem[{{Abbott} {et~al.}(2020){Abbott}, {Abbott}, {Abbott}, {Abraham},
  {Acernese}, {Ackley}, {Adams}, {Adhikari}, {Adya}, {Affeldt}, {Agathos},
  {Agatsuma}, {Aggarwal}, {Aguiar}, {Aiello}, {Ain}, {Ajith}, {Allen},
  {Allocca}, {Aloy}, {Altin}, {Amato}, {Anand}, {Ananyeva}, {Anderson},
  {Anderson}, {Angelova}, {Antier}, {Appert}, {Arai}, {Araya}, {Areeda},
  {Ar{\`e}ne}, {Arnaud}, {Aronson}, {Arun}, {Ascenzi}, {Ashton}, {Aston},
  {Astone}, {Aubin}, {Aufmuth}, {AultONeal}, {Austin}, {Avendano},
  {Avila-Alvarez}, {Babak}, {Bacon}, {Badaracco}, {Bader}, {Bae}, {Baird},
  {Baker}, {Baldaccini}, {Ballardin}, {Ballmer}, {Bals}, {Banagiri},
  {Barayoga}, {Barbieri}, {Barclay}, {Barish}, {Barker}, {Barkett}, {Barnum},
  {Barone}, {Barr}, {Barsotti}, {Barsuglia}, {Barta}, {Bartlett}, {Bartos},
  {Bassiri}, {Basti}, {Bawaj}, {Bayley}, {Baylor}, {Bazzan}, {B{\'e}csy},
  {Bejger}, {Belahcene}, {Bell}, {Beniwal}, {Benjamin}, {Berger}, {Bergmann},
  {Bernuzzi}, {Berry}, {Bersanetti}, {Bertolini}, {Betzwieser}, {Bhandare},
  {Bidler}, {Biggs}, {Bilenko}, {Bilgili}, {Billingsley}, {Birney},
  {Birnholtz}, {Biscans}, {Bischi}, {Biscoveanu}, {Bisht}, {Bitossi},
  {Bizouard}, {Blackburn}, {Blackman}, {Blair}, {Blair}, {Blair}, {Bloemen},
  {Bobba}, {Bode}, {Boer}, {Boetzel}, {Bogaert}, {Bondu}, {Bonnand}, {Booker},
  {Boom}, {Bork}, {Boschi}, {Bose}, {Bossilkov}, {Bosveld}, {Bouffanais},
  {Bozzi}, {Bradaschia}, {Brady}, {Bramley}, {Branchesi}, {Brau}, {Breschi},
  {Briant}, {Briggs}, {Brighenti}, {Brillet}, {Brinkmann}, {Brockill},
  {Brooks}, {Brooks}, {Brown}, {Brunett}, {Buikema}, {Bulik}, {Bulten},
  {Buonanno}, {Buskulic}, {Buy}, {Byer}, {Cabero}, {Cadonati}, {Cagnoli},
  {Cahillane}, {Calder{\'o}n Bustillo}, {Callister}, {Calloni}, {Camp},
  {Campbell}, {Canepa}, {Cannon}, {Cao}, {Cao}, {Carapella}, {Carbognani},
  {Caride}, {Carney}, {Carullo}, {Casanueva Diaz}, {Casentini}, {Caudill},
  {Cavagli{\`a}}, {Cavalier}, {Cavalieri}, {Cella}, {Cerd{\'a}-Dur{\'a}n},
  {Cesarini}, {Chaibi}, {Chakravarti}, {Chamberlin}, {Chan}, {Chao},
  {Charlton}, {Chase}, {Chassande-Mottin}, {Chatterjee}, {Chaturvedi},
  {Chatziioannou}, {Cheeseboro}, {Chen}, {Chen}, {Chen}, {Cheng}, {Cheong},
  {Chia}, {Chiadini}, {Chincarini}, {Chiummo}, {Cho}, \& {Cho}}]{GW190425}
---. 2020, \apjl, 892, L3, \dodoi{10.3847/2041-8213/ab75f5}

\bibitem[{Altaha~Motahar {et~al.}(2017)Altaha~Motahar, Bl\'azquez-Salcedo,
  Kleihaus, \& Kunz}]{AltahaMotahar:2017ijw}
Altaha~Motahar, Z., Bl\'azquez-Salcedo, J.~L., Kleihaus, B., \& Kunz, J. 2017,
  Phys. Rev. D, 96, 064046, \dodoi{10.1103/PhysRevD.96.064046}

\bibitem[{{Antoniadis} {et~al.}(2013){Antoniadis}, {Freire}, {Wex}, {Tauris},
  {Lynch}, {van Kerkwijk}, {Kramer}, {Bassa}, {Dhillon}, {Driebe}, {Hessels},
  {Kaspi}, {Kondratiev}, {Langer}, {Marsh}, {McLaughlin}, {Pennucci}, {Ransom},
  {Stairs}, {van Leeuwen}, {Verbiest}, \& {Whelan}}]{Antoniadis2013}
{Antoniadis}, J., {Freire}, P. C.~C., {Wex}, N., {et~al.} 2013, Science, 340,
  448, \dodoi{10.1126/science.1233232}

\bibitem[{{Archibald} {et~al.}(2018){Archibald}, {Gusinskaia}, {Hessels},
  {Deller}, {Kaplan}, {Lorimer}, {Lynch}, {Ransom}, \&
  {Stairs}}]{Archibald2018}
{Archibald}, A.~M., {Gusinskaia}, N.~V., {Hessels}, J. W.~T., {et~al.} 2018,
  \nat, 559, 73, \dodoi{10.1038/s41586-018-0265-1}

\bibitem[{{Baade} \& {Zwicky}(1934)}]{bz_1934}
{Baade}, W., \& {Zwicky}, F. 1934, Proceedings of the National Academy of
  Science, 20, 259, \dodoi{10.1073/pnas.20.5.259}

\bibitem[{{Bagchi} {et~al.}(2025){Bagchi}, {Abbate}, {Balakrishnan},
  {Colom~i~Bernadich}, {Bhattacharyya}, {Dutta}, {Freire}, {Halley}, {Hessels},
  {Kumari}, {Lorimer}, {Possenti}, {Nag}, {Ransom}, {Ridolfi},
  {Venkatraman~Krishnan}, {Zhu}, \& {{The SKA Pulsar Science Working
  Group}}}]{Bagchi2025_SKA_GlobClust}
{Bagchi}, M., {Abbate}, F., {Balakrishnan}, V., {et~al.} 2025

\bibitem[{{Balakrishnan} {et~al.}(2022){Balakrishnan}, {Champion}, {Barr},
  {Kramer}, {Venkatraman Krishnan}, {Eatough}, {Sengar}, \&
  {Bailes}}]{Balakrishnan:2022}
{Balakrishnan}, V., {Champion}, D., {Barr}, E., {et~al.} 2022, \mnras, 511,
  1265, \dodoi{10.1093/mnras/stab3746}

\bibitem[{{Barr} {et~al.}(2024){Barr}, {Dutta}, {Freire}, {Cadelano}, {Gautam},
  {Kramer}, {Pallanca}, {Ransom}, {Ridolfi}, {Stappers}, {Tauris}, {Venkatraman
  Krishnan}, {Wex}, {Bailes}, {Behrend}, {Buchner}, {Burgay}, {Chen},
  {Champion}, {Chen}, {Corongiu}, {Geyer}, {Men}, {Padmanabh}, \&
  {Possenti}}]{Barr2024}
{Barr}, E.~D., {Dutta}, A., {Freire}, P. C.~C., {et~al.} 2024, Science, 383,
  275, \dodoi{10.1126/science.adg3005}

\bibitem[{{Basu} {et~al.}(2025){Basu}, {Graber}, {Lower}, {Antonelli},
  {Antonopoulou}, {Bagchi}, {Char}, {Freire}, {Haskell}, {Hu}, {Jones},
  {Mukhopadhyay}, {Oertel}, {Rea}, {Sagun}, {Shaw}, {Singha}, {Stappers},
  {Thongmeearkom}, {Watts}, {Weltevrede}, \& {{The SKA Pulsar Science Working
  Group}}}]{Basu2025_SKA_EOS}
{Basu}, A., {Graber}, V., {Lower}, M.~E., {et~al.} 2025

\bibitem[{{Batrakov} {et~al.}(2024){Batrakov}, {Hu}, {Wex}, {Freire},
  {Venkatraman Krishnan}, {Kramer}, {Guo}, {Guillemot}, {McKee}, {Cognard}, \&
  {Theureau}}]{Batrakov2024}
{Batrakov}, A., {Hu}, H., {Wex}, N., {et~al.} 2024, \aap, 686, A101,
  \dodoi{10.1051/0004-6361/202245246}

\bibitem[{{Ben-Salem} \& {Hackmann}(2022)}]{Ben-Salem2022}
{Ben-Salem}, B., \& {Hackmann}, E. 2022, \mnras, 516, 1768,
  \dodoi{10.1093/mnras/stac2337}

\bibitem[{{Berti} {et~al.}(2015){Berti}, {Barausse}, {Cardoso}, {Gualtieri},
  {Pani}, {Sperhake}, {Stein}, {Wex}, {Yagi}, {Baker}, {Burgess}, {Coelho},
  {Doneva}, {Felice}, {Ferreira}, {Freire}, {Healy}, {Herdeiro}, {Horbatsch},
  {Kleihaus}, {Klein}, {Kokkotas}, {Kunz}, {Laguna}, {Lang}, {Li},
  {Littenberg}, {Matas}, {Mirshekari}, {Okawa}, {Radu}, {O'Shaughnessy},
  {Sathyaprakash}, {Broeck}, {Winther}, {Witek}, {Aghili}, {Alsing}, {Bolen},
  {Bombelli}, {Caudill}, {Chen}, {Degollado}, {Fujita}, {Gao}, {Gerosa},
  {Kamali}, {Silva}, {Rosa}, {Sadeghian}, {Sampaio}, {Sotani}, \&
  {Zilhao}}]{Berti2015}
{Berti}, E., {Barausse}, E., {Cardoso}, V., {et~al.} 2015, Classical and
  Quantum Gravity, 32, 243001, \dodoi{10.1088/0264-9381/32/24/243001}

\bibitem[{{Bertotti} {et~al.}(2003){Bertotti}, {Iess}, \&
  {Tortora}}]{Bertotti2003}
{Bertotti}, B., {Iess}, L., \& {Tortora}, P. 2003, \nat, 425, 374,
  \dodoi{10.1038/nature01997}

\bibitem[{{Blandford} \& {Teukolsky}(1976)}]{BT1976}
{Blandford}, R., \& {Teukolsky}, S.~A. 1976, \apj, 205, 580,
  \dodoi{10.1086/154315}

\bibitem[{{Breton} {et~al.}(2008){Breton}, {Kaspi}, {Kramer}, {McLaughlin},
  {Lyutikov}, {Ransom}, {Stairs}, {Ferdman}, {Camilo}, \&
  {Possenti}}]{Breton2008}
{Breton}, R.~P., {Kaspi}, V.~M., {Kramer}, M., {et~al.} 2008, Science, 321,
  104, \dodoi{10.1126/science.1159295}

\bibitem[{{Burgay} {et~al.}(2003){Burgay}, {D'Amico}, {Possenti}, {Manchester},
  {Lyne}, {Joshi}, {McLaughlin}, {Kramer}, {Sarkissian}, {Camilo}, {Kalogera},
  {Kim}, \& {Lorimer}}]{Burgay2003}
{Burgay}, M., {D'Amico}, N., {Possenti}, A., {et~al.} 2003, \nat, 426, 531,
  \dodoi{10.1038/nature02124}

\bibitem[{{Chatterjee} {et~al.}(2009){Chatterjee}, {Brisken}, {Vlemmings},
  {Goss}, {Lazio}, {Cordes}, {Thorsett}, {Fomalont}, {Lyne}, \&
  {Kramer}}]{Chatterjee2009}
{Chatterjee}, S., {Brisken}, W.~F., {Vlemmings}, W.~H.~T., {et~al.} 2009, \apj,
  698, 250, \dodoi{10.1088/0004-637X/698/1/250}

\bibitem[{{Cordes} {et~al.}(2004){Cordes}, {Kramer}, {Lazio}, {Stappers},
  {Backer}, \& {Johnston}}]{Cordes:2004}
{Cordes}, J.~M., {Kramer}, M., {Lazio}, T.~J.~W., {et~al.} 2004, \nar, 48,
  1413, \dodoi{10.1016/j.newar.2004.09.040}

\bibitem[{Corman \& East(2024)}]{Corman:2024vlk}
Corman, M., \& East, W.~E. 2024, Phys. Rev. D, 110, 084065,
  \dodoi{10.1103/PhysRevD.110.084065}

\bibitem[{{Damour} \& {Deruelle}(1986)}]{DD86}
{Damour}, T., \& {Deruelle}, N. 1986, Annales de L'Institut Henri Poincare
  Section (A) Physique Theorique, 44, 263

\bibitem[{Damour \& Esposito-Farese(1992)}]{Damour:1992ah}
Damour, T., \& Esposito-Farese, G. 1992, Phys. Rev. D, 46, 4128,
  \dodoi{10.1103/PhysRevD.46.4128}

\bibitem[{{Damour} \& {Esposito-Far{\`e}se}(1992)}]{Damour:1992PhRvD}
{Damour}, T., \& {Esposito-Far{\`e}se}, G. 1992, \prd, 46, 4128,
  \dodoi{10.1103/PhysRevD.46.4128}

\bibitem[{{Damour} \& {Esposito-Farese}(1993)}]{DEF93}
{Damour}, T., \& {Esposito-Farese}, G. 1993, \prl, 70, 2220,
  \dodoi{10.1103/PhysRevLett.70.2220}

\bibitem[{{Damour} \& {Sch{\"a}fer}(1991)}]{ds:1991}
{Damour}, T., \& {Sch{\"a}fer}, G. 1991, Journal of Mathematical Physics, 32,
  127, \dodoi{10.1063/1.529135}

\bibitem[{{Damour} \& {Taylor}(1992)}]{DamourTaylor1992}
{Damour}, T., \& {Taylor}, J.~H. 1992, \prd, 45, 1840,
  \dodoi{10.1103/PhysRevD.45.1840}

\bibitem[{{de Rham} {et~al.}(2017){de Rham}, {Deskins}, {Tolley}, \&
  {Zhou}}]{deRham:2017}
{de Rham}, C., {Deskins}, J.~T., {Tolley}, A.~J., \& {Zhou}, S.-Y. 2017,
  Reviews of Modern Physics, 89, 025004, \dodoi{10.1103/RevModPhys.89.025004}

\bibitem[{{de Rham} {et~al.}(2013){de Rham}, {Tolley}, \& {Wesley}}]{de:2013}
{de Rham}, C., {Tolley}, A.~J., \& {Wesley}, D.~H. 2013, \prd, 87, 044025,
  \dodoi{10.1103/PhysRevD.87.044025}

\bibitem[{{Deller} {et~al.}(2019){Deller}, {Goss}, {Brisken}, {Chatterjee},
  {Cordes}, {Janssen}, {Kovalev}, {Lazio}, {Petrov}, {Stappers}, \&
  {Lyne}}]{Deller19}
{Deller}, A.~T., {Goss}, W.~M., {Brisken}, W.~F., {et~al.} 2019, \apj, 875,
  100, \dodoi{10.3847/1538-4357/ab11c7}

\bibitem[{{Desvignes} {et~al.}(2019){Desvignes}, {Kramer}, {Lee}, {van
  Leeuwen}, {Stairs}, {Jessner}, {Cognard}, {Kasian}, {Lyne}, \&
  {Stappers}}]{Desvignes2019}
{Desvignes}, G., {Kramer}, M., {Lee}, K., {et~al.} 2019, Science, 365, 1013,
  \dodoi{10.1126/science.aav7272}

\bibitem[{{Ding} {et~al.}(2023){Ding}, {Deller}, {Stappers}, {Lazio}, {Kaplan},
  {Chatterjee}, {Brisken}, {Cordes}, {Freire}, {Fonseca}, {Stairs},
  {Guillemot}, {Lyne}, {Cognard}, {Reardon}, \& {Theureau}}]{Ding23}
{Ding}, H., {Deller}, A.~T., {Stappers}, B.~W., {et~al.} 2023, \mnras, 519,
  4982, \dodoi{10.1093/mnras/stac3725}

\bibitem[{Doneva \& Yazadjiev(2016)}]{Doneva:2016xmf}
Doneva, D.~D., \& Yazadjiev, S.~S. 2016, JCAP, 11, 019,
  \dodoi{10.1088/1475-7516/2016/11/019}

\bibitem[{{Doroshenko} \& {Kopeikin}(1995)}]{Doroshenko1995}
{Doroshenko}, O.~V., \& {Kopeikin}, S.~M. 1995, \mnras, 274, 1029,
  \dodoi{10.1093/mnras/274.4.1029}

\bibitem[{{Eardley}(1975)}]{Eardley1975}
{Eardley}, D.~M. 1975, \apjl, 196, L59, \dodoi{10.1086/181744}

\bibitem[{{Edwards} {et~al.}(2006){Edwards}, {Hobbs}, \& {Manchester}}]{ehm06}
{Edwards}, R.~T., {Hobbs}, G.~B., \& {Manchester}, R.~N. 2006, \mnras, 372,
  1549, \dodoi{10.1111/j.1365-2966.2006.10870.x}

\bibitem[{{Einstein}(1915)}]{Einstein1915}
{Einstein}, A. 1915, Sitzungsberichte der K\&ouml;niglich Preussischen Akademie
  der Wissenschaften, 844

\bibitem[{{Event Horizon Telescope Collaboration}(2019)}]{EHT2019_M87}
{Event Horizon Telescope Collaboration}. 2019, The Astrophysical Journal
  Letters, 875, L1, \dodoi{10.3847/2041-8213/ab0ec7}

\bibitem[{{Event Horizon Telescope Collaboration}(2022)}]{EHT2022_SgrA}
---. 2022, The Astrophysical Journal Letters, 930, L12,
  \dodoi{10.3847/2041-8213/ac6674}

\bibitem[{Everitt {et~al.}(2011)Everitt, DeBra, Parkinson, Turneaure, Conklin,
  Heifetz, Keiser, Silbergleit, Holmes, Kolodziejczak, Al-Meshari, Mester,
  Muhlfelder, Solomonik, Stahl, Worden, Bencze, Buchman, Clarke, Al-Jadaan,
  Al-Jibreen, Li, Lipa, Lockhart, Al-Suwaidan, Taber, \& Wang}]{Everitt2011}
Everitt, C. W.~F., DeBra, D.~B., Parkinson, B.~W., {et~al.} 2011, Phys. Rev.
  Lett., 106, 221101, \dodoi{10.1103/PhysRevLett.106.221101}

\bibitem[{{Faucher-Gigu{\`e}re} \& {Loeb}(2011)}]{fl10}
{Faucher-Gigu{\`e}re}, C.-A., \& {Loeb}, A. 2011, \mnras, 415, 3951,
  \dodoi{10.1111/j.1365-2966.2011.19019.x}

\bibitem[{{Ferdman} {et~al.}(2020){Ferdman}, {Freire}, {Perera}, {Pol},
  {Camilo}, {Chatterjee}, {Cordes}, {Crawford}, {Hessels}, {Kaspi},
  {McLaughlin}, {Parent}, {Stairs}, \& {van Leeuwen}}]{Ferdman2020}
{Ferdman}, R.~D., {Freire}, P.~C.~C., {Perera}, B.~B.~P., {et~al.} 2020, \nat,
  583, 211, \dodoi{10.1038/s41586-020-2439-x}

\bibitem[{{Finn} \& {Sutton}(2002)}]{Finn:2002}
{Finn}, L.~S., \& {Sutton}, P.~J. 2002, \prd, 65, 044022,
  \dodoi{10.1103/PhysRevD.65.044022}

\bibitem[{{Fonseca} {et~al.}(2014){Fonseca}, {Stairs}, \&
  {Thorsett}}]{Fonseca2014}
{Fonseca}, E., {Stairs}, I.~H., \& {Thorsett}, S.~E. 2014, \apj, 787, 82,
  \dodoi{10.1088/0004-637X/787/1/82}

\bibitem[{{Freire} {et~al.}(2012{\natexlab{a}}){Freire}, {Kramer}, \&
  {Wex}}]{fkw:2012}
{Freire}, P. C.~C., {Kramer}, M., \& {Wex}, N. 2012{\natexlab{a}}, Classical
  and Quantum Gravity, 29, 184007, \dodoi{10.1088/0264-9381/29/18/184007}

\bibitem[{{Freire} \& {Wex}(2024)}]{FreireWex2024}
{Freire}, P. C.~C., \& {Wex}, N. 2024, Living Reviews in Relativity, 27, 5,
  \dodoi{10.1007/s41114-024-00051-y}

\bibitem[{{Freire} {et~al.}(2012{\natexlab{b}}){Freire}, {Wex},
  {Esposito-Far{\`e}se}, {Verbiest}, {Bailes}, {Jacoby}, {Kramer}, {Stairs},
  {Antoniadis}, \& {Janssen}}]{Freire2012}
{Freire}, P. C.~C., {Wex}, N., {Esposito-Far{\`e}se}, G., {et~al.}
  2012{\natexlab{b}}, \mnras, 423, 3328,
  \dodoi{10.1111/j.1365-2966.2012.21253.x}

\bibitem[{{Guo} {et~al.}(2021){Guo}, {Freire}, {Guillemot}, {Kramer}, {Zhu},
  {Wex}, {McKee}, {Deller}, {Ding}, {Kaplan}, {Stappers}, {Cognard}, {Miao},
  {Haase}, {Keith}, {Ransom}, \& {Theureau}}]{Guo2021}
{Guo}, Y.~J., {Freire}, P.~C.~C., {Guillemot}, L., {et~al.} 2021, \aap, 654,
  A16, \dodoi{10.1051/0004-6361/202141450}

\bibitem[{Gupta {et~al.}(2021)Gupta, Herrero-Valea, Blas, Barausse, Cornish,
  Yagi, \& Yunes}]{Gupta:2021vdj}
Gupta, T., Herrero-Valea, M., Blas, D., {et~al.} 2021, Class. Quant. Grav., 38,
  195003, \dodoi{10.1088/1361-6382/ac1a69}

\bibitem[{{Hofmann} \& {M{\"u}ller}(2018)}]{LLR}
{Hofmann}, F., \& {M{\"u}ller}, J. 2018, Classical and Quantum Gravity, 35,
  035015, \dodoi{10.1088/1361-6382/aa8f7a}

\bibitem[{Hu {et~al.}(2020)Hu, Kramer, Wex, Champion, \& Kehl}]{Hu:2020ubl}
Hu, H., Kramer, M., Wex, N., Champion, D.~J., \& Kehl, M.~S. 2020, Mon. Not.
  Roy. Astron. Soc., 497, 3118, \dodoi{10.1093/mnras/staa2107}

\bibitem[{{Hu} {et~al.}(2022){Hu}, {Kramer}, {Champion}, {Wex},
  {Parthasarathy}, {Pennucci}, {Porayko}, {van Straten}, {Venkatraman
  Krishnan}, {Burgay}, {Freire}, {Manchester}, {Possenti}, {Stairs}, {Bailes},
  {Buchner}, {Cameron}, {Camilo}, \& {Serylak}}]{Hu2022}
{Hu}, H., {Kramer}, M., {Champion}, D.~J., {et~al.} 2022, \aap, 667, A149,
  \dodoi{10.1051/0004-6361/202244825}

\bibitem[{{Hulse} \& {Taylor}(1975)}]{ht75}
{Hulse}, R.~A., \& {Taylor}, J.~H. 1975, \apjl, 195, L51,
  \dodoi{10.1086/181708}

\bibitem[{Juli\'e {et~al.}(2025)Juli\'e, Pompili, \& Buonanno}]{Julie:2024fwy}
Juli\'e, F.-L., Pompili, L., \& Buonanno, A. 2025, Phys. Rev. D, 111, 024016,
  \dodoi{10.1103/PhysRevD.111.024016}

\bibitem[{Kleihaus {et~al.}(2016)Kleihaus, Kunz, Mojica, \&
  Zagermann}]{Kleihaus:2016dui}
Kleihaus, B., Kunz, J., Mojica, S., \& Zagermann, M. 2016, Phys. Rev. D, 93,
  064077, \dodoi{10.1103/PhysRevD.93.064077}

\bibitem[{{Kopeikin} \& {Sch{\"a}fer}(1999)}]{Kopeikin1999}
{Kopeikin}, S.~M., \& {Sch{\"a}fer}, G. 1999, \prd, 60, 124002,
  \dodoi{10.1103/PhysRevD.60.124002}

\bibitem[{{Kramer} {et~al.}(2004){Kramer}, {Backer}, {Cordes}, {Lazio},
  {Stappers}, \& {Johnston}}]{Kramer:2004}
{Kramer}, M., {Backer}, D.~C., {Cordes}, J.~M., {et~al.} 2004, \nar, 48, 993,
  \dodoi{10.1016/j.newar.2004.09.020}

\bibitem[{{Kramer} \& {Wex}(2009)}]{Kramer:2009}
{Kramer}, M., \& {Wex}, N. 2009, Classical and Quantum Gravity, 26, 073001,
  \dodoi{10.1088/0264-9381/26/7/073001}

\bibitem[{{Kramer} {et~al.}(2006){Kramer}, {Stairs}, {Manchester},
  {McLaughlin}, {Lyne}, {Ferdman}, {Burgay}, {Lorimer}, {Possenti}, {D'Amico},
  {Sarkissian}, {Hobbs}, {Reynolds}, {Freire}, \& {Camilo}}]{Kramer2006}
{Kramer}, M., {Stairs}, I.~H., {Manchester}, R.~N., {et~al.} 2006, Science,
  314, 97, \dodoi{10.1126/science.1132305}

\bibitem[{{Kramer} {et~al.}(2021){Kramer}, {Stairs}, {Manchester}, {Wex},
  {Deller}, {Coles}, {Ali}, {Burgay}, {Camilo}, {Cognard}, {Damour},
  {Desvignes}, {Ferdman}, {Freire}, {Grondin}, {Guillemot}, {Hobbs}, {Janssen},
  {Karuppusamy}, {Lorimer}, {Lyne}, {McKee}, {McLaughlin}, {M{\"u}nch},
  {Perera}, {Pol}, {Possenti}, {Sarkissian}, {Stappers}, \&
  {Theureau}}]{Kramer2021}
---. 2021, Physical Review X, 11, 041050, \dodoi{10.1103/PhysRevX.11.041050}

\bibitem[{Kulkarni {et~al.}(1993)Kulkarni, Hut, \& McMillan}]{khm93}
Kulkarni, S.~R., Hut, P., \& McMillan, S. 1993, 364, 421,
  \dodoi{10.1038/364421a0}

\bibitem[{{Kyutoku} {et~al.}(2019){Kyutoku}, {Nishino}, \&
  {Seto}}]{Kyutoku:2019}
{Kyutoku}, K., {Nishino}, Y., \& {Seto}, N. 2019, \mnras, 483, 2615,
  \dodoi{10.1093/mnras/sty3322}

\bibitem[{{Lai} \& {Rafikov}(2005)}]{Rafikov2005}
{Lai}, D., \& {Rafikov}, R.~R. 2005, \apjl, 621, L41, \dodoi{10.1086/429146}

\bibitem[{{Lau} {et~al.}(2020){Lau}, {Mandel}, {Vigna-G{\'o}mez}, {Neijssel},
  {Stevenson}, \& {Sesana}}]{Lau2020}
{Lau}, M. Y.~M., {Mandel}, I., {Vigna-G{\'o}mez}, A., {et~al.} 2020, MNRAS,
  492, 3061, \dodoi{10.1093/mnras/staa002}

\bibitem[{{Lazarus} {et~al.}(2016){Lazarus}, {Freire}, {Allen}, {Aulbert},
  {Bock}, {Bogdanov}, {Brazier}, {Camilo}, {Cardoso}, {Chatterjee}, {Cordes},
  {Crawford}, {Deneva}, {Eggenstein}, {Fehrmann}, {Ferdman}, {Hessels},
  {Jenet}, {Karako-Argaman}, {Kaspi}, {Knispel}, {Lynch}, {van Leeuwen},
  {Machenschalk}, {Madsen}, {McLaughlin}, {Patel}, {Ransom}, {Scholz},
  {Seymour}, {Siemens}, {Spitler}, {Stairs}, {Stovall}, {Swiggum},
  {Venkataraman}, \& {Zhu}}]{Lazarus2016ApJ}
{Lazarus}, P., {Freire}, P.~C.~C., {Allen}, B., {et~al.} 2016, \apj, 831, 150,
  \dodoi{10.3847/0004-637X/831/2/150}

\bibitem[{{Lense} \& {Thirring}(1918)}]{Lense1918}
{Lense}, J., \& {Thirring}, H. 1918, Physikalische Zeitschrift, 19, 156

\bibitem[{Liu {et~al.}(2014)Liu, Eatough, Wex, \& Kramer}]{Liu:2014uka}
Liu, K., Eatough, R.~P., Wex, N., \& Kramer, M. 2014, Mon. Not. Roy. Astron.
  Soc., 445, 3115, \dodoi{10.1093/mnras/stu1913}

\bibitem[{{Liu} {et~al.}(2014){Liu}, {Eatough}, {Wex}, \& {Kramer}}]{lewk14}
{Liu}, K., {Eatough}, R.~P., {Wex}, N., \& {Kramer}, M. 2014, 445, 3115,
  \dodoi{10.1093/mnras/stu1913}

\bibitem[{{Liu} {et~al.}(2020){Liu}, {Guillemot}, {Istrate}, {Shao}, {Tauris},
  {Wex}, {Antoniadis}, {Chalumeau}, {Cognard}, {Desvignes}, {Freire}, {Kehl},
  \& {Theureau}}]{Liu:2020}
{Liu}, K., {Guillemot}, L., {Istrate}, A.~G., {et~al.} 2020, \mnras, 499, 2276,
  \dodoi{10.1093/mnras/staa2993}

\bibitem[{Lorimer(2005)}]{LorimerStairsLRR}
Lorimer, D.~R. 2005, Living Reviews in Relativity, 8, 7,
  \dodoi{10.12942/lrr-2005-7}

\bibitem[{{Lorimer} \& {Kramer}(2004)}]{lk04}
{Lorimer}, D.~R., \& {Kramer}, M. 2004, {Handbook of Pulsar Astronomy}, Vol.~4
  (Cambridge: Cambridge University Press)

\bibitem[{{Lower} {et~al.}(2024){Lower}, {Kramer}, {Shannon}, {Breton}, {Wex},
  {Johnston}, {Bailes}, {Buchner}, {Hu}, {Venkatraman Krishnan}, {Blackmon},
  {Camilo}, {Champion}, {Freire}, {Geyer}, {Karastergiou}, {van Leeuwen},
  {McLaughlin}, {Reardon}, \& {Stairs}}]{Lower2024}
{Lower}, M.~E., {Kramer}, M., {Shannon}, R.~M., {et~al.} 2024, \aap, 682, A26,
  \dodoi{10.1051/0004-6361/202347857}

\bibitem[{{Lyne} {et~al.}(2004){Lyne}, {Burgay}, {Kramer}, {Possenti},
  {Manchester}, {Camilo}, {McLaughlin}, {Lorimer}, {D'Amico}, {Joshi},
  {Reynolds}, \& {Freire}}]{Lyne2004}
{Lyne}, A.~G., {Burgay}, M., {Kramer}, M., {et~al.} 2004, Science, 303, 1153,
  \dodoi{10.1126/science.1094645}

\bibitem[{Meng {et~al.}(2024)}]{Meng:2024yyh}
Meng, L., {et~al.} 2024, Astrophys. J., 966, 46,
  \dodoi{10.3847/1538-4357/ad381c}

\bibitem[{{Miao} {et~al.}(2019){Miao}, {Shao}, \& {Ma}}]{Miao:2019}
{Miao}, X., {Shao}, L., \& {Ma}, B.-Q. 2019, \prd, 99, 123015,
  \dodoi{10.1103/PhysRevD.99.123015}

\bibitem[{{Miao} {et~al.}(2021){Miao}, {Xu}, {Shao}, {Liu}, \& {Ma}}]{Miao2021}
{Miao}, X., {Xu}, H., {Shao}, L., {Liu}, C., \& {Ma}, B.-Q. 2021, \apj, 921,
  114, \dodoi{10.3847/1538-4357/ac1d48}

\bibitem[{{Miao} {et~al.}(2020){Miao}, {Zhao}, {Shao}, {Wex}, {Kramer}, \&
  {Ma}}]{Miao:2020}
{Miao}, X., {Zhao}, J., {Shao}, L., {et~al.} 2020, \apj, 898, 69,
  \dodoi{10.3847/1538-4357/ab9dfe}

\bibitem[{{Misner} {et~al.}(1973){Misner}, {Thorne}, \& {Wheeler}}]{mtw73}
{Misner}, C.~W., {Thorne}, K.~S., \& {Wheeler}, J.~A. 1973, {Gravitation} (San
  Francisco: W. H. Freeman and Company)

\bibitem[{{Nordtvedt}(1968)}]{Nordtvedt1968}
{Nordtvedt}, K. 1968, Physical Review, 169, 1014,
  \dodoi{10.1103/PhysRev.169.1014}

\bibitem[{{Nordtvedt}(1987)}]{Nordtvedt:1987}
---. 1987, \apj, 320, 871, \dodoi{10.1086/165603}

\bibitem[{{Oppenheimer} \& {Volkoff}(1939)}]{ov_1939}
{Oppenheimer}, J.~R., \& {Volkoff}, G.~M. 1939, Physical Review, 55, 374,
  \dodoi{10.1103/PhysRev.55.374}

\bibitem[{{Oswald} {et~al.}(2025){Oswald}, {Basu}, {Chakraborty}, {Joshi},
  {Lewandowska}, {Liu}, {Lower}, {Philippov}, {Song}, {Tarafdar},
  {van~Leeuwen}, {Watts}, {Weltevrede}, {Wright}, \& {{The SKA Pulsar Science
  Working Group}}}]{Oswald2025_MAG}
{Oswald}, L.~S., {Basu}, A., {Chakraborty}, M., {et~al.} 2025

\bibitem[{Pani {et~al.}(2011)Pani, Berti, Cardoso, \& Read}]{Pani:2011xm}
Pani, P., Berti, E., Cardoso, V., \& Read, J. 2011, Phys. Rev. D, 84, 104035,
  \dodoi{10.1103/PhysRevD.84.104035}

\bibitem[{{Penrose}(1979)}]{pen79}
{Penrose}, R. 1979, in General Relativity: An Einstein centenary survey, ed.
  {S.~W.~Hawking \& W.~Israel}, Vol.~1 (Cambridge; New York: Cambridge
  University Press), 581--638

\bibitem[{{Peters}(1964)}]{Pet64}
{Peters}, P.~C. 1964, Physical Review, 136, 1224,
  \dodoi{10.1103/PhysRev.136.B1224}

\bibitem[{Pfahl {et~al.}(2005)Pfahl, Podsiadlowski, \& Rappaport}]{ppr05}
Pfahl, E., Podsiadlowski, P., \& Rappaport, S. 2005, \apj, 628, 343

\bibitem[{{Psaltis} {et~al.}(2016){Psaltis}, {Wex}, \& {Kramer}}]{Psaltis2016}
{Psaltis}, D., {Wex}, N., \& {Kramer}, M. 2016, \apj, 818, 121,
  \dodoi{10.3847/0004-637X/818/2/121}

\bibitem[{{Rafikov} \& {Lai}(2006{\natexlab{a}})}]{Rafikov2006}
{Rafikov}, R.~R., \& {Lai}, D. 2006{\natexlab{a}}, \prd, 73, 063003,
  \dodoi{10.1103/PhysRevD.73.063003}

\bibitem[{{Rafikov} \& {Lai}(2006{\natexlab{b}})}]{Rafikov2006b}
---. 2006{\natexlab{b}}, \apj, 641, 438, \dodoi{10.1086/500346}

\bibitem[{{Ransom} {et~al.}(2003){Ransom}, {Cordes}, \&
  {Eikenberry}}]{Ransom2003}
{Ransom}, S.~M., {Cordes}, J.~M., \& {Eikenberry}, S.~S. 2003, \apj, 589, 911,
  \dodoi{10.1086/374806}

\bibitem[{{Ransom} {et~al.}(2014){Ransom}, {Stairs}, {Archibald}, {Hessels},
  {Kaplan}, {van Kerkwijk}, {Boyles}, {Deller}, {Chatterjee},
  {Schechtman-Rook}, {Berndsen}, {Lynch}, {Lorimer}, {Karako-Argaman}, {Kaspi},
  {Kondratiev}, {McLaughlin}, {van Leeuwen}, {Rosen}, {Roberts}, \&
  {Stovall}}]{Ransom2014}
{Ransom}, S.~M., {Stairs}, I.~H., {Archibald}, A.~M., {et~al.} 2014, \nat, 505,
  520, \dodoi{10.1038/nature12917}

\bibitem[{{Ridolfi} {et~al.}(2022){Ridolfi}, {Freire}, {Gautam}, {Ransom},
  {Barr}, {Buchner}, {Burgay}, {Abbate}, {Venkatraman Krishnan}, {Vleeschower},
  {Possenti}, {Stappers}, {Kramer}, {Chen}, {Padmanabh}, {Champion}, {Bailes},
  {Levin}, {Keane}, {Breton}, {Bezuidenhout}, {Grie{\ss}meier}, {K{\"u}nkel},
  {Men}, {Camilo}, {Geyer}, {Hugo}, {Jameson}, {Parthasarathy}, \&
  {Serylak}}]{Ridolfi2022}
{Ridolfi}, A., {Freire}, P.~C.~C., {Gautam}, T., {et~al.} 2022, \aap, 664, A27,
  \dodoi{10.1051/0004-6361/202143006}

\bibitem[{{Saffer} {et~al.}(2025){Saffer}, {Fonseca}, {Ransom}, {Stairs},
  {Lynch}, {Good}, {Masui}, {McKee}, {Meyers}, {Patil}, \& {Tan}}]{Saffer2025}
{Saffer}, A., {Fonseca}, E., {Ransom}, S., {et~al.} 2025, \apjl, 983, L20,
  \dodoi{10.3847/2041-8213/adc25e}

\bibitem[{S\"anger {et~al.}(2024)}]{Sanger:2024axs}
S\"anger, E.~M., {et~al.} 2024, arXiv e-prints,
  \dodoi{10.48550/arXiv.2406.03568}

\bibitem[{{Savchenko} {et~al.}(2017){Savchenko}, {Ferrigno}, {Kuulkers},
  {Bazzano}, {Bozzo}, {Brandt}, {Chenevez}, {Courvoisier}, {Diehl}, {Domingo},
  {Hanlon}, {Jourdain}, {von Kienlin}, {Laurent}, {Lebrun}, {Lutovinov},
  {Martin-Carrillo}, {Mereghetti}, {Natalucci}, {Rodi}, {Roques}, {Sunyaev}, \&
  {Ubertini}}]{GW170817_Integral}
{Savchenko}, V., {Ferrigno}, C., {Kuulkers}, E., {et~al.} 2017, \apjl, 848,
  L15, \dodoi{10.3847/2041-8213/aa8f94}

\bibitem[{Schiff(1960)}]{Schiff1960}
Schiff, L.~I. 1960, Phys. Rev. Lett., 4, 215, \dodoi{10.1103/PhysRevLett.4.215}

\bibitem[{{Schneider}(1990)}]{Schneider1990}
{Schneider}, J. 1990, \aap, 232, 62

\bibitem[{{Seymour} \& {Yagi}(2018)}]{SeymourYagi2018}
{Seymour}, B., \& {Yagi}, K. 2018, \prd, 98, 124007,
  \dodoi{10.1103/PhysRevD.98.124007}

\bibitem[{{Shao} {et~al.}(2013){Shao}, {Caballero}, {Kramer}, {Wex},
  {Champion}, \& {Jessner}}]{Shao:2013}
{Shao}, L., {Caballero}, R.~N., {Kramer}, M., {et~al.} 2013, Classical and
  Quantum Gravity, 30, 165019, \dodoi{10.1088/0264-9381/30/16/165019}

\bibitem[{Shao {et~al.}(2017)Shao, Sennett, Buonanno, Kramer, \&
  Wex}]{Shao:2017gwu}
Shao, L., Sennett, N., Buonanno, A., Kramer, M., \& Wex, N. 2017, Phys. Rev. X,
  7, 041025, \dodoi{10.1103/PhysRevX.7.041025}

\bibitem[{Shao \& Wex(2012)}]{Shao:2012eg}
Shao, L., \& Wex, N. 2012, Class. Quant. Grav., 29, 215018,
  \dodoi{10.1088/0264-9381/29/21/215018}

\bibitem[{{Shao} \& {Wex}(2012)}]{Shao:2012}
{Shao}, L., \& {Wex}, N. 2012, Classical and Quantum Gravity, 29, 21.5018,
  \dodoi{10.1088/0264-9381/29/21/215018}

\bibitem[{{Shao} \& {Wex}(2013)}]{Shao_Wex:2013}
---. 2013, Classical and Quantum Gravity, 30, 165020,
  \dodoi{10.1088/0264-9381/30/16/165020}

\bibitem[{Shao {et~al.}(2015)Shao, Wex, \& Kramer}]{Shao:2012xb}
Shao, L., Wex, N., \& Kramer, M. 2015, in {13th Marcel Grossmann Meeting on
  Recent Developments in Theoretical and Experimental General Relativity,
  Astrophysics, and Relativistic Field Theories}, 1704--1706,
  \dodoi{10.1142/9789814623995_0261}

\bibitem[{Shao {et~al.}(2018)Shao, Wex, \& Kramer}]{Shao:2018klg}
Shao, L., Wex, N., \& Kramer, M. 2018, Phys. Rev. Lett., 120, 241104,
  \dodoi{10.1103/PhysRevLett.120.241104}

\bibitem[{{Shao} {et~al.}(2020){Shao}, {Wex}, \& {Zhou}}]{Shao:2020PhRvD}
{Shao}, L., {Wex}, N., \& {Zhou}, S.-Y. 2020, \prd, 102, 024069,
  \dodoi{10.1103/PhysRevD.102.024069}

\bibitem[{{Shao} {et~al.}(2015){Shao}, {Stairs}, {Antoniadis}, {Deller},
  {Freire}, {Hessels}, {Janssen}, {Kramer}, {Kunz}, {Laemmerzahl}, {Perlick},
  {Possenti}, {Ransom}, {Stappers}, \& {van Straten}}]{Shao2015}
{Shao}, L., {Stairs}, I., {Antoniadis}, J., {et~al.} 2015, in Advancing
  Astrophysics with the Square Kilometre Array (AASKA14), 42,
  \dodoi{10.22323/1.215.0042}

\bibitem[{{Shapiro}(1964)}]{Shapiro1964}
{Shapiro}, I.~I. 1964, \prl, 13, 789, \dodoi{10.1103/PhysRevLett.13.789}

\bibitem[{Silva {et~al.}(2021)Silva, Holgado, C\'ardenas-Avenda\~no, \&
  Yunes}]{Silva:2020acr}
Silva, H.~O., Holgado, A.~M., C\'ardenas-Avenda\~no, A., \& Yunes, N. 2021,
  Phys. Rev. Lett., 126, 181101, \dodoi{10.1103/PhysRevLett.126.181101}

\bibitem[{{Sipior} {et~al.}(2004){Sipior}, {Portegies Zwart}, \&
  {Nelemans}}]{spn04}
{Sipior}, M.~S., {Portegies Zwart}, S., \& {Nelemans}, G. 2004, \mnras, 354,
  L49, \dodoi{10.1111/j.1365-2966.2004.08373.x}

\bibitem[{Stairs {et~al.}(2004)Stairs, Thorsett, \& Arzoumanian}]{Stairs2003}
Stairs, I.~H., Thorsett, S.~E., \& Arzoumanian, Z. 2004, Phys. Rev. Lett., 93,
  141101, \dodoi{10.1103/PhysRevLett.93.141101}

\bibitem[{{Stovall} {et~al.}(2018){Stovall}, {Freire}, {Chatterjee},
  {Demorest}, {Lorimer}, {McLaughlin}, {Pol}, {van Leeuwen}, {Wharton},
  {Allen}, {Boyce}, {Brazier}, {Caballero}, {Camilo}, {Camuccio}, {Cordes},
  {Crawford}, {Deneva}, {Ferdman}, {Hessels}, {Jenet}, {Kaspi}, {Knispel},
  {Lazarus}, {Lynch}, {Parent}, {Patel}, {Pleunis}, {Ransom}, {Scholz},
  {Seymour}, {Siemens}, {Stairs}, {Swiggum}, \& {Zhu}}]{Stovall:2018ApJ}
{Stovall}, K., {Freire}, P.~C.~C., {Chatterjee}, S., {et~al.} 2018, ApJL, 854,
  L22, \dodoi{10.3847/2041-8213/aaad06}

\bibitem[{{Taylor} {et~al.}(1979){Taylor}, {Fowler}, \& {McCulloch}}]{tfm79}
{Taylor}, J.~H., {Fowler}, L.~A., \& {McCulloch}, P.~M. 1979, \nat, 277, 437,
  \dodoi{10.1038/277437a0}

\bibitem[{{Taylor} \& {Weisberg}(1982)}]{TaylorWeisberg1982}
{Taylor}, J.~H., \& {Weisberg}, J.~M. 1982, \apj, 253, 908,
  \dodoi{10.1086/159690}

\bibitem[{{Taylor} \& {Weisberg}(1989)}]{TaylorWeisberg1989}
---. 1989, \apj, 345, 434, \dodoi{10.1086/167917}

\bibitem[{{Thrane} {et~al.}(2020){Thrane}, {Os{\l}owski}, \&
  {Lasky}}]{Thrane2020}
{Thrane}, E., {Os{\l}owski}, S., \& {Lasky}, P.~D. 2020, MNRAS, 493, 5408,
  \dodoi{10.1093/mnras/staa593}

\bibitem[{{Tolman}(1939)}]{tol_1939}
{Tolman}, R.~C. 1939, Physical Review, 55, 364, \dodoi{10.1103/PhysRev.55.364}

\bibitem[{{Venkatraman Krishnan} {et~al.}(2020){Venkatraman Krishnan},
  {Bailes}, {van Straten}, {Wex}, {Freire}, {Keane}, {Tauris}, {Rosado},
  {Bhat}, {Flynn}, {Jameson}, \& {Os{\l}owski}}]{VenkatramanKrishnan2020}
{Venkatraman Krishnan}, V., {Bailes}, M., {van Straten}, W., {et~al.} 2020,
  Science, 367, 577, \dodoi{10.1126/science.aax7007}

\bibitem[{{Voisin} {et~al.}(2020){Voisin}, {Cognard}, {Freire}, {Wex},
  {Guillemot}, {Desvignes}, {Kramer}, \& {Theureau}}]{Voisin2020}
{Voisin}, G., {Cognard}, I., {Freire}, P.~C.~C., {et~al.} 2020, \aap, 638, A24,
  \dodoi{10.1051/0004-6361/202038104}

\bibitem[{{Voisin} {et~al.}(2024){Voisin}, {Cognard}, {Saillenfest}, {Tauris},
  {Wex}, {Guillemot}, {Theureau}, {Freire}, \& {Kramer}}]{Voisin2024}
{Voisin}, G., {Cognard}, I., {Saillenfest}, M., {et~al.} 2024, arXiv e-prints,
  arXiv:2411.10066, \dodoi{10.48550/arXiv.2411.10066}

\bibitem[{{Voss} \& {Tauris}(2003)}]{vt03}
{Voss}, R., \& {Tauris}, T.~M. 2003, \mnras, 342, 1169,
  \dodoi{10.1046/j.1365-8711.2003.06616.x}

\bibitem[{{Watts} {et~al.}(2015){Watts}, {Espinoza}, {Xu}, {Andersson},
  {Antoniadis}, {Antonopoulou}, {Buchner}, {Datta}, {Demorest}, {Freire},
  {Hessels}, {Margueron}, {Oertel}, {Patruno}, {Possenti}, {Ransom}, {Stairs},
  \& {Stappers}}]{wxe+14}
{Watts}, A., {Espinoza}, C.~M., {Xu}, R., {et~al.} 2015, in Advancing
  Astrophysics with the Square Kilometre Array (AASKA14), 43,
  \dodoi{10.22323/1.215.0043}

\bibitem[{{Weisberg} \& {Huang}(2016)}]{WeisbergHuang2016}
{Weisberg}, J.~M., \& {Huang}, Y. 2016, \apj, 829, 55,
  \dodoi{10.3847/0004-637X/829/1/55}

\bibitem[{Wex \& Kopeikin(1999)}]{wk99}
Wex, N., \& Kopeikin, S. 1999, 514, 388

\bibitem[{{Wex} \& {Kopeikin}(1999)}]{WexKopeikin1999}
{Wex}, N., \& {Kopeikin}, S.~M. 1999, \apj, 514, 388, \dodoi{10.1086/306933}

\bibitem[{{Wex} \& {Kramer}(2007)}]{Wex:2007}
{Wex}, N., \& {Kramer}, M. 2007, \mnras, 380, 455,
  \dodoi{10.1111/j.1365-2966.2007.12093.x}

\bibitem[{{Wex} \& {Kramer}(2020)}]{Wex2020Univ}
---. 2020, Universe, 6, 156, \dodoi{10.3390/universe6090156}

\bibitem[{{Will}(1993{\natexlab{a}})}]{will93}
{Will}, C.~M. 1993{\natexlab{a}}, {Theory and Experiment in Gravitational
  Physics} (Cambridge: Cambridge University Press),
  \dodoi{10.1017/CBO9780511564246}

\bibitem[{{Will}(1993{\natexlab{b}})}]{Will1993}
---. 1993{\natexlab{b}}, {Theory and Experiment in Gravitational Physics}
  (Cambridge: Cambridge University Press), \dodoi{10.1017/CBO9780511564246}

\bibitem[{{Will}(2018)}]{Will2018}
---. 2018, {Theory and Experiment in Gravitational Physics} (Cambridge:
  Cambridge University Press), \dodoi{10.1017/9781316338612}

\bibitem[{Will(2018)}]{Will:2018bme}
Will, C.~M. 2018, {Theory and Experiment in Gravitational Physics} (Cambridge
  University Press)

\bibitem[{{Will} \& {Nordtvedt}(1972)}]{will1972}
{Will}, C.~M., \& {Nordtvedt}, Jr., K. 1972, \apj, 177, 757,
  \dodoi{10.1086/151754}

\bibitem[{Xu {et~al.}(2020)Xu, Gao, \& Shao}]{Xu:2020vbs}
Xu, R., Gao, Y., \& Shao, L. 2020, Phys. Rev. D, 102, 064057,
  \dodoi{10.1103/PhysRevD.102.064057}

\bibitem[{Yagi {et~al.}(2014{\natexlab{a}})Yagi, Blas, Barausse, \&
  Yunes}]{Yagi:2013ava}
Yagi, K., Blas, D., Barausse, E., \& Yunes, N. 2014{\natexlab{a}}, Phys. Rev.
  D, 89, 084067, \dodoi{10.1103/PhysRevD.89.084067}

\bibitem[{Yagi {et~al.}(2014{\natexlab{b}})Yagi, Blas, Yunes, \&
  Barausse}]{Yagi:2013qpa}
Yagi, K., Blas, D., Yunes, N., \& Barausse, E. 2014{\natexlab{b}}, Phys. Rev.
  Lett., 112, 161101, \dodoi{10.1103/PhysRevLett.112.161101}

\bibitem[{Yagi {et~al.}(2016)Yagi, Stein, \& Yunes}]{Yagi:2015oca}
Yagi, K., Stein, L.~C., \& Yunes, N. 2016, Phys. Rev. D, 93, 024010,
  \dodoi{10.1103/PhysRevD.93.024010}

\bibitem[{Yagi {et~al.}(2013)Yagi, Stein, Yunes, \& Tanaka}]{Yagi:2013mbt}
Yagi, K., Stein, L.~C., Yunes, N., \& Tanaka, T. 2013, Phys. Rev. D, 87,
  084058, \dodoi{10.1103/PhysRevD.87.084058}

\bibitem[{Yordanov {et~al.}(2024)Yordanov, Staykov, Yazadjiev, \&
  Doneva}]{Yordanov:2024lfk}
Yordanov, P.~Y., Staykov, K.~V., Yazadjiev, S.~S., \& Doneva, D.~D. 2024,
  Astron. Astrophys., 687, A17, \dodoi{10.1051/0004-6361/202449679}

\bibitem[{Yunes \& Spergel(2009)}]{Yunes:2008ua}
Yunes, N., \& Spergel, D.~N. 2009, Phys. Rev. D, 80, 042004,
  \dodoi{10.1103/PhysRevD.80.042004}

\bibitem[{{Yungelson} \& {Portegies Zwart}(1998)}]{yp98}
{Yungelson}, L., \& {Portegies Zwart}, S.~F. 1998, arXiv:astro-ph/9801127

\bibitem[{{Zhao} {et~al.}(2022){Zhao}, {Freire}, {Kramer}, {Shao}, \&
  {Wex}}]{Zhao2022}
{Zhao}, J., {Freire}, P. C.~C., {Kramer}, M., {Shao}, L., \& {Wex}, N. 2022,
  Classical and Quantum Gravity, 39, 11LT01, \dodoi{10.1088/1361-6382/ac69a3}

\bibitem[{Zhu {et~al.}(2019)}]{Zhu:2018etc}
Zhu, W.~W., {et~al.} 2019, Mon. Not. Roy. Astron. Soc., 482, 3249,
  \dodoi{10.1093/mnras/sty2905}

\end{thebibliography}

\end{document}